\documentclass[acmsmall,screen]{acmart}

\usepackage{graphicx}
\usepackage{float}
\usepackage{subfigure}
\usepackage{multirow}
\usepackage{makecell}

\usepackage{amssymb}
\usepackage{array}
\usepackage{threeparttable}
\usepackage{longtable}
\usepackage{geometry}

\newcommand{\hk}[1]{\textcolor{black}{#1}}

\newcommand{\hkr}[1]{\textcolor{black}{#1}}

\AtBeginDocument{%
  \providecommand\BibTeX{{%
    \normalfont B\kern-0.5em{\scshape i\kern-0.25em b}\kern-0.8em\TeX}}}

\setcopyright{acmcopyright}
\copyrightyear{2022}
\acmYear{2022}
\acmDOI{XXXXXXX.XXXXXXX}

\acmJournal{CSUR}
\acmVolume{37}
\acmNumber{4}
\acmArticle{111}
\acmMonth{8}

\acmSubmissionID{CSUR-2022-0516}

\begin{document}

\title{A Survey on Automated Program Repair Techniques}



\author{Kai Huang}
\email{huangk@nipc.org.cn}
\affiliation{%
  \institution{Xidian University}
  \city{Xi'an}
  \country{China}
}

\author{Zhengzi Xu}
\email{zhengzi.xu@ntu.edu.sg}
\affiliation{%
  \institution{Nanyang Technological University}
  \country{Singapore}
}

\author{Su Yang}
\email{yangs@nipc.org.cn}
\affiliation{%
  \institution{University of Chinese Academy of Sciences}
  \state{{Bejing}}
  \country{China}
}

\author{Hongyu Sun}
\email{sunhy@nipc.org.cn}
\affiliation{%
  \institution{Xidian University}
  \city{Xi'an}
  \country{China}
}

\author{Xuejun Li}
\email{aluckydd@mail.xidian.edu.cn}
\affiliation{%
  \institution{Xidian University}
  \city{Xi'an}
  \country{China}
}

\author{Zheng Yan}
\email{zyan@xidian.edu.cn}
\affiliation{%
  \institution{Xidian University}
  \city{Xi'an}
  \country{China}
}
\affiliation{%
  \institution{Aalto University}
  \state{Espoo}
  \country{Finland}
}

\author{Yuqing Zhang}
\authornote{Yuqing Zhang is the corresponding author.}
\email{zhangyq@nipc.org.cn}
\affiliation{%
  \institution{University of Chinese Academy of Sciences}
  \city{Beijing}
  \country{China}
}
\affiliation{%
  \institution{Xidian University}
  \city{Xi'an}
  \country{China}
}
\affiliation{%
  \institution{Hainan University}
  \city{Haikou}
  \country{China}
}

\thanks{This work was supported in part by the National Natural Science Foundation of China under Grant 62072351; in part by the Academy of Finland under Grant 345072 and Grant 350464.}

\renewcommand{\shortauthors}{Huang, et al.}

\begin{abstract}
With the rapid development and large-scale popularity of program software, modern society increasingly relies on software systems.
However, the problems exposed by software have also come to the fore. Software defect has become an important factor troubling developers. 
In this context, Automated Program Repair (APR) techniques have emerged, aiming to automatically fix software defect problems and reduce manual debugging work.
In particular, benefiting from the advances in deep learning, numerous learning-based APR techniques have emerged in recent years, which also bring new opportunities for APR research.
To give researchers a quick overview of APR techniques' complete development and future opportunities, we revisit the evolution of APR techniques and discuss in depth the latest advances in APR research. 
In this paper, the development of APR techniques is introduced in terms of four different patch generation schemes: search-based, constraint-based, template-based, and learning-based. 
Moreover, we propose a uniform set of criteria to review and compare each APR tool, summarize the advantages and disadvantages of APR techniques, and discuss the current state of APR development. Furthermore, we introduce the research on the related technical areas of APR that have also provided a strong motivation to advance APR development. 
Finally, we analyze current challenges and future directions, especially highlighting the critical opportunities that large language models bring to APR research.
\end{abstract}



\maketitle

\section{Introduction}
\label{sec:intro}

The defects in software development, testing, and maintenance processes are a widespread and costly category of serious problems in the software lifecycle, making software engineers spend a lot of time fixing software defects to ensure that the software works properly. 
Software defects have long caused significant and persistent harm to various aspects of human activities. 
For example, in 2017, attackers exploited the Eternal Blue series of vulnerabilities to create a widespread WannaCry ransomware incident~\cite{1}. 
Between 2018 and 2020, software design flaws at Boeing caused two Boeing 737 Max planes to crash and jeopardized a critical test flight of the Starliner spacecraft~\cite{2}. 
According to a recent report by the Consortium for Information \& Software Quality (CISQ)~\cite{3}, in 2020, the U.S. spent approximately \$2.08 trillion due to low-quality software, with software failures being the leading cause of the Cost of Poor Software Quality (CPSQ) (approximately \$1.56 trillion). 
Thus, the defect problem has attracted much attention. 
Software defects are mostly historical problems left over from the development process. Therefore, the impact of software defects can only be mitigated by testing and patching, which makes defect repair a tedious and complex task. 

Nowadays, the creation and use of automation tools have become a trend in industrial development.
People use automation tools in industry and academia to promote productivity and research. 
In this context, Automated Program Repair (APR)~\cite{4,5,6} techniques have emerged to shift the heavy manual software maintenance work to precise and efficient automated repair work.
Over the past decade, APR techniques have developed rapidly and received much attention from researchers.
Many APR works have been done around this research area in academia~\cite{18}. APR techniques have a broad impact covering software engineering (SE), system security (SS), and artificial intelligence (AI). 
Besides, APR research is also accompanied by many related research topics, such as APR evaluation~\cite{40}, fault localization~\cite{175}, patch assessment~\cite{202}, etc. 
These related works laid the foundation for the further development of APR.
In recent years, APR has also received attention from the industry. Companies such as Meta~\cite{9,10}, Fujitsu~\cite{11,13,14}, and Alibaba~\cite{15,16} are actively exploring the deployment of APR techniques in real industrial environments.

As a hot research field, APR techniques have been reviewed by many scholars~\cite{4,5,6}. 
Today, APR techniques are still evolving rapidly, with many data-driven APR efforts emerging in recent years in particular.
\hk{However, most prior surveys have not yet covered these recent technical advances.}
\hk{Therefore, our work aims to provide a more comprehensive survey to fill this gap.}
\hkr{Furthermore, we hope to revisit the development process of APR techniques to highlight the emerging trends in the data-driven era of APR.
Overall, unlike previous works, we focus on the latest research advances, particularly the critical role of deep learning in the APR field. Additionally, we propose a set of metrics to present and compare the differences between APR tools uniformly in multiple dimensions, which also reflects the continuous progress in the development of APR techniques. More importantly, we provide insightful discussions on recent APR techniques, highlighting potential problems and future technology trends.}

Next, the main work and contributions of this paper are shown below.

\begin{itemize}
\item{
\textbf{Systematic taxonomy and review:}
We classify APR techniques into four categories: search-based, constraint-based, template-based, and learning-based. 
Importantly, we propose a uniform set of criteria to review and compare APR tools at a fine-grained level (Section~\ref{sec:2-3-evaluation criteria}).
Then, we systematically review the development of APR techniques in time order (Section~\ref{sec:Search-based}-\ref{sec:Learning-based}).
Specially, we review the learning-based techniques in terms of different defect problems (i.e., bug, error, vulnerability), which were not covered in the previous works.
}

\item{
\textbf{Insightful summary and discussion:}
We summarize the advantages and disadvantages of each type of repair solution (Section~\ref{summary}). More importantly, we provide an in-depth discussion of advanced APR techniques (Section~\ref{discussion}). During the detailed discussion, we reveal many potential problems, such as unfair comparisons of experiments, dataset overlap, and dataset quality. Additionally, we analyze the latest technology trends, such as the critical role of large language model (LLM) in future APR research, exploring new paradigms of program repair, and focusing on both repair quality and efficiency.
}


\item{
\textbf{More forward-looking directional guidance:} 
We analyze the latest research progress and propose current pressing issues and forward-looking directions (Section~\ref{sec:Challenges and Directions}). 
In particular, we analyze the challenges of APR techniques for industrial applications, and in addition, we highlight future research opportunities for LLM and the necessity of more empirical studies. These were also not covered in detail in previous surveys.
}
\end{itemize}

The remainder of this article is organized as follows. 
Section~\ref{sec:paper select} describes the criteria and process of selecting papers for this survey.
Section~\ref{sec:Background} introduces the background knowledge of APR techniques. 
Section~\ref{sec:2-3-evaluation criteria} proposes a uniform set of evaluation criteria for reviewing and comparing APR tools.
Section~\ref{sec:APR techniques} provides an all-around review of the development of APR and summarizes and discusses the progress of APR techniques.
Section~\ref{sec:Related Work} introduces related research efforts on APR. Section~\ref{sec:Challenges and Directions} points out challenges and directions. Section~\ref{sec:Conclusion} concludes the whole paper.

\section{Paper Selection}
\label{sec:paper select}

\begin{figure*}
    \centering
    \includegraphics[width=1\textwidth]{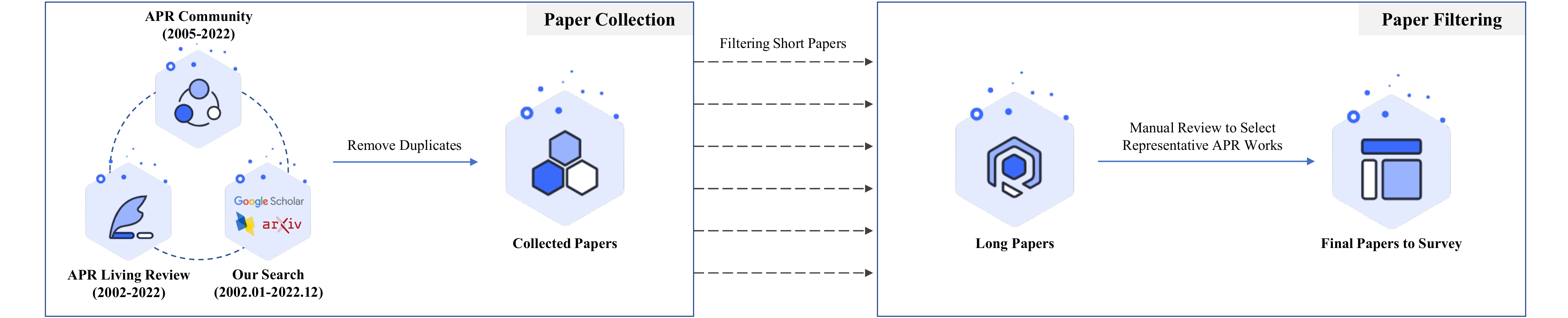}
    \caption{The workflow for conducting paper selection.}
    \label{fig:PaperSec}
    \vspace{-3ex}
\end{figure*}

This section describes the workflow (Fig.~\ref{fig:PaperSec}) for conducting paper selection. Details are as follows.

\textbf{Paper Collection}.
As this survey aims to revisit the development process of APR techniques, we needed to search for papers covering all phases of APR works. We would like to thank the APR community and Monperrus for their comprehensive collection of APR works, which provided us with direct access to relevant APR works. During the paper collection, we mainly focused on papers published up to December 2022. According to the statistics provided by the APR community\footnote{http://program-repair.org/}, 405 APR papers were published between 2005 and 2022. In the living review by Monperrus (August 2022)~\cite{18}, a total of 401 APR papers were included. Additionally, to follow recent research progress, we searched articles from Google Scholar, DBLP, and arXiv for the period of 2022.01-2022.12 using the keyword "Automated/Automatic Program Repair."
Finally, we aggregated and removed duplicates from the papers retrieved by the APR community, Monperrus living review, and our own search, forming the basis for this survey's paper selection.

\textbf{Paper Filtering}. 
Due to page limitations, we cannot review all APR works in detail, so we need to filter out some representative works for presentation. In the filtering process, we first excluded short papers of no more than 6 pages, and then we manually reviewed the remaining papers to select a few typical works. (Please note that the term "typical works" here refers to those that represent the characteristics of certain types of techniques in the development of APR or reflect the development trend of APR techniques. Due to page limitations and possible omissions from the manual review, it is impossible to cover all typical works in our survey. However, since our goal is to review the evolution of APR, it is sufficient to select some representative works for presentation.) Furthermore, given that traditional APR techniques have been reviewed in previous works~\cite{4,5,6} and that our focus is on data-driven APR techniques in recent years, we have retained more data-driven works in the manual selection process. Ultimately, we primarily selected \hkr{140} APR works for this paper, which will form the main part of our survey.

\section{APR General Process}
\label{sec:Background}

The concepts of bug, error, fault, vulnerability, etc., are all referred to as “defect” in a long study, so we will call them “defects” here for the sake of unity. 
After defining the defect problem, the details of APR in the fixing process still need to be thought about and analyzed. Most APR techniques use a generic repair framework process~\cite{40,41,51} (shown in Fig.~\ref{fig:fig6}): fault localization (FL), patch generation, and patch validation.
In the process of APR, a defective program is usually given, which can be a piece of code, and then handed over to the APR tool for repair, resulting in one or more patches. We briefly introduce the general repair process of typical APR techniques:

\begin{figure*}
    \centering
    \includegraphics[width=1\textwidth]{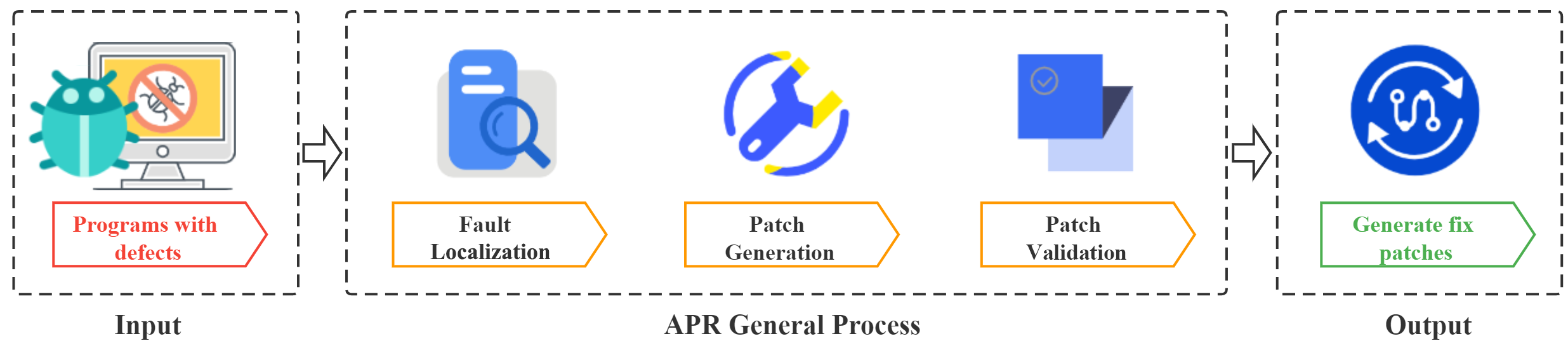}
    \caption{APR general process.}
    \label{fig:fig6}
    \vspace{-3ex}
\end{figure*}

{\textbf{Fault Localization:}} 
\hk{This step aims to find and locate defects in the program so that the APR tool can make exact fixes.
Researchers have developed many FL tools or methods~\cite{175}, and APR tools often draw on these off-the-shelf FL techniques to locate potential fault locations.}

\textbf{Patch Generation:} 
After obtaining the specific defect location,
APR tools use patch generation techniques to generate candidate fixes, where the “patch” can be a program patch in the actual sense or a specific suggestion to guide the fix. 

\textbf{Patch Validation:} A large number of candidate fixes will be generated in the patch generation stage, but not all candidate fixes can repair defects. 
Some candidate fixes may be incorrect or invalid patches.
The patch validation phase aims to detect them (i.e., candidate patches) and find plausible patches and correct patches~\cite{13}.


Currently, FL techniques~\cite{175} are known as a separate research area. While in APR techniques, patch generation and validation (G\&V) are usually the main focus of researchers. 
In particular, patch generation techniques are at the core of the APR tool. In view of this, this paper will adopt the techniques used in the patch generation phase to classify APR research.

\section{Evaluation Criteria for APR Tools}
\label{sec:2-3-evaluation criteria}

In this section, we propose a uniform set of evaluation criteria for reviewing and comparing different APR tools (see Table~\ref{tab:table23}).
This set of criteria reflects APR tools in four dimensions: 
\textbf{Approach Design}, \textbf{Repair Ability}, \textbf{Repair Quality}, and \textbf{Repair Efficiency}.
These four dimensions contain 12 evaluation metrics, 
\hkr{which can reflect the similarities and differences between different APR tools from different perspectives.}
It should be noted that the first two dimensions (i.e., \textbf{Approach Design} and \textbf{Repair Ability}) analyze APR tools from a qualitative perspective, and the last two dimensions (i.e., \textbf{Repair Quality} and \textbf{Repair Efficiency}) analyze APR tools from a quantitative perspective. 
Since this paper is a survey rather than an empirical study, we analyze the comparison of APR tools only from a qualitative perspective. Our work is to summarize these obvious features from the existing literature. 
In addition, because APR tools often use different test benchmarks and few APR tools explore repair efficiency, we are unable to provide a uniform view of these quantitative metrics.
As for the quantitative analysis, this should be done by dedicated empirical studies. And they should conduct large-scale experiments to reflect these data details. Also, these quantitative metrics provided in this paper can provide an important reference for subsequent APR work.

\begin{table}[]
\caption{A uniform set of evaluation criteria on APR techniques}
\label{tab:table23}
\tiny
\resizebox{1.0\columnwidth}{!}{
\begin{tabular}{|p{3.0cm}|p{10.7cm}|}
\hline
\multicolumn{1}{|c|}{\textbf{4 Dimensions}} &
  \multicolumn{1}{c|}{\textbf{12 Metrics}} \\ \hline
\multirow{3}{3.0cm}{\textbf{Approach Design}. 
APR techniques aim to automate the resolution of software defects,
and APR tools need to have the high level of automation to reduce human workload. The differences in the approach design of each APR tool reflect the degree of advancement in technical details.
Therefore, we would like to reflect the superiority of APR tools at the technical level by showing the details of their approach design.
Overall, we use three metrics to reflect the differences in the approach design.} &
  \textbf{(1) Context-Aware (CA)}. This metric is used to indicate whether the APR tool employs the context-aware patch generation strategy. Specifically, approaches that are context-aware take into account the contextual information of the source code when generating fixes, which allows them to generate patches that work better with the original code and avoid introducing side effects. This results in higher repair success rate as well as higher repair quality. \\ \cline{2-2} 
 &
  \textbf{(2) Fault Localization (FL)}. This metric is used to indicate whether the APR tool is specifically designed and implemented with the FL function. Because some APR tools integrate FL in their design process, they are specifically designed or improved for FL, which allows the repair process to be free from dependency on external FL tools. On the other hand, many APR tools focus only on the patch G\&V process, and they tend to use existing FL tools or directly use the assumption of perfect fault localization, which makes it difficult for them to effectively link localization and repair closely. The quality of FL will also affect the quality of the final fix. 
  Therefore, we use whether the APR tool integrates a specialized FL function as a factor to reflect differences in approach design.
  \\ \cline{2-2} 
 &
  \textbf{(3) Empirical Knowledge (EK)}. This metric is used to indicate whether the APR tool is able to automate the acquisition (learning) of empirical knowledge for defect repair. For traditional heuristics, such as GP-based APR tools that generate patch spaces by means of mutations, etc., such methods do not inherently learn the empirical knowledge used for defect repair. Constraint-based or template-based APR tools may need to manually define a set of constraint rules or repair templates, which are the empirical knowledge to guide defect repair. The manual extraction of the empirical knowledge still means that the tool does not have the ability to automate learning. Other tools, such as learning-based data-driven APR tools, can automate the learning of empirical knowledge from a large number of repair samples, 
  \hkr{which means that they can continuously enhance the repair potential through automated learning.}
  \\ \hline
\multirow{4}{3.0cm}{\textbf{Repair Ability}. The repair ability of an APR tool is used to indicate the size or applicability of the repair scope of the tool on defect repair problems. APR tools aim to solve defect problems of different types, languages, scales, etc. Different tools may target different repair goals at the beginning of their design, so their repair ability will vary. 
Here, we reflect the repair ability of APR tools in four aspects.} &
  \textbf{(4) Multi-Type (MT)} Fixes. This metric reflects whether the APR tool supports generic type (multi-type) defect repair. Some APR tools may be designed for specific defect type fixes, and they do not extend to a wider range of other types of defect fixing problems, so there is a limit to their repair ability. Some APR tools, on the other hand, can address multiple types of defects, or even most of the general defect types. For example, data-driven APR tools can solve a wider range of defects after learning about multiple types of defect fixes, so they have better scalability and generalizability. \\ \cline{2-2} 
 &
  \textbf{(5) Multi-Language (ML)} Fixes. This metric reflects whether the APR tool supports multiple programming languages (PLs). Some APR tools work only on a single programming language, and they cannot be extended to fix problems in other languages. Thus, their repair ability is limited by the type of language. On the other hand, some APR tools support fixes for multiple programming languages. For example, most data-driven APR tools can be extended to support multiple programming languages, which will further enhance the overall repair ability of the tool. 
  \hkr{Note that since most APR techniques (approaches) are language agnostic, their specific implementations (APR tools) can be designed for different PLs. For example, the original implementation of GenProg~\cite{62} was evaluated on C, and its alternative implementation, jGenProg~\cite{ASTOR}, supports Java language. Therefore the multi-language (or multi-lingual) repair ability we discuss here is explored only for specific APR tools.}
  \\ \cline{2-2} 
 &
  \textbf{(6) Multi-Hunk (MH)} Fixes. Also called Multi-Line/Edit/Location/Statement fixes~\cite{13,DEAR,112,VulRepair,197}. This metric reflects whether the APR tool can resolve more complex multi-hunk defects. Different APR tools differ in their repair ability. Some APR tools do not scale to large-scale program repair problems; they only support simple single-line fixes, which limits their repair ability in real-world environments. Some APR tools can fix more complex multi-hunk bugs and even handle longer input sequences, which allows them to scale to complex program repair problems and have greater repair ability. \\ \cline{2-2} 
 &
  \textbf{(7) Multi-Fault (MF)} Fixes. This metric reflects whether the APR tool can effectively handle programs that contain multiple faults at the same time. Since multiple faults may be contained in the same program at the same time \cite{66}, the occurrence of multiple faults may affect each other, which makes it difficult for some APR tools to address such multiple-fault program fixes problems. Therefore, APR tools that are well enough to handle this class of more complex repair problems can be considered to have stronger repair ability. \\ \hline
\multirow{3}{3.0cm}{\textbf{Repair Quality}. A good APR tool should generate high-quality fixes, which can be understood in several ways.} &
  \textbf{(8) Repair Reliability (RR)}. This metric reflects the likelihood of APR tools introducing new defects during the repair process. Since APR tools are highly susceptible to introducing new defects during the repair process \cite{77}, the quality of such a repair is a cause for concern. A high-quality fix should not only fix the current defect, but also avoid introducing new defects \cite{62}. Therefore, we can evaluate the repair quality in terms of the number of new defects introduced during the repair process. \\ \cline{2-2} 
 &
  \textbf{(9) Patch Acceptability (PA)}. (maintainability~\cite{77}, understandability~\cite{74}, or interpretability). This metric reflects the degree to which the patches generated by the APR tool are accepted by developers \cite{73}. That is, whether the generated patches are closer to the quality of human writing or are easily accepted by developers. The literature \cite{77} describes that program repairs should resemble those done by humans, and therefore, repairs should not reduce maintainability. If an APR tool generates a patch that is obscure and overly complex and redundant, then the patch may have performance differences from a human-written patch or even be difficult to integrate into the original system. Instead, a high-quality patch should be more accessible to developers, work with the entire development process, be easier to maintain, and even make the generated fixes interpretable. \\ \cline{2-2} 
 &
  \textbf{(10) Patch Effectiveness (PE)}. This metric reflects the degree to which APR tools are effective in fixing defects. Currently, test case based APR tools suffer from patch overfitting. These APR tools produce plausible patches (patches that pass all test cases, but are not necessarily the correct patches, and they may be completely ineffective at fixing defects). Developers need to further sift through these plausible patches to get the correct ones that actually work. Therefore, the higher the percentage of correct patches / plausible patches generated by a high-quality APR tool, the more correct patches are generated. We consider the patches generated by the APR tool to be relatively high quality. For data-driven APR tools, they tend to evaluate repair performance on large-scale datasets, which are usually not equipped with complete test cases, so such tools can use repair accuracy to measure their effectiveness. \\ \hline
\multirow{2}{3.0cm}{\textbf{Repair Efficiency}. Automated repair should consider the repair cost issue. 
Here we divide the repair efficiency into time cost and money cost.} &
  \textbf{(11) Time Cost (TC)}. This metric reflects the time cost spent by the APR tool to repair the defect \cite{62}. A good APR tool should generate patches in a limited time. 
  We believe that the shorter the time spent by an APR tool to fix a defect, the lower the time cost will be, i.e., the higher the repair efficiency.
  \\ \cline{2-2} 
 &
  \textbf{(12) Money Cost (MC)}. This metric reflects the money cost spent by the APR tool to repair the defect. We propose to deploy these APR tools under a unified experimental environment to fairly compare the monetary cost behind fixing defects with these APR tools, e.g., some works deploy APR tools in an online cloud environment to evaluate the money cost spent on fixing defects \cite{127,71}. \\ \hline
\end{tabular}
}
\end{table}

\section{APR Techniques Research Progress}
\label{sec:APR techniques}

\begin{figure*}
    \centering
    \includegraphics[width=1\textwidth]{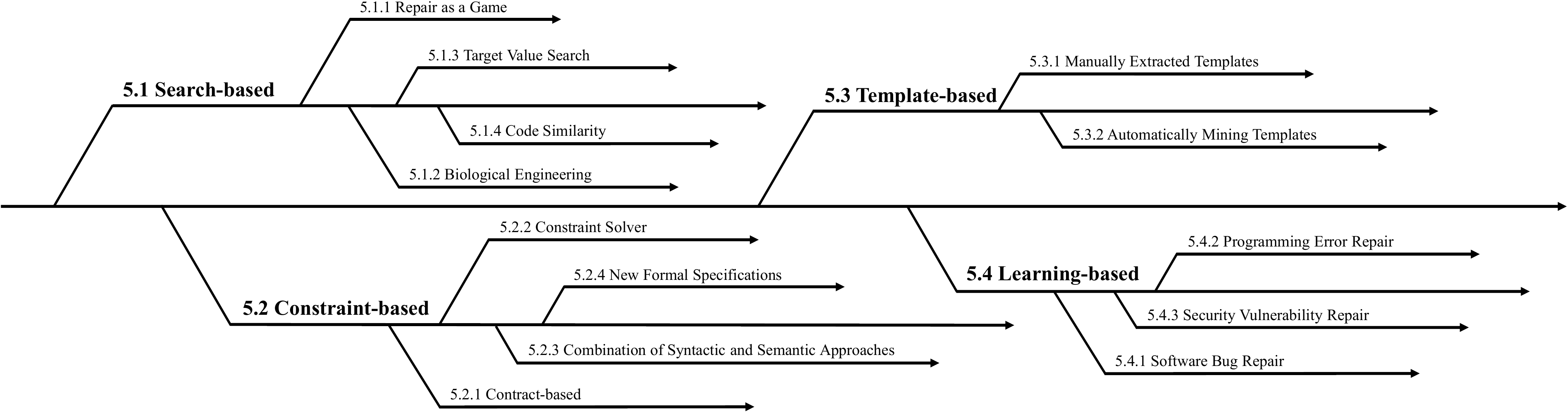}
    \caption{A taxonomy of APR techniques.}
    \label{fig:fig7}
\end{figure*}

We use the approach of the patch generation phase as the basis for APR techniques taxonomy. 
We combine the latest APR research~\cite{41,49,50,51} and classify APR techniques into four categories: search-based, constraint-based, template-based, and learning-based. We provide a taxonomy and overview of the development process of the four types of APR techniques in Fig.~\ref{fig:fig7}. 
\hkr{Here, we analyze these APR works based on the qualitative perspective proposed in Section~\ref{sec:2-3-evaluation criteria},
and Table~\ref{tab:table2-3-2} provides a uniform view to present the differences in approach design and repair ability of different APR tools.}
Next, we will review the development of four types of APR techniques in time order and finally summarize the pros and cons of these techniques and discuss 
the current state of APR development.

\begin{table}[]
\caption{A summary and comparison of different APR tools}
\label{tab:table2-3-2}
\resizebox{1.0\columnwidth}{!}{
} \\ \hline
\end{tabular}
}
\begin{tablenotes}
\tiny
\item{
Note that the term "APR Tool" refers to the specific implementation of the APR technique in the experimental evaluation phase of the cited paper.}
\end{tablenotes}
\end{table}

\subsection{Search-based}
\label{sec:Search-based}

\hk{Search-based methods are also known as heuristics solutions.}
The core idea of search-based techniques is to search for the correct patch in a predefined patch space. 
The techniques first look in the search space to identify the most likely locations of buggy code in the program, usually with the help of heuristics to generate candidate fixes, and then search for suitable fix patches by means of mutation-selection, test execution, traversal strategies, and etc. 
Heuristic methods emerged as early as 2005~\cite{56}. 
Later, the seminal GenProg~\cite{62} used genetic programming (GP) to generate program variants. The recent SimFix~\cite{57} used advanced code search for patch generation. 
\hk{Next, we will categorize and review these APR works according to the search strategies used.}


\subsubsection{Repair as a Game}
\label{Repair as a Game}

In early APR techniques, researchers proposed a heuristic approach to automatically repair faults in a finite state program by treating the repair problem as a game. Back in 2005, Jobstmann \textit{et al.}~\cite{56} proposed a method for program repair as a game strategy, which consists of a modified version of the program and an automaton representing a linear time logic (LTL) specification. 
Each winning finite state strategy of the game corresponds to a successful repair, and the opposite case represents a failure strategy if the repair is unsuccessful. 
In 2006, Griesmayer \textit{et al.}~\cite{59} extended the work~\cite{56} by showing how to find and repair errors in Boolean programs. This approach has been applied to fix Windows device drivers by SLAM as an abstraction.

Early APR works treat repair as a game and explore the repair problem from a novel perspective. But, they revealed shortcomings such as inefficient fixes, limited types of fixes, and repetitive fixing behavior. 
Overall, early works did not focus on the details of the approach design, and they have significant limitations in repair ability. They do not support multi-type, multi-language fixes, and it remains unclear whether they can address complex multi-hunk fixes as well as multi-fault fixes.

\subsubsection{Biological Engineering}
\label{Biological Engineering}

In the history of APR techniques, bioengineering-based ideas have been widely used in repair work, especially evolutionary computation (EC)~\cite{60} and genetic programming~\cite{48,61}. 
EC is a guided stochastic search method that draws on the mechanisms and terminology of biological evolutionary processes. 
GP is a learning algorithm inspired by biological evolution that mimics the operations of variation and crossover in biological evolution and selects the most suitable individual program for a specific task by computing an adaptation function. 
In 2012, the representative GP-based work GenProg~\cite{62} was born. 
GenProg uses a fine-grained abstract syntax tree (AST) level search.
Genetic operators were also localized to statements executed on failed test cases, alleviating the problem of too large a search space making it difficult to find the correct fix. 
In the final experiments, successful fixes were achieved on 16 C programs and eight types of defects. 
In short, GenProg supports multi-type fixes and has designed a weighted path-based FL scheme. 
However, \hkr{this APR tool} does not support multi-language fixes, and it also does not explore the multi-hunk fixes.
By 2014, GP has been widely used in APR~\cite{48,63,68,69,66,60,25,62,70,71,72}. However, the search efficiency of GP methods is still a challenge.
In 2014, Qi \textit{et al.}~\cite{76} proposed RSRepair.
RSRepair uses random search and discards the failed test cases when it encounters them, which means it executes fewer test cases to speed up the processing. 
Final results show that RSRepair outperforms GenProg in repair effectiveness and efficiency.
Besides, RSRepair supports multi-type fixes. However, this APR tool does not explore multi-language fixes and multi-hunk fixes, and it uses GenProg's FL scheme directly.


In addition to the GP and EC, in 2015, Stelios \textit{et al.}~\cite{78} proposed an approach based on gene transfer - Code Phage (CP). 
CP transfers the correct code from donor program to recipient program, thus eliminating defects in the recipient program. 
However, it relies on the code provider,
which prevents code migration fixes if a defect lacks correct representation in other systems. 
Final results show that it can fix multiple types of faults but does not support multi-language fixes. It is also not shown whether it can fix multi-hunk and multi-fault, and no FL scheme is designed.

In 2017, Yuan \textit{et al.}~\cite{47} proposed ARJA. 
It improves the effectiveness of GP repair by improving the search strategy. 
They viewed APR as a multi-objective optimization problem, and developed the optimization in terms of the size of patch-to-code changes and the number of patch tests. 
Later, in 2019, they added the template-based idea (see Section~\ref{sec:Template-based}) to ARJA and proposed ARJA-e~\cite{217}. 
ARJA-e takes from PAR~\cite{73} which extends seven templates, allowing ARJA-e to show greater advantages than a single repair approach.
Notably, they support multi-line fixes and multi-type fixes, and ARJA-e employs a light-weight contextual analysis strategy.
However, they do not support multi-language and use existing FL tools. It is unclear whether they can support multi-fault fixes.

In 2018, Wen \textit{et al.}~\cite{83} proposed CapGen, the first context-aware based patch generation technique. This technique makes it improve search efficiency and repair quality.
CapGen also supports multi-type fixes, but cannot handle multi-hunk fixes and multi-language fixes. Moreover, it does not focus on FL design, but directly uses off-the-shelf FL solutions.
By 2020, Wen \textit{et al.}~\cite{58} proposed a new repair strategy by learning how bugs are introduced (\textit{bug-inducing commits}, BICs). 
By inferring from a large number of variation operators and repair components needed to fix the bug, 
such an approach outperforms CapGen.
This suggests that incorporating \textit{bug-inducing commits} is an effective direction to facilitate APR research.


Finally, it should be added that none of the work reviewed in this section has automated learning capabilities from the approach design perspective. They cannot automatically learn the empirical knowledge to guide defect repair.
In addition, most works do not explicitly state whether they can support complex multi-hunk and multi-fault fixes.

\subsubsection{Target Value Search}
\label{Target Value Search}

In 2015, Long \textit{et al.}~\cite{52} present the staged program repair (SPR) strategy.  
SPR uses three key techniques: parameterized transformation schemas, target value search, and conditional synthesis. 
According to these three techniques, SPR can be classified into template-based, search-based, and constraint-based, respectively. 
\hk{Since this section focus on the search strategy, we discuss its target value search.}
In repairing, given a parametrized transformation schema, SPR uses the target value search to quickly determine if there are any parameter values that would allow the schema to generate a successful fix.
Unlike prior works~\cite{62,71}, 
which only reuse statements from the program, SPR defines a novel set of modification operators that allow SPR to run on a search space containing meaningful and useful fixes.
The contribution of SPR is the proposal of staged program repair to selectively perform space search, thus improving the search efficiency.
Notably, SPR presents its own FL algorithm and supports multi-type fixes. However, it does not support multi-language fixes, nor does it have automated learning capabilities. Its experimental part does not state whether it supports multi-hunk and multi-fault fixes.

\subsubsection{Code Similarity}
\label{Code Similarity}

The code similarity-based (or redundancy-based~\cite{87}) techniques emerged to solve the search space problem~\cite{88}. 
By reusing these codes, repair techniques can avoid generating a large amount of code and thus reduce the search space. 
Among them, both syntactic (e.g., ssFix~\cite{82}) and semantic (e.g., SearchRepair~\cite{89}) search solutions were generated, and we will describe the semantic constraint-based approach in Section~\ref{sec:Constraint-based}. 
Then later, work has also emerged on reasoning about code similarity through deep learning (e.g., DeepRepair~\cite{87}), which will not be expanded here (see Section~\ref{DL for bug}). Here, we focus on heuristic syntactic search methods. 

In 2016, Ji \textit{et al.}~\cite{91} proposed SCRepair. 
They proposed a reusability metric for similar code fragments used for program repair. By combining similarity at the AST level, the reusability metric can help to select the most suitable candidate fixes. 
This work laid the foundation for later work.
In 2017, Wang \textit{et al.}~\cite{92} proposed CRSearcher. 
CRSearcher uses the token-based code similarity calculation method. 
It searches similar code from code repositories and recommends code snippets as repair suggestions. 
Specifically, the experimental part of CRSearcher also analyzes the repair efficiency.
The results show that CRSearcher does help to generate high-quality patches and reduce the repair time
(about 50\% of the repair time cost savings).
In short, both of the above works can repair multiple types of defects but do not support multi-language fixes. It is not clear whether multi-hunk and multi-fault fixes are supported. And they both directly use existing FL strategies.

In 2017, Xin \textit{et al.}~\cite{82} proposed ssFix.
Unlike the token-based CRSearcher, ssFix identifies the bug statement and context as a code block for similarity search. 
These improvements reduced search space and generate high-quality fixes.
In terms of repair ability, ssFix is consistent with the previous SCRepair and CRSearcher. However, it considers code context to guide code search and patch generation, so it is superior in approach design.

Previous works help repair by relying on patch information or code information. 
Specifically, some works~\cite{73,54,125} analyze existing patches to select those most likely modifications to guide fixes; others~\cite{62,89} analyze source code to search for fix components used to generate patches (more details can refer to~\cite{57}).
Combining the two may allow exploring the search space more effectively.
In 2018, Jiang \textit{et al.}~\cite{57} proposed SimFix.
They first mined a search space from the patch information and explored the search space further with the help of code similarity.
Finally, SimFix presents a superior performance than ssFix~\cite{82}. 
SimFix fixes 14 more errors than ssFix~\cite{82}.
And in terms of repair ability, SimFix shows potential for multi-line bug fixes. It also supports multi-type fixes but does not extend to multi-language fixes. And it also does not declare whether it supports multi-fault fixes. In terms of approach design, it directly uses the existing FL algorithm and does not emphasize context-awareness.

In 2019, Xin \textit{et al.}~\cite{88} proposed sharpFix, which follows the basic ideas of ssfix~\cite{82} 
but with improvements in code search and code reuse.
It uses statement-level code blocks for local search and method-level comparisons for global search. 
Besides, it uses token-level matching compared to the inflexible approach of ssFix, which uses rule and tree matching.
Overall, sharpFix and ssFix are identical in approach design and repair ability, 
but the improvements in sharpFix can fix more bugs.

In 2021, Wong \textit{et al.}~\cite{90} used similarity information from patch tests to speed up the search, and they applied variational execution to propose an APR tool, VarFix. 
By observing which combinations of edit operations pass the test, the combinations of edits that satisfy all the test-passing conditions represent a reasonable patch. Experiments show that VarFix can fix multi-line and multi-type bugs, and performs well in large-scale programs.
However, it does not support multi-language fixes and has not explored multi-fault fixing capabilities.
In terms of approach design, it directly uses GenProg's FL strategy and does not focus on context-aware capabilities.

Finally, it should be added that while the work reviewed in this section draws on existing knowledge bases (code bases) to guide the fixes, they do not automate the learning of empirical knowledge used for defect fixes. They only make use of this knowledge to help with the repair; the tools themselves do not have the ability to learn the knowledge.

\subsubsection{Summary}
\label{Search Summary}

The search-based methods are an early class of APR techniques, among which bioengineering ideas have had a profound impact. 
This class of methods generates suitable candidate fixes by establishing search spaces and developing search strategies.
In the process, many researchers have proposed a series of improvements to GP and how to establish efficient patch search strategies.
At the same time, these methods have also exposed some problems, such as the explosive nature of the search space and the \textit{patch overfitting problem} (see Section~\ref{Patch Overfitting Problem}). 
Besides, due to the uncertainty of mutation, the validity of the patches generated is not guaranteed, and the generation of invalid and incorrect patches will increase the system overhead. 
These limit the quality and efficiency of APR.
The core of the similarity information-based approach is how to compute code similarity to find corresponding fixes. 
And it has undergone a series of similarity elements in its development from the token level only to code segment context information to patch information, which inspired the later template-based (see Section~\ref{sec:Template-based}) approach. 
It exhibits a similar problem to the template approach in that if similarity information cannot be found (no repair pattern can be found), no valid repair can be generated. 
In summary, the heuristic search approach has been a great impetus in the APR process, which has provided preliminaries for the generation of other repair approaches.

\subsection{Constraint-based}
\label{sec:Constraint-based}

The core idea of constraint-based (or semantic constraint) approaches guides the repair process by developing a set of constraint specifications.
Such techniques transform the program repair problem into a constraint solver problem and use formal specifications to quickly prune infeasible parts to find expression-level variations that satisfy constraints collected from tests by program analysis, thus facilitating patch generation. 
For example, Nopol~\cite{106} and SemFix~\cite{98} use test cases as implicit program specifications to guide the patch synthesis process.
\hk{Next, we will review these APR works according to different forms of constraints.}


\subsubsection{Contract-based}
\label{Contract-based}

Between 2010 and 2014, the contract-based AutoFix project emerged~\cite{99,100,101}. Contracts provide a set of correct behavior specifications.
AutoFix project guides the FL and repair process by using contracts. 
AutoFix-E~\cite{99} uses dynamic execution to build an abstract model to guide the repair. 
AutoFix-E2~\cite{100} adds static analysis by analyzing the expressions in the program. 
AutoFix~\cite{101} combines AutoFix-E and AutoFix-E2 to improve flexibility and generality. 
Here we discuss the latest AutoFix, which not only combines static and dynamic analysis to design FL strategy but also supports multi-type fixes. However, it does not support multi-language fixes, and it is not clear whether it can fix multi-hunk and multi-fault issues.

Although the contract-based technique has achieved effective fixes, it still has limitations in flexibility. 
This method requires developers to write the correct contract and only works on the Eiffel language.
To address the limitation, in 2017, Chen \textit{et al.}~\cite{102,103} proposed JAID. 
JAID is able to construct abstract states from the source code, thus replacing the original contract and supporting the Java. 
These improvements make JAID more adaptable than the AutoFix project, eliminating the tedious manual annotation of contracts and extending the scope of fixes to mainstream programming languages. 
In addition, JAID shows the potential for multi-line fixes compared to AutoFix. However, in terms of approach design, the contract-based approach cannot automate learning to empirical knowledge and does not focus on the context-aware patch generation strategy.

\subsubsection{Constraint Solvers}
\label{Constraint Solvers}

Boolean satisfiability (SAT) is a problem of checking whether a propositional logic formula can be evaluated as true. Satisfiability modulo theories (SMT) generalize Boolean satisfiability.
In these theories, SMT solvers are tools used to determine whether formulas are satisfiable or not. 
This approach is also used in APR techniques to perform program constraints.


There are some test-based (test-driven or test case driven) works that use test cases to generate constraints.
In 2013, Hoang \textit{et al.}~\cite{98} proposed SemFix to fix single-line bugs. 
They use symbolic execution to generate repair constraints.
In 2015, the team proposed DirectFix~\cite{21}. 
DirectFix optimizes the patch generation process by using constraint solving and component-based program synthesis. 
It has the ability to fix multi-line bugs, but at the expense of some scalability.
In 2016, they proposed another work Angelix~\cite{112}.
It uses a novel lightweight constraint (i.e., angelic forest) that addresses the scalability issue of DirectFix while retaining the multi-line bug fixing capability, which allows for scaling to large programs.
In general, the above works support multi-type fixes, but they only work on C programs, and it is not clear if they can support multi-fault fixes. They are also not concerned with context-aware strategies in approach design. 
Note that DirectFix innovatively integrates fault localization and repair into a single step to provide simple fixes.

In 2016, Xuan \textit{et al.}~\cite{106} proposed Nopol for fixing if-conditional bugs using SMT.
Nopol builds on SemFix~\cite{98}, they are both test-driven solutions, but they have significant differences.
Instead of symbolic execution, Nopol uses value replacement.
Crucially, Nopol shows potential for large-scale program repair.
Note that Nopol does not support multi-type, multi-language, multi-hunk, and multi-fault fixes; it is designed for specific bug types.
In approach design, while Nopol is not concerned with context-awareness, it uses value replacement to design the FL function.

Besides, the semantic code similarity scheme has also been widely used.
In 2015, 
SearchRepair~\cite{89} encoded codebase data into SMT constraints, and it uses constraint solving and semantic code search to find potential fixes.
It supports multi-type fixes but does not extend to multi-language fixes, and it is not clear the ability on multi-hunk and multi-fault fixes. In terms of approach design, it uses spectrum-based FL techniques.
In 2019, SOSRepair~\cite{110} performed semantic code search at a finer granularity level and was ahead of prior work~\cite{89} in repair effectiveness.
Compared to SearchRepair, SOSRepair uses a context-aware patch generation strategy and can apply it to large real-world programs.



\subsubsection{Combination of Syntactic and Semantic Approaches}
\label{Combination of Syntactic and Semantic Approaches}

In the development of APR, both heuristic search and semantic constraint methods have produced effective fixes for defects. 
\hk{Consequently, combining these two types of approaches to facilitate fixes became a useful research direction.}
In 2016, Le \textit{et al.}~\cite{113} combined syntactic (GP) and semantic (deductive verification) approaches. 
Later in 2017, they presented S3~\cite{114}. It uses syntactic and semantic features to guide patch generation, 
which helps generate high-quality patches and mitigate the \textit{patch overfitting problem}.
S3 supports multi-type and multi-line fixes, but it only focuses on Java bugs and does not explore multi-fault fixes.
In 2021, Campos \textit{et al.}~\cite{35} implemented a VSCode extension pAPRika. It uses a mutation-based solution~\cite{66} and uses unit tests as constraints to restrict these mutations.
This work provides a new solution by developing real-time IDE plug-ins that can quickly generate fixes.


\subsubsection{New Formal Specifications}
\label{New formal specifications}

\hk{Prior works~\cite{106,98,21,112} used test cases to generate constraints to guide patch generation, and this solution is known as the test-driven approach.}
But this approach can produce overfitting patches and does not apply to bugs that are difficult to test deterministically. 
To address this challenge, researchers have proposed formal specifications to guide repair~\cite{101,111,117,118}, which can be contracts, behavioral specifications, deductive synthesis, etc.
In recent years, some new approaches to formal specification also attracted attention.

In 2018, Rijnard \textit{et al.}~\cite{119} proposed FootPatch.
It uses Separation Logic~\cite{120} to build formal specifications.
The significance of this work is that the use of static analysis techniques frees APR from the constraints of test cases, making it possible to find more potential defects and fix them. 
Note that FootPatch is designed to fix faults related to general pointer safety properties, it supports multi-language and multi-line fixes.
But it is not clear whether it can fix multi-fault programs. It is also not concerned with context-aware strategies and uses a existing static analysis-based FL tool.

In 2019, Nguyen \textit{et al.}~\cite{122} also followed the formal specification and proposed Maple.
It first generates template patches and then generates constraints for these template fixes and performs constraint solving.
The addition of template-based ideas further improves repair efficiency.
In general, Maple supports multi-type fixes and has designed a formal verification FL strategy. However, it does not support multi-language fixes and is not applied in multi-hunk/fault fix scenarios.


In 2021, Gao \textit{et al.}~\cite{196} applied constraint-based techniques to C vulnerability repair by proposing the APR tool ExtractFix. Unlike bug fixing tasks, vulnerability repair usually lacks complete test suites, so using test cases to guide fault localization and constraint synthesis is challenging. Based on this, ExtractFix uses execution information of vulnerability exploits to synthesize constraints, alleviating the patch overfitting problem caused by weak test cases. In addition, ExtractFix guides FL by combining control/data dependency analysis, which provides an effective FL strategy for real fix scenarios. In addition, it supports multi-hunk fixes, but it does not explore multi-fault fixes.

\subsubsection{Summary}
\label{Constraint Summary}

Constraint-based approaches improve the repair quality~\cite{112,114,102,103,110} by developing constraint rules to transform the patch generation problem into a constraint solving problem to reduce search space
and generate a single or limited number of candidate fixes. 
Through the review, we note that a lot of innovations have been made in semantic works. Some works proposed syntactic-semantic combination~\cite{113,114,35}, some freed APR from test cases by using formal specifications~\cite{119,122,196}, and some proposed real-time APR tools~\cite{35}. All of which were revolutionary in the APR process and opened new research approaches. 
Compared with search-based heuristics, constraint-based methods do not require extensive search and therefore have much lower computational resources and costs and higher repair efficiency. 
However, such methods rely heavily on the formulated specification constraints to guide the program repair, so it lacks flexibility.
In addition, constraint-based APR tools still cannot automate the acquisition of empirical knowledge for software defect repair.

\subsection{Template-based}
\label{sec:Template-based}

In the history of APR, heuristic search methods have developed mature solutions, but they have huge search spaces that seriously affect the repair efficiency and make it costly to fix defects. 
The method of semantic constraints also suffers shortcomings, usually unable to establish a suitable constraint specification for each type of bug, and the reliance on constraints makes it exist many limitations.
In order to alleviate these problems, template-based APR techniques were born. 
Template-based methods utilize a predefined program fix template~\cite{73} (also known as "fix pattern"~\cite{44} or "transformation schema"~\cite{124}) to generate repair patches. 
The fix templates may be manually extracted (e.g., kPAR~\cite{43}) or automatically mined (e.g., HDRepair~\cite{125}). 
After FL, the template-based APR tool will target these defects to select the corresponding fix templates to generate candidate patches. 
In a recent empirical study~\cite{51}, it was shown that template-based repair tools are the most effective compared to search-based and constraint-based schemes.
\hk{Next, we will review these efforts according to different template extraction schemes.}


\subsubsection{Manually Extracted Templates}
\label{sec:Manually Extracted Templates}

In its early development history, templates were almost always summarized manually. 
The method was explored by Kim’s team back in 2009~\cite{126}. 
Later in 2013, they presented Pattern-based Automatic program Repair (PAR)~\cite{73}. 
They first examined patches in open-source projects, and created ten repair templates. 
These templates are used as automatic program editing scripts to guide specific repair operations.
Experimental results demonstrate that PAR can generate more acceptable patches than GenProg~\cite{62,70}. 
\hkr{This technique alleviates the problem that GP solutions may generate nonsensical patches~\cite{73} and opens up a new path for APR.}

In 2015, Tan \textit{et al.}~\cite{128} proposed relifix, a repair tool using software change history to solve regression errors. 
They obtained a set of code transformations from manual inspection of 73 real software regressions. 
Here, this set of manually extracted code transformations can be understood as templates for program repair.
Unlike PAR’s approach of extracting templates from patches, relifix’s templates (code transformations) are syntax information extracted from program versions and test execution histories. 
And they open the history-driven template-based approaches.

The above works are very similar regarding repair ability and solution design; none of them support multi-language fixes, nor do they explore multi-hunk, multi-fault repair capabilities. Moreover, relifix is only used to repair specific defect types. These works use existing FL tools directly and do not focus on context-aware patch generation strategies. Notably, these early works relied on manual extraction of repair templates and were not equipped with automated learning capabilities.

\subsubsection{Automatically Mining Templates}
\label{Automatically Generated Templates}

To improve the efficiency of template generation, 
automated template mining started to emerge. 
In 2016, HDRepair~\cite{125}
used graph mining techniques to automate the extraction of fix patterns from project histories. 
The contribution of HDRepair is to pioneer automated pattern extraction. 
Note that while HDRepair claims to fix multi-line bugs, it does not explore multi-line repair ability.

By 2017, automated pattern generation had entered a phase of further optimization. 
Increasing the number of template types will help to fix a broader range of bug types. 
Long \textit{et al.}~\cite{54} proposed Genesis.
Compared to PAR~\cite{73}, which used only ten manually refined templates, Genesis inferred 108 code transformations from 577 sampled transformations. 
In addition, Genesis has designed the FL algorithm and supports multi-type fixes. However, it does not explore multi-hunk and multi-fault fixes, nor does it focus on context-aware strategies.
Meanwhile, pattern mining has been explored in the work of Liu \textit{et al.}~\cite{129}. 
They proposed to use of convolutional neural network (CNN) to learn features and combine X-means clustering algorithm to generate repair templates.
The significance of this work lies in the use of deep learning techniques to advance the pattern mining process.
Later, their work AVATAR~\cite{44} incorporates this approach into APR tools for automated repair.
Compared to Genesis, AVATAR uses existing FL tools.
Although partially fixes 11 bugs that have multiple faulty code fragments, it does not discuss multi-hunk fixes ability in depth.

In 2018, Koyuncu \textit{et al.}~\cite{42} proposed FixMiner, an automated method for mining repair patterns, and implemented it in an APR tool, PARFixMiner. 
It uses a context-aware patch generation strategy and supports multi-type fixes. The tool does not focus on FL techniques and multi-language fixes, and it is unclear if it can support multi-hunk and multi-fault fixes.
The method captures code changes and contextual information at the AST level. 
Compared with previous works, FixMiner mines richer characterization information, making PARFixMiner have better performance.

In 2019, Koyuncu \textit{et al.}~\cite{134} developed iFixR, an APR tool driven by bug reports. 
iFixR builds on kPAR~\cite{43} (a template-based APR tool) and uses its repair templates.
Specially, it uses bug reports as the main input to guide program repair. However, there are shortcomings. The quality of the bug reports will directly affect patch generation, and low-quality or even incorrect bug reports may threaten the final fixing results.
Similar to iFixR~\cite{134},
Liu \textit{et al.}~\cite{45} also summarized 35 repair templates from previous template-based works and proposed an APR tool TBar.
They found that the selection of repair patterns is important. A single fault may apply to multiple repair templates. It may generate multiple patches that seem reasonable but have differing performance, so prioritizing the selected patterns avoids this problem and prevents overfitting from occurring. 
In general, iFixR and TBar are similar in approach design. They both employ the context-aware patch generation strategy and use existing FL tools directly. In terms of repair ability, they both support multi-type fixes, work only on Java, and do not explore multi-hunk and multi-fault repair capabilities.


Note that the repair templates can be viewed as empirical knowledge to guide repair. It means that the above automated template mining efforts can be automated to learn empirical knowledge. There are two exceptions, iFixR~\cite{134} and TBar~\cite{45}, as they directly use templates from other tools.
In addition, multi-hunk and multi-language fixes have received attention in the development of template-based techniques, which we will describe next.

\hkr{
Given that the above template-based efforts are not specifically designed for multi-hunk bug fixing scenarios.
Therefore, some researchers have launched targeted research for multi-hunk bug fixing. In 2019, Saha \textit{et al.}~\cite{13} proposed the APR tool HERCULES, a multi-hunk repair solution. They still followed the template-based technical foundation. The innovation of this work is that it associates multiple bug locations to achieve simultaneous fixes, which is a capability not available in the previous multi-hunk repair tool, Angelix~\cite{112}. In summary, HERCULES has made an important contribution to APR research by furthering the mitigation of the multi-hunk repair problem, which will accelerate the process of moving APR tools from academic research to industrial deployment.}

\hkr{Specifically, multi-language repair has also been investigated in traditional APR techniques. In 2019, Ghanbari \textit{et al.}~\cite{166} proposed PraPR, a repair tool that supports multiple JVM languages. Unlike previous works that mostly perform the repair at the source code level, PraPR works at the bytecode level. This approach allows PraPR to support multiple JVM languages and has higher scalability. Overall, bytecode-level repair techniques provide a new way for APR research, which helps to free the repair behavior from the source code level to explore deeper root causes of defects and extend to multi-language repair scenarios.}

\subsubsection{Summary}
\label{Template Summary}

By combing the development of template-based APR techniques, it has evolved from the early manual extraction to the current automated mining.
Different data sources (such as patch information~\cite{73}, project history~\cite{125}, defect report~\cite{134}, etc.) are used in the template generation approach. 
In patch generation strategy optimization, factors such as template donor quality and template priority order~\cite{45} are also considered. 
These are all effective ways to improve the quality of fixes. 
Notably, template-based techniques have made great contributions to the field by alleviating the problems of search space explosion, invalid patch generation, and patch generation efficiency. 
However, template extraction is affected by noisy data, and there are also problems, such as the trade-off between template granularity and defect type coverage. 
Another research dilemma of template-based approaches is that they cannot fix defects beyond the basic patterns due to the lack of continuous learning capability. 
Such approaches tend to focus on fixing more general, repetitive defect classes.
On the other hand, they neglect to focus on complex, special, non-universal defect classes. 
Future work can continue to improve template mining, selecting and matching strategies, and extending the template-based approach to repair more defect classes. 


\subsection{Learning-based}
\label{sec:Learning-based}

With the above review, template-based APR techniques show great potential in that they can automate template mining to obtain empirical knowledge (repair templates) used to guide defect repair.
And early learning-based approaches are also based on template-based methods. They set up probabilistic models on the distribution of repair patterns to guide better selective fix patterns, such as early works by Martinez \textit{et al.}~\cite{138}, Long \textit{et al.}~\cite{139}, and Saha \textit{et al.}~\cite{11}. 
\hkr{However, the addition of probabilistic models only optimizes the detailed steps (i.e., template selection and matching), and these formalized repair templates do not mean that template-based methods learn high-level repair knowledge. 
It is crucial to advance the automation of the repair process and the learning of high-level repair knowledge. The emergence of deep learning technologies provides the possibility of automating the learning and repair implementation.}
\hkr{With the help of neural network models, APR tools can automate the learning of empirical knowledge from a large number of defect repair samples that can be used to guide defect repair. For this reason, learning-based APR techniques are also known as data-driven repair solutions.}
Learning-based APR techniques have become the hottest avenue of research today. 
\hkr{Unlike traditional APR techniques (Section~\ref{sec:Search-based}-\ref{sec:Template-based}) that focus more on software bug fixing, researchers have explored different types of defect problems (e.g., software bugs, programming errors, and security vulnerabilities) separately in learning-based techniques.}
\hk{Next, we will review these SOTA works from the perspective of different defect problems.}

\subsubsection{Software Bug Repair}
\label{DL for bug}

\hkr{Software bugs have been a major concern in software engineering for a long time, and researchers hope to address this class of defects in real-world software projects with the help of learning-based APR techniques. Software bugs are mostly logical defect problems that require specific test cases for defect triggering and patch validation. Therefore, researchers typically use two metrics to evaluate specific fixes: for test benchmarks with test suites, "\textit{correct patches/plausible patches}" is often used to measure fixes; for large-scale datasets without test suites, the \textit{repair accuracy} 
 (prediction fix exact match correct fix) is used to measure the repair results.}

The development of data-driven bug-fixing techniques was first inspired by tasks related to natural language processing (NLP). Researchers viewed repair tasks as translation tasks, i.e., translating a buggy statement into a correct statement.
Back in 2018, Tufano \textit{et al.}~\cite{145} used Neural Machine Translation (NMT) techniques to fix bugs.
They extracted bug codes and fix codes to obtain bug-fix pairs (BFPs). Then they build an NMT model to learn repair knowledge from BFPs.
This work had a profound impact as a pioneer in learning-based bug repair techniques, introducing the NMT task framework for program repair tasks and laying the foundation for later learning-based APR works.

In 2019, Chen \textit{et al.}~\cite{49} proposed SequenceR, which still follows NMT techniques.
Specifically, it uses the copy mechanism to overcome the large vocabulary problem. 
Moreover, it motivates the model to focus on and distinguish between bug code and contextual content for targeted fixes by adding special markers to bug locations. 
This idea, in turn, opens the door to context-aware repair techniques.
In addition, in 2019, work emerged to address specific bugs and give birth to hybrid technical solutions. 
Vasic \textit{et al.}~\cite{149} proposed a solution for the variable-misuse bug.  
This work utilizes a multi-headed pointer network that can be promoted to locate and fix bugs simultaneously. 
White \textit{et al.}~\cite{87} proposed DeepRepair. It relies on deep learning to infer code similarity to select repair ingredients from code fragments.
Overall, all three efforts focus on bug fixes, but they have many differences.
Vasic \textit{et al.} focus on specific bug types, while DeepRepair and SequenceR support multi-type fixes. In addition, Vasic \textit{et al.} jointly localize and repair, which is an end-to-end solution. Whereas DeepRepair and SequenceR do not focus on FL, they are a patch generation strategy. Similarly, all three works are not designed for multi-hunk and multi-fault fixes.

In 2020, DL techniques continued to fuel the APR, researchers began to explore the limitations.

Li \textit{et al.}~\cite{136} found that DL-based APR tools still have limitations. 
Strict code structure features can easily be overlooked.
So, they proposed DLFix.
Unlike the representational form of sequences used by SequenceR~\cite{49}, DLFix uses tree (AST) to record the code structure.
This allows DLFix to work better considering the context and produces better fixes than previous NMT approaches.
Besides, Lutellier \textit{et al.}~\cite{137} proposed CoCoNuT, which uses CNNs to extract hierarchical features, which helps to capture rich (multiple levels of) semantic information. 
Moreover, it uses a new context-aware NMT architecture, which allows it to handle buggy code and context separately, as DLFix does.
CoCoNut also applies ensemble learning, further improving the repair performance beyond prior works~\cite{49,136}. 
However, these tools are limited to fixing single-hunk bugs.

Data-driven solutions use many repair samples to drive the repair. 
Whether this process can incorporate domain knowledge to better guide the repair, with this in mind, Huq \textit{et al.}~\cite{154} used code review to aid the APR process and proposed Review4Repair. 
The input to the model is no longer a single code, but the code review information is added. 
By leveraging code review information, Review4Repair achieves better results than SequenceR~\cite{49}.

Overall, in terms of approach design, DLFix, CoCoNut, and Review4Repair all use context-aware patch generation strategies, especially Review4Repair adds code review comments as context. But none of them is concerned with FL design, and they can work together with the help of existing FL tools. In terms of repair ability, they both support multi-type fixes. 
Also, none of them explore the multi-hunk and multi-fault repair capability.

In 2021, researchers began to focus on improving models to further enhance their ability to capture and understand PL features to better learn PL knowledge. Further, the idea of using pre-trained PL models also started to emerge gradually, which ushered in the era of large language models (LLMs) for APR research.

Most previous works~\cite{145,49,137} have adopted NLP techniques, such as NMT.
However, these models were not designed for PL scenarios
(i.e., previous NMT-based APR tools were not designed for PLs), the patches they generated contained many uncompilable or incorrect patches.
Therefore, some researchers have started to focus on improving the models.
Tang \textit{et al.}~\cite{157} found that previous Seq2Seq models often suffer from grammar and syntax errors.
So they proposed a grammar-based rule-to-rule model for the characteristics of PL. 
These grammar rules are used to ensure the grammatical correctness of model-generated fixes.
It is shown that this model incorporating grammar-constrained outperforms the previous NMT model~\cite{145}.
Ye \textit{et al.}~\cite{160} focused on the fact that previous NMT approaches were not optimized for patch quality, 
so they proposed a model RewardRepair can learn programming knowledge.
They made targeted improvements to the loss function. 
It adds program compilation and test execution information to the loss function, rewarding the model for generating compilable and overfitting-free patches. 
This method improves the patch quality, outperforming SequenceR~\cite{49}, CoCoNut~\cite{137} and DLFix~\cite{136}. Similarly, Zhu \textit{et al.}~\cite{Recoder} have improved the model architecture. They proposed Recoder, a syntax-guided edit decoder. Recoder's improvements allow the decoder to incorporate knowledge of the syntactic structure of the programming language to guide the generation of syntactically correct repair patches.
Notably, Recoder is the first DL-based APR approach that has outperformed traditional APR techniques on the Defects4J benchmark.

Jiang \textit{et al.}~\cite{161} also improved on the shortcomings of previous NMT-based schemes by proposing an APR tool, CURE. 
To enable the NMT model to incorporate PL knowledge, they pre-trained a GPT PL model and combined it with the NMT model. 
Besides, they propose a code-aware beam-search strategy to explore the patch space more efficiently and use byte-pair encoding (BPE) techniques to address the out-of-vocabulary (OOV) issue.
These improvements make CURE superior to previous APR tools~\cite{137,136,49}.
Besides, TFix~\cite{162} took a different line of thought than CURE~\cite{161}. 
TFix uses a pre-trained natural language model (not a PL model), T5, and fine-tunes it on the programming language task.
It also uses multi-task learning to repair multiple error types and outperforms SequenceR~\cite{49} and CoCoNuT~\cite{137}. 
Importantly, the above works~\cite{161,162} introduced the workflow of pre-training and fine-tuning to APR, which also provided inspiration for later solving the small sample problem of vulnerability repair~\cite{197,SeqTrans} (see Section~\ref{DL for vul} for details).

Through the above works, we observe that how to make models better capture and understand PL knowledge becomes a key issue. 
To better adapt to PL tasks rather than NL tasks, we believe that using pre-trained PL models for APR tools will be a direction that can continue to be explored.
In summary, these works \cite{157,160,161,162,Recoder} aim to address general bugs and do not focus on FL research, using existing FL tools or perfect fault localization in the experimental part. Also, most works considered context-aware strategies. In terms of repair ability, these works support multi-type fixes, but they are designed for single-line/hunk fixes.


In 2022, learning-based APR techniques entered the era of LLMs, with researchers integrating them (LLMs) into APR tools to deeply exploit the repair potential of LLMs. Additionally, multi-language and multi-hunk repair have been further investigated, and hybrid techniques have been used to facilitate traditional APR techniques.

Unlike previous works that train and tune models to gain knowledge of fixes, LLMs provide an alternative route for APR research that directly draws on the code understanding and generation capabilities of the LLMs to generate fixes. This way was explored by Xia \textit{et al.}~\cite{AlphaRepair}, who proposed AlaphaRepair. 
Unlike the NMT approach that treats bug fixing as a translation task, AlphaRepair directly uses CodeBERT~\cite{CodeBert} to predict correct fixes by code context. Notably, AlaphaRepair outperforms all previous APR tools in repair results and is inherently multi-language capable.
Subsequently, researchers have explored the repair capabilities of LLMs in the zero-shot / few-shot learning paradigm through large-scale empirical studies~\cite{LLM2APR_fan, LLM2APR_xia}.
This signifies the great potential of LLMs to facilitate APR research.

In other works, researchers have explored applying hybrid techniques to enhance APR tools. 
Li \textit{et al.}~\cite{DEAR} have combined the DL with the traditional FL technique and then proposed a novel FL technique suitable for multi-hunk repair scenarios. They improved on DLFix to implement the new APR tool DEAR, which can solve multi-hunk bugs using a divide-and-conquer strategy.
In addition, Meng \textit{et al.}'s work~\cite{TRANSFER} not only uses DL to improve FL but also guides template selection by training a neural network model. Their tool, called TRANSFER, brings together learning-based and template-based techniques to enhance APR tools. In general, combining DL with other techniques is an effective path to further optimize APR techniques.

At the same time, multi-language repair has also been explored by researchers. Yuan \textit{et al.}~\cite{CIRCLE} combined the ideas of continual learning to train CIRCLE to obtain cross-language repair ability. Notably, CIRCLE implements ML fixes in a single model, which addresses the limitation of previous APR tools that require training multiple models to repair multiple languages.

The above review shows that applying LLMs and developing hybrid techniques will be the future trend in APR research. The direct use of off-the-shelf LLMs will reduce the training cost of APR~\cite{AlphaRepair}, while the development of hybrid techniques will compensate for the shortcomings of a single technique~\cite{TRANSFER}.
In general, the above works~\cite{AlphaRepair,DEAR,TRANSFER,CIRCLE} is designed for generic bug fixing. In particular, DEAR~\cite{DEAR} has multi-hunk repair ability, while the other tools are designed for single-hunk repair. Moreover, AlphaRepair~\cite{AlphaRepair} and CIRCLE~\cite{CIRCLE} can support multi-language fixes with a single model, while other tools need to train different models for different languages. In terms of FL, DEAR~\cite{DEAR} and TRANSFER~\cite{TRANSFER} have designed novel FL techniques, while the other APR tools do not improve on FL. In addition, all the above tools rely on contextual information to guide patch generation, yet none explore the ability of multi-fault repair.

\subsubsection{Programming Error Repair}
\label{DL for error}
Unlike real-world software bugs, programming errors are simple types of defects, usually syntactic or semantic errors, that may prevent a program from compiling successfully and are therefore known as compilation errors~\cite{TransR}.
The researchers want to use APR techniques to automate the repair of such simple errors caused by inexperience in programming.
Since programming errors can be detected by the compiler and therefore do not require special FL tools and test cases for fault localization, 
the researchers mainly use \textit{repair accuracy} to measure the repair effectiveness.
Note that error repair usually has two evaluation metrics ~\cite{TransR} when calculating repair accuracy: 
First, \textit{Single Repair}. It is used to evaluate if the generated statement exactly matches the ground-truth associated with a broken statement.  
Second, \textit{Full Repair}. It is calculated as the percentage of generated programs that pass the compiler successfully.

Back in 2016, Bhatia \textit{et al.}~\cite{140} and Pu \textit{et al.}~\cite{141} have provided solutions to programming errors by using recurrent neural networks (RNN). 
They don't focus on FL because the compiler can detect these simple errors. But such a practice is inaccurate.
Therefore, Gupta \textit{et al.}~\cite{143} proposed an end-to-end solution DeepFix, which integrated the error localization. 
It uses an attention based RNN model to capture long term dependencies, which allows it to iteratively fix multiple errors in a single program.
Note that early works~\cite{140,141,142,143} implement repair only for errors and not software bugs. But they demonstrate the effectiveness of DL for program repair tasks, which also provides the basis for extending DL to bug fixing. 

In 2019, the researchers began specific optimization of the model training phase.
Gupta \textit{et al.}~\cite{150} presented RLAssist, which employs deep reinforcement learning to repair syntactic errors.
Most previous works used supervised learning. 
Instead, they used unsupervised learning. 
The approach relies on the number of error messages to design the reward function. This makes the model more inclined to learn the correct fixes (not incorrect fixes) during the learning process.
Besides, Hajipour \textit{et al.}~\cite{152} proposed SampleFix. 
They argue that given a corrupted program, there are multiple repair ways. 
In other words, the problem of program repair is modeled as a one-to-many problem in which the model learns the distribution of potential fixes through a generative model conditional on the incorrect input program. 
In their experiment, SampleFix achieves up to 45\% of error fixes and generates multiple diverse fixes. 
In general, these tools focus on syntax errors and support FL functions by themselves. In particular, RLAssist considers the code context and can fix multiple errors, but SampleFix does not describe these.

In 2020, Yasunaga \textit{et al.}~\cite{163,164} began to focus on the impact that the training process has on the repair effect. 
In 2020, they proposed DrRepair~\cite{163}.
They expanded the training data by corrupting correctly fixed examples to obtain more code with errors, and they used a graph neural network (GNN) to record the code dependencies. 
These two innovations enabled DrRepair to achieve a repair success rate of 68.2\% on the DeepFix dataset~\cite{143}. 
However, this data generated by adding perturbations differed from real-world errors.
So they propose a new training method Break-It-Fix-It (BIFI)~\cite{164} in 2021. They train a breaker by using realistic bad code so that the breaker can generate bad examples close to real-world data. 
In the experimental part, the best result on the DeepFix dataset was further improved by combining BIFI with DrRepair, achieving 71.7\% repair accuracy. 
This suggests that BIFI is an effective data enhancement to improve repair performance.
These works are dedicated to improving the training process to enhance the repair effect, which deserves to be explored and applied in-depth in learning-based schemes.

In addition, some works have tried to incorporate some external information to assist the repair process.
Abhinav \textit{et al.}~\cite{165} proposed RepairNet. 
The major contribution of this work is the encoding of error messages as additional contexts in the model to fix errors. With the combined compiler error information, RepairNet achieved a 66.4\% repair accuracy on the DeepFix dataset~\cite{143}. 
This indicates that the error information provided by the compiler is a vital feature to guide the repair.

Note that DrRepair~\cite{163,164} and RepairNet~\cite{165} support multi-line fixes. In particular, RepairNet does not use the FL function but uses the error messages provided by the compiler. DrRepair divides multiple error types in its experiments,
and it supports repairing multi-errors in a program.


In 2022, error repair began to benefit from bug repair research, and some of the ideas applied to bug repair are applied to error repair. And error repair is also ushering in the era of LLMs.

Context-aware repair solutions have previously been widely used in bug repair~\cite{136,137,161}, and this has also inspired error repair. Based on this, Li \textit{et al.} proposed TransRepair~\cite{TransR}, a context-aware error repair tool. Similar to the bug repair task, the erroneous statement and its context are treated as different inputs to be given to the neural network for processing, which enables the model to combine rich contextual information for accurate repair. By combining the advanced Transformer architecture and other technical details, TransRepair outperforms previous works~\cite{163,143,150,152} of its kind in terms of fixing results.
In particular, TransRepair incorporates MLP to design an FL strategy and supports multi-type, multi-line, and multi-error repair.

In addition, LLMs are also used for error-fixing tasks. Similarly to AlphaRepair, Zhang \textit{et al.}~\cite{MMAPR} used LLM to fix programming errors and proposed an APR system MMAPR based on CodeX.
MMAPR ensembles multi-modal prompts to generate complementary repair candidates.
Similar work was proposed by Joshi \textit{et al.}~\cite{RING}, who implemented a repair tool RING based on Codex, and it implemented multi-language repair. 
RING and MMAPR use similar technical paths (prompt learning~\cite{Prompt-Learning}), 
and although multi-hunk and multi-fault fixes have emerged in the MMAPR fixes, their repair ability has not been discussed in depth.

The above works~\cite{TransR,MMAPR,RING} show that solutions from different problem domains can be cross-pollinated to facilitate APR research and that LLMs have great potential for error repair tasks.

\subsubsection{Security Vulnerability Repair}
\label{DL for vul}
Vulnerabilities (security bugs), a subset of software bugs, are more damaging and require more urgent fixes, making it critical to automate vulnerability fixes. However, vulnerability repair has not been independently studied in the previous software engineering APR community, and both security and non-security bugs have been fixed as the same kind of problem, which makes it hard to know how effective these APR tools are in fixing vulnerabilities. For this reason, learning-based vulnerability repair, a major concern in system security, has received targeted research in recent years.

Early vulnerability repair efforts also received inspiration from NMT techniques. However, the task framework of NMT requires a large amount of annotated data to support the training of Seq2Seq models. 
To mitigate against this challenge,
Harer \textit{et al.}~\cite{146} proposed to perform adversarial learning (ADL) using Generative Adversarial Network (GAN), which allows training without pairing (BFPs). 

Given the many advances that have been made in bug repair problems by researchers in the field of software engineering, researchers also wish to draw on this empirical knowledge to further contribute to the vulnerability repair problem. Among them, recent representative works are VRepair~\cite{197} and SeqTrans~\cite{SeqTrans}, which both rely on the pre-training + fine-tuning learning paradigm by pre-training on large-scale bug fixing datasets to gain empirical knowledge on bug fixing and then using transfer learning to fine-tune on small-sample vulnerability fixing datasets to improve the vulnerability fixing capability. Unlike VRepair, which improves the code representation by allowing multiple bug locations to be tagged to support multi-hunk bug fixing, SeqTrans only works in single-line repair scenarios. However, SeqTrans adds the program's data flow dependencies as features, further improving the model's ability to capture deep semantic information for understanding.
Overall, these works draw on empirical knowledge of software bug fixing and use transfer learning to alleviate the limitations of the small sample problem in the vulnerability repair domain.

Recently, researchers have also explored the capabilities of LLMs for vulnerability repair through a large-scale empirical study~\cite{LLM2Vul}, which signifies the same potential of large language models for vulnerability repair.
In a recent data-driven vulnerability repair effort, VulRepair~\cite{VulRepair} fine-tuned it with the help of the LLM, CodeT5~\cite{CodeT5}. 
Thanks to the LLM's powerful knowledge base and ability to capture and understand features, VulRepair achieved superior repair results compared to VRepair.
The results demonstrate that LLM fine-tuning is more efficient and faster than traditional model training and effectively improves the repair result of APR tools. In the future, vulnerability repair research will enter the era of LLMs. The powerful code understanding and generation capabilities of LLMs will further promote the development of the vulnerability repair community.

From the above review, we observe that vulnerability repair is facilitated by bug repair, with SeqTrans~\cite{SeqTrans}, VRepair~\cite{197}, and VulRepair~\cite{VulRepair} all leveraging context-aware patch generation strategies and supporting multi-type repair, and VRpair and VulRepair supporting multi-hunk repair. However, vulnerability repair research has not yet delved into vulnerability localization, and multi-defect repair has not been explored in depth. With the advent of LLMs, the use of LLMs to facilitate FL and further improve repair ability will be a future chance.

\subsubsection{Summary}
\label{Learning Summary}

The learning-based approaches learn empirical knowledge of program repair from large amounts of data in open source software (OSS) communities, which addresses many limitations in previous APR techniques and further facilitates the automation of APR techniques. 
DL’s powerful learning capabilities offer the possibility of multi-language repair. 
APR tools have even gained the ability to continuously learn and repair. 
More importantly, high-quality data, more advanced models, and better representation methods all provide strong support for APR studies. 
However, APR tools are still challenging to apply these techniques due to some limitations of current AI technologies: 
\hkr{For example, neural networks cannot handle long sequences well~\cite{VulRepair,145,49,143}; 
the OOV problem limits the ability of models to generate new tokens\cite{VulRepair,197}; 
the strict syntactic structural features and complex semantic dependencies of PLs are difficult to effectively capture and understand by models~\cite{161}, and so on.
Fortunately, AI is developing rapidly, and the emergence of LLMs provides new opportunities for APR research.
Therefore, We hope that more efforts in the future will transfer these latest AI techniques to the APR domain. 
Moreover, combining DL solutions with traditional APR techniques is a valuable direction.}

\subsection{APR Development Summary and Discussion}
\label{sec:Summary and Discussion}

In this section, we will summarize and discuss these APR techniques so that the reader can quickly learn the pros and cons of these techniques and the current state of development of APR research. 

\subsubsection{Summary}
\label{summary}

Here, we summarize the development of four classes of APR techniques. 
We also provide a summary of the pros and cons of APR techniques using different approaches in Table~\ref{tab:table2-3-1}.


\begin{itemize}
    \item The search-based approach was the first repair technique explored, and it shows extremely many possibilities for program repair, but the huge search space and high computational resources are important factors that limit its development.
    \item The constraint-based approach has advantages in terms of repair efficiency, but its reliance on correct rule constraints makes it less flexible in repair, and the formulation of constraints may require manual definition, which undoubtedly increases labor costs.
    \item The template-based approach demonstrates efficiency in repairing specific classes of defects, but it cannot repair defects that are not summarized by templates.
    \item The learning-based approach shows great flexibility and scalability in repairing multiple languages and types of defects. 
    However, its repair capability is vulnerable to training data and thus benchmark overfitting, and the patch interpretability remains to be explored.
\end{itemize}

In general, the development of APR has been a continuous process of integration.
The search-based approach, which first emerged, used various heuristics to generate and find candidate fixes in the search space.
The constraint-based approach added formal specifications and other constraints to limit the search space and improve the efficiency of the search-based approach.
The template-based approach adds repair templates to guide patch generation, solving the shortcomings of the constraint-based approach's lack of flexibility and the heuristic approach's lack of guidance on program variant generation.
The learning-based approach further solves the tedious work of the template-based approach on template extraction and obtains richer program syntax and semantic-level features through learning. 
It also has the scalability and flexibility of multi-language and multi-type repair, thus making it possible to automate APR in the true sense. 
In this process, various techniques reinforce and complement each other. 
However, there are still many limitations. As shown in Table~\ref{tab:table2-3-2}, efforts are still needed to design efficient FL functions and closely link localization and repair; in addition, APR tools still need efforts in multi-hunk and multi-fault fixes.
We believe that the future trend will be to integrate multiple types of techniques that draw on each other, 
compensating for the single technique's inherent shortcomings and helping propose new solutions.

\begin{center}
\tiny
\begin{longtable}[]{llllll}
\caption{A summary of the pros and cons of different classes of APR techniques}
\label{tab:table2-3-1}
\\
\hline
\multicolumn{1}{|c|}{\textbf{Sec.}} &
  \multicolumn{1}{c|}{\textbf{Typical Work}} &
  \multicolumn{1}{c|}{\textbf{Time}} &
  \multicolumn{1}{c|}{\textbf{Approach}} &
  \multicolumn{1}{c|}{\textbf{Pros}} &
  \multicolumn{1}{c|}{\textbf{Cons}}
  \\ \hline
  \endfirsthead
  
  \multicolumn{6}{c}{Continued}\\
  \hline
  \multicolumn{1}{|c|}{\textbf{Sec.}} &
  \multicolumn{1}{c|}{\textbf{Typical Work}} &
  \multicolumn{1}{c|}{\textbf{Time}} &
  \multicolumn{1}{c|}{\textbf{Approach}} &
  \multicolumn{1}{c|}{\textbf{Pros}} &
  \multicolumn{1}{c|}{\textbf{Cons}}
  \\ \hline
  \endhead
  
  \multicolumn{6}{r}{(Continued on next page)}\\
  \endfoot
  
  \endlastfoot

\multicolumn{6}{|c|}{\ref{sec:Search-based} Search-based} \\ \hline
\multicolumn{1}{|c|}{\ref{Repair as a Game}} &
  \multicolumn{1}{m{1.4cm}|}{\cite{56,59}} &
  \multicolumn{1}{m{0.5cm}|}{2005-2006} &
  \multicolumn{1}{m{1.5cm}|}{Repair as a Game} &
  \multicolumn{1}{m{4.1cm}|}{Partial fixes do not affect the structure and logic of the source program.} &
  \multicolumn{1}{m{4.1cm}|}{Repair Efficiency Problem and Generic Fixes Problem: The approach revealed shortcomings such as inefficient fixes, limited types of fixes, and repetitive fixing behavior.} \\ \hline
\multicolumn{1}{|c|}{\multirow{4}{*}{\ref{Biological Engineering}}} &
  \multicolumn{1}{m{1.4cm}|}{\cite{48,60,62,63,68,69,70,71,72,73}} &
  \multicolumn{1}{m{0.5cm}|}{2009-2014} &
  \multicolumn{1}{m{1.5cm}|}{Program repair using EC and GP} &
  \multicolumn{1}{m{4.1cm}|}{First, GP can be extended to large program repair. Second, the flexible mutation capability of the GP approach allows the implementation of multiple types of defect repair. Third, mutation at multiple locations allows for multi-line bug fixes.} &
  \multicolumn{1}{m{4.1cm}|}{Repair Efficiency Problem and Repair Quality Problem: The presence of a large search space in the GP strategy leads to inefficient repair. Mutation process may generate a large number of invalid or incorrect fixes. 
  } \\ \cline{2-6} 
\multicolumn{1}{|c|}{} &
  \multicolumn{1}{m{1.4cm}|}{\cite{76}} &
  \multicolumn{1}{m{0.5cm}|}{2013-2014} &
  \multicolumn{1}{m{1.5cm}|}{Program repair using random search} &
  \multicolumn{1}{m{4.1cm}|}{Random search strategy is simpler and outperforms GP method (GenProg) in repair effectiveness and efficiency.} &
  \multicolumn{1}{m{4.1cm}|}{Generic Fixes Problem: This work has only explored the case of fixing a program that contains a single bug, and has not explored the case of a program that contains multiple bugs.} \\ \cline{2-6} 
\multicolumn{1}{|c|}{} &
  \multicolumn{1}{m{1.4cm}|}{\cite{78}} &
  \multicolumn{1}{m{0.5cm}|}{2015} &
  \multicolumn{1}{m{1.5cm}|}{Program repair using HGT} &
  \multicolumn{1}{m{4.1cm}|}{This approach can automatically leverage the combined knowledge integrated across multiple software systems to improve individual applications, and it avoids the resource consumption of the GP method mutation process.} &
  \multicolumn{1}{m{4.1cm}|}{Repair Quality Problem: It relies on the code knowledge of the code provider, which prevents code migration fixes if a defect lacks correct representation in other systems. In addition, mixed code is more likely to introduce new defects.} \\ \cline{2-6} 
\multicolumn{1}{|c|}{} &
  \multicolumn{1}{m{1.4cm}|}{\cite{47,217}} &
  \multicolumn{1}{m{0.5cm}|}{2017-2019} &
  \multicolumn{1}{m{1.5cm}|}{Program repair using GP and template-based idea} &
  \multicolumn{1}{m{4.1cm}|}{This work combines redundancy assumption and repair templates to generate repair statements to further improve repair efficiency, reduce the search space, and alleviate the patch overfitting problem.} &
  \multicolumn{1}{m{4.1cm}|}{Large-scale Project Bug Fixes: This approach also suffers from template coverage, and the lack of repair templates leads to possible limitations in the repair of large-scale programs.} \\ \hline
\multicolumn{1}{|c|}{\ref{Target Value Search}} &
  \multicolumn{1}{m{1.4cm}|}{\cite{52}} &
  \multicolumn{1}{m{0.5cm}|}{2015} &
  \multicolumn{1}{m{1.5cm}|}{Program repair using staged program repair} &
  \multicolumn{1}{m{4.1cm}|}{Staged program repair can reduce the size of the search space and explore large search spaces more efficiently.} &
  \multicolumn{1}{m{4.1cm}|}{Generic Fixes Problem: SPR will only apply one transformation each time a candidate fix is generated, and cannot combine multiple transformations. It means that it has limitations when fixing complex bugs that require multiple fix patterns.} \\ \hline
\multicolumn{1}{|c|}{\multirow{4}{*}{\ref{Code Similarity}}} &
  \multicolumn{1}{m{1.4cm}|}{\cite{91,92,82,88}} &
  \multicolumn{1}{m{0.5cm}|}{2016-2019} &
  \multicolumn{1}{m{1.5cm}|}{Program repair using code block similarity} &
  \multicolumn{1}{m{4.1cm}|}{The code similarity-based approach effectively reduces the search space compared to the GP-based and random search approaches. In addition, using ready-made code from open-source code repositories for fixes helps to obtain high-quality fixes and avoid introducing new bugs.} &
  \multicolumn{1}{m{4.1cm}|}{Repair Efficiency Problem: Limited by the size of the code base and the search capability of the search algorithm, the APR tool may not be able to search for available similarity codes in all cases.} \\ \cline{2-6} 
\multicolumn{1}{|c|}{} &
  \multicolumn{1}{m{1.4cm}|}{\cite{57}} &
  \multicolumn{1}{m{0.5cm}|}{2018} &
  \multicolumn{1}{m{1.5cm}|}{Program repair using patches and similar code} &
  \multicolumn{1}{m{4.1cm}|}{Compared to previous work using only code similarity, adding patch information can further narrow the final search space and help explore the search space more efficiently.} &
  \multicolumn{1}{m{4.1cm}|}{Repair Quality Problem: The quality of existing patches can easily have an impact on the quality of the final fix, and low quality patches can lead to low quality or incorrect fixes.} \\ \cline{2-6} 
\multicolumn{1}{|c|}{} &
  \multicolumn{1}{m{1.4cm}|}{\cite{90}} &
  \multicolumn{1}{m{0.5cm}|}{2021} &
  \multicolumn{1}{m{1.5cm}|}{Program repair using similarity information from patch tests} &
  \multicolumn{1}{m{4.1cm}|}{Using variational execution techniques to obtain similarity information from patch tests allows for more efficient exploration of large search spaces.} &
  \multicolumn{1}{m{4.1cm}|}{Repair Efficiency Problem: The variational execution used in this work is a dynamic analysis technique with a large resource overhead, which leads to a larger repair time cost.} \\ \hline
\multicolumn{6}{|c|}{\ref{sec:Constraint-based} Constraint-based} \\ \hline
\multicolumn{1}{|c|}{\multirow{2}{*}{\ref{Contract-based}}} &
  \multicolumn{1}{m{1.4cm}|}{\cite{99,100,101}} &
  \multicolumn{1}{m{0.5cm}|}{2010-2014} &
  \multicolumn{1}{m{1.5cm}|}{Program repair using contracts} &
  \multicolumn{1}{m{4.1cm}|}{The approach does not require user-written test cases as compared to the test case based solution. In addition, the approach using a contract requires lower computational resources, resulting in lower repair costs.} &
  \multicolumn{1}{m{4.1cm}|}{Generic Fixes Problem: It still has limitations in flexibility. This class of methods requires developers to write the correct contract during programming, which does not support current mainstream programming languages.} \\ \cline{2-6} 
\multicolumn{1}{|c|}{} &
  \multicolumn{1}{m{1.4cm}|}{\cite{102,103}} &
  \multicolumn{1}{m{0.5cm}|}{2017-2021} &
  \multicolumn{1}{m{1.5cm}|}{Program repair using contract-based idea without contracts} &
  \multicolumn{1}{m{4.1cm}|}{The method inherits the advantages of contract-based repair techniques, alleviates the patch overfitting problem, and supports mainstream programming languages.} &
  \multicolumn{1}{m{4.1cm}|}{Generic Fixes Problem: The fixes generated by this tool tend to be small and unable to generate syntactically complex fixes, which also means that it is not advantageous for fixing multi-line bug tasks that require more changes.} \\ \hline
\multicolumn{1}{|c|}{\multirow{3}{*}{\ref{Constraint Solvers}}} &
  \multicolumn{1}{m{1.4cm}|}{\cite{95}} &
  \multicolumn{1}{m{0.5cm}|}{2011} &
  \multicolumn{1}{m{1.5cm}|}{Program repair using SAT to solve specification constraints} &
  \multicolumn{1}{m{4.1cm}|}{Using rich behavior specifications can help fix programs with complex structures.} &
  \multicolumn{1}{m{4.1cm}|}{Repair Quality Problem: The accuracy of the fixes generated by this technique is affected by the completeness and accuracy of the constraints and is limited by the specification-based bounded verification techniques. In addition, it generates only near-accurate repair suggestions, which still need to be verified by the user.} \\ \cline{2-6} 
\multicolumn{1}{|c|}{} &
  \multicolumn{1}{m{1.4cm}|}{\cite{98,21,106,112}} &
  \multicolumn{1}{m{0.5cm}|}{2013-2016} &
  \multicolumn{1}{m{1.5cm}|}{Program repair using test case driven semantic constraint-based approach} &
  \multicolumn{1}{m{4.1cm}|}{The semantic constraint-based approach has a higher repair success rate and shorter repair time than the syntax search-based approach (such as GenProg)~\cite{98}. \hkr{This is thanks to the fact that they alleviate the huge search space and avoid the uncertainty of the mutation process.}} &
  \multicolumn{1}{m{4.1cm}|}{Repair Quality Problem and Generic Fixes Problem: The test case driven solution lacks flexibility compared to the approach that uses behavioral specifications. Besides, the generation of repair constraints depends on the execution of test cases, and therefore the quality of the test cases affects the quality of the repair.
  (Note that most generate-and-validate repair techniques are affected by the quality of the test cases~\cite{208}.)
  } \\ \cline{2-6} 
\multicolumn{1}{|c|}{} &
  \multicolumn{1}{m{1.4cm}|}{\cite{89,110}} &
  \multicolumn{1}{m{0.5cm}|}{2015-2019} &
  \multicolumn{1}{m{1.5cm}|}{Program repair using semantic code search} &
  \multicolumn{1}{m{4.1cm}|}{Using a human-written code base and semantic code search can provide high-quality fixes.} &
  \multicolumn{1}{m{4.1cm}|}{Repair Quality Problem: The quality of the artificially constructed code repository will affect repair quality. Especially the artificial nature of the repository construction may bias the repair.} \\ \hline
\multicolumn{1}{|c|}{\ref{Combination of Syntactic and Semantic Approaches}} &
  \multicolumn{1}{m{1.4cm}|}{\cite{113,114}} &
  \multicolumn{1}{m{0.5cm}|}{2016-2017} &
  \multicolumn{1}{m{1.5cm}|}{Program repair using syntactic and semantic features} &
  \multicolumn{1}{m{4.1cm}|}{Combining the two techniques can overcome the shortcomings of using a single technique, can provide high-quality fixes, and mitigate patch overfitting problems.} &
  \multicolumn{1}{m{4.1cm}|}{Generic Fixes Problem: The semantic-based approach can only modify the expressions and the constraint solving technique still cannot handle some special types of variables. These inherent limitations lead to many types of defects that cannot be repaired, which limits the overall repair capability.} \\ \hline
\multicolumn{1}{|c|}{\ref{New formal specifications}} &
  \multicolumn{1}{m{1.4cm}|}{\cite{119,122}} &
  \multicolumn{1}{m{0.5cm}|}{2018-2020} &
  \multicolumn{1}{m{1.5cm}|}{Program repair using formal specifications} &
  \multicolumn{1}{m{4.1cm}|}{Compared with the test suite based approach, the formal specification based approach provides better coverage of the program input/output relationships and avoids the test case coverage problem. In terms of fixing results, using the formal specification approach mitigates the patch overfitting problem caused by the test case based approach and helps improve the quality of fixes.} &
  \multicolumn{1}{m{4.1cm}|}{Generic Fixes Problem: Like other constraint-based techniques, the use of formal specifications sacrifices flexibility and scalability, and most of these efforts are designed for specific defect types and cannot be generalized to broad generic defect fixes.} \\ \hline
\multicolumn{6}{|c|}{\ref{sec:Template-based} Template-based} \\ \hline
\multicolumn{1}{|c|}{\ref{sec:Manually Extracted Templates}} &
  \multicolumn{1}{m{1.4cm}|}{\cite{126,73,128}} &
  \multicolumn{1}{m{0.5cm}|}{2009-2015} &
  \multicolumn{1}{m{1.5cm}|}{Program repair using manually extracted repair templates} &
  \multicolumn{1}{m{4.1cm}|}{Compared with search-based GP techniques, the use of repair templates avoids aimless random mutations and makes the program repair behavior toward a given template, which avoids generating a large number of nonsensical patches and improves repair efficiency and patch quality.} &
  \multicolumn{1}{m{4.1cm}|}{Repair Efficiency Problem and Generic Fixes Problem: 
  \hkr{Manual mining is tedious, error-prone, and cannot scale~\cite{42}.
  First, it requires additional labor costs to summarize the repair templates.}
  In addition, the manual work cannot generate a large number of templates quickly, which suffers from the problem of template coverage, and many defects cannot be fixed due to the lack of corresponding repair templates.} \\ \hline
\multicolumn{1}{|c|}{\ref{Automatically Generated Templates}} &
  \multicolumn{1}{m{1.4cm}|}{\cite{125,54,129,44,42,134,45}} &
  \multicolumn{1}{m{0.5cm}|}{2016-2021} &
  \multicolumn{1}{m{1.5cm}|}{Program repair using automated template mining method} &
  \multicolumn{1}{m{4.1cm}|}{Compared to manual extraction of templates, the automated approach improves efficiency and generates more templates to address more defect types.} &
  \multicolumn{1}{m{4.1cm}|}{Repair Quality Problem: Although the number of templates is improved, the balance of granularity and coverage of templates is an issue. And the automated extraction approach is often not as precise as human work, which may introduce some noise in the repair templates.} \\ \hline
\multicolumn{6}{|c|}{\ref{sec:Learning-based} Learning-based} \\ \hline
\multicolumn{1}{|c|}{\multirow{6}{*}{\ref{sec:Learning-based}}} &
  \multicolumn{1}{m{1.4cm}|}{\cite{145,146}} &
  \multicolumn{1}{m{0.5cm}|}{2018-2019} &
  \multicolumn{1}{m{1.5cm}|}{Program repair using NMT techniques} &
  \multicolumn{1}{m{4.1cm}|}{Compared to test-driven techniques, NMT techniques using deep learning do not require the help of test cases to guide patch generation. Compared to other previous techniques, data-driven solutions have higher fixing efficiency and lower fixing cost because the training of the model is a one-time cost. In addition, data-driven solutions can be extended to repair tasks in multiple PLs by replacing components such as lexer, parser, etc., and can repair multiple defect types, solving the generic fixes problem with higher scalability.} &
  \multicolumn{1}{m{4.1cm}|}{Repair Efficiency Problem and Generic Fixes Problem: NMT systems are indeed data-hungry system. It also requires a significant training time cost, and model training is a time-consuming process. In addition, early works still cannot handle excessively long sentences, making them unable to fix complex bugs with excessively long sentences and thus unable to effectively address large program fixes.} \\ \cline{2-6} 
\multicolumn{1}{|c|}{} &
  \multicolumn{1}{m{1.4cm}|}{\cite{49}} &
  \multicolumn{1}{m{0.5cm}|}{2019} &
  \multicolumn{1}{m{1.5cm}|}{Program repair using NMT techniques and Seq2Seq learning} &
  \multicolumn{1}{m{4.1cm}|}{Compared to previous works by Tufano \textit{et al.}~\cite{145}, SequenceR uses the copy mechanism to overcome the unlimited vocabulary problem. In addition, it constructs abstract bug contexts, which simplifies the complexity of the input sequence, allowing it to capture long range dependencies that are required for the fix.} &
  \multicolumn{1}{m{4.1cm}|}{Generic Fixes Probllem and Repair Quality Problem: SequenceR is designed for single-line bug fixing, it does not yet address multi-line bug fixing. The sequence representations and sequence-based DL models are not suitable for capturing code structures, and this could lead to imprecise mappings and eventually incorrect fixes.} \\ \cline{2-6} 
\multicolumn{1}{|c|}{} &
  \multicolumn{1}{m{1.4cm}|}{\cite{136}} &
  \multicolumn{1}{m{0.5cm}|}{2020} &
  \multicolumn{1}{m{1.5cm}|}{Program repair using NMT techniques and tree-based RNN} &
  \multicolumn{1}{m{4.1cm}|}{Compared to previous works by Tufano \textit{et al.}~\cite{145} and SequenceR~\cite{49}, DLFix uses a tree structure to capture the syntactic structural features of the code and uses the tree RNN to learn code transformations. 
  This helps to make the model generate fixes that conform to the rules of the PL, further improving the repair quality.
  } &
  \multicolumn{1}{m{4.1cm}|}{Generic Fixes Problem: While the tree-based feature representation and learning approach is effective, it also poses significant limitations. DLFix can only fix single-line bugs which have the same bug location as the fix location, and it does not handle multiple bugs in a method very well.} \\ \cline{2-6} 
\multicolumn{1}{|c|}{} &
  \multicolumn{1}{m{1.4cm}|}{\cite{161}} &
  \multicolumn{1}{m{0.5cm}|}{2021} &
  \multicolumn{1}{m{1.5cm}|}{Program repair using NMT techniques and PL model} &
  \multicolumn{1}{m{4.1cm}|}{Compared to previous NMT models, they have difficulty understanding the strict code syntactic features and thus generate some incorrect or non-compilable patches. Instead, CURE uses a pre-trained PL model, which allows it to incorporate the correct knowledge of the source code and significantly improve the repair quality.} &
  \multicolumn{1}{m{4.1cm}|}{Generic Fixes Problem: CURE is designed to fix single-hunk bugs. Besides, pre-training PL models is an extremely time-consuming process.} \\ \cline{2-6}
\multicolumn{1}{|c|}{} &
  \multicolumn{1}{m{1.4cm}|}{\cite{VulRepair}} &
  \multicolumn{1}{m{0.5cm}|}{2022} &
  \multicolumn{1}{m{1.5cm}|}{Program repair using NMT techniques and LLM fine-tuning} &
  \multicolumn{1}{m{4.1cm}|}{Fine-tuning the LLM can speed up model convergence. Importantly, the LLM has already learned about the PL knowledge during the pre-training phase, so fine-tuning on top of it can further improve the repair. For example, VulRepair~\cite{VulRepair} obtains higher repair accuracy than VRepair~\cite{197}.} &
  \multicolumn{1}{m{4.1cm}|}{Generic Fixes Problem: LLM fine-tuning requires a certain number of bug-fix pairs to support the training process, so the fine-tuning workflow cannot be applied for some scenarios that lack training data.} \\ \cline{2-6}
\multicolumn{1}{|c|}{} &
  \multicolumn{1}{m{1.4cm}|}{\cite{AlphaRepair,LLM2APR_fan,LLM2APR_xia,LLM2Vul,MMAPR,RING}} &
  \multicolumn{1}{m{0.5cm}|}{2022} &
  \multicolumn{1}{m{1.5cm}|}{Program repair using LLM prompt learning (zero / few-shot learning)} &
  \multicolumn{1}{m{4.1cm}|}{
  Compared to fine-tuning workflows, prompt learning requires only a small amount of training data or no training data to achieve repair predictions. This process aims to emphasize the LLM's code understanding and generation capabilities.
  } &
  \multicolumn{1}{m{4.1cm}|}{Repair Quality Problem: Since the pre-training of LLMs is done in a large amount of source code. Therefore, LLM may generate incorrect repairs due to the pre-training phase learning defective code samples (incorrect repair knowledge).} \\ \hline
\end{longtable}
\end{center}

\vspace{-20px}
\subsubsection{Discussion}
\label{discussion}

Further, while reviewing the APR techniques, 
we also hope to discuss the current state of technical development to seek some interesting findings.
To thoroughly survey state-of-the-art techniques and seek valuable findings, we have added more details on some representative works in Appendix~\ref{sec:APR work}.
In Appendix~\ref{sec:APR work}, we selected six representative works for search-based, constraint-based, and template-based APR, respectively. 
Given that deep learning techniques have been widely used in recent years, we focus on learning-based techniques and present 26 representative works. 
Next, we discuss some interesting findings by combining a review of previous works (Table~\ref{tab:table2-3-2}) and an analysis of these details (Appendix~\ref{sec:APR work}). 
These findings may reflect current problems and future research trends, and we will discuss this content.

\textbf{Finding 1:} 
\textbf{\textit{Java and C are the main research targets supported by APR tools, and there are differences between learning-based APR techniques and traditional APR techniques in terms of multi-language repair ability.}} 
The reason for this phenomenon is that Java and C are still the dominant PLs.
Therefore, researchers are also keener to focus on these PLs, which are more representative. 
However, this also reflects the lack of current research, and other PLs also deserve our attention. 
And it also shows that many APR tools are deficient in scalability, as they usually support only a single PL. 
It is worth noting that some learning-based APR techniques implement support for multiple PLs in a single APR tool~\cite{CIRCLE,AlphaRepair}, while most of the traditional APR techniques need to implement different APR tools for different PLs. Therefore, it is an interesting direction to combine DL with traditional APR techniques to improve APR tools' repair ability.

\textbf{Finding 2:} 
\textbf{\textit{Looking at the defect scenarios fixed by APR tools, not only software bugs but also programming errors are included in the scope of problems. }}
Because in the early research, bug repair was the initial concern of software engineering. 
And later on, researchers found that the essence of defect repair is a process of code error correction, which means that APR techniques can also solve some typical programming errors. 
The researchers who focus on this aspect are mostly in the field of AI, and they hope to study deeper code problems with the help of DL. 
In addition, research on vulnerability repair appears to be in its infancy. 
Researchers in security are leveraging previous APR ideas to address these security bugs.
It can be seen that cross-fertilization of different problem areas is the future trend, and we emphasize the migratory nature of different techniques.
The multidisciplinary intersection of software engineering, artificial intelligence, system security, etc., will be a research trend that will jointly help APR techniques solve various code problems.

\textbf{Finding 3:} 
\textbf{\textit{In terms of the complexity of fixing defects, single-hunk bugs seem to be of more concern, and most APR techniques only support the fixing of single-hunk bugs.}}
This phenomenon reflects that the multi-hunk fixes still is a challenge. 
Even though the single-hunk repair tool can iteratively fix multi-hunk bugs, it may use the context with the bugs, so this is not an appropriate solution~\cite{DEAR}.
In recent years, although HERCULES~\cite{13} has emerged, a repair tool designed explicitly for multi-hunk fixes scenarios, it addresses only a specific class of multi-hunk fixes, namely ones with substantially similar patches for each hunk. 
In addition, while learning-based solutions are now beginning to focus on multi-hunk fixes~\cite{DEAR}, there are many limitations, which we will discuss in detail in \textbf{Finding 18}.
Overall, the multi-hunk fixes problem is always a challenge, and the complex dependencies between multiple lines of code make the repair even more difficult. 
In the long run, multi-hunk bug fixing is still a direction for future researchers.

\textbf{Finding 4:} 
\textbf{\textit{From the benchmark datasets used by each APR tool, Defects4J is the most widely used benchmark for software bugs, DeepFix dataset is the most commonly used benchmark for programming errors, CVEfixes and Big-Vul are two of the latest vulnerability fix datasets.}} 
But the amount of data in these datasets is still limited.
Defects4J 2.0 contains only 835 bugs. CVEfixes and Big-Vul contain only thousands of real CVE vulnerabilities.
In addition, benchmark overfitting may occur when using a single benchmark for APR performance evaluation~\cite{169}. 
We recommend evaluating multiple benchmarks when possible during experimental evaluation. 
We believe that future research should aim to present more diverse and comprehensive benchmarks.
Benchmarks for different PLs and different defect types should be an ongoing effort in the future. 
With more datasets, the data basis of the APR research area will be further enriched.

\begin{table}[]
\tiny
\caption{The repair result of APR tools with/without perfect fault faultlization on Defects4J V1.2.0}
\label{tab:tableRepairResult}
\begin{tabular}{|l|ccccc|cccc|}
\hline
\textbf{Technique Type} &
  \multicolumn{5}{c|}{Learning-based APR techniques} &
  \multicolumn{4}{c|}{Traditional APR techniques} \\ \hline
\textbf{APR Tool} &
  \multicolumn{1}{c|}{AlphaRepair} &
  \multicolumn{1}{c|}{Recoder} &
  \multicolumn{1}{c|}{CURE} &
  \multicolumn{1}{c|}{CoCoNuT} &
  DLFix &
  \multicolumn{1}{c|}{TBar} &
  \multicolumn{1}{c|}{AVATAR} &
  \multicolumn{1}{c|}{FixMiner} &
  jGenProg \\ \hline
\textbf{X/Y (with PFL)} &
  \multicolumn{1}{c|}{\textbf{74/109}} &
  \multicolumn{1}{c|}{\textbf{65/112}} &
  \multicolumn{1}{c|}{\textbf{57/104}} &
  \multicolumn{1}{c|}{\textbf{44/85}} &
  \textbf{40/68} &
  \multicolumn{1}{c|}{\textbf{68/95}} &
  \multicolumn{1}{c|}{\textbf{29/50}} &
  \multicolumn{1}{c|}{\textbf{34/62}} &
  \textbf{6/16} \\
\textbf{X/Y (without PFL)} &
  \multicolumn{1}{c|}{50/90} &
  \multicolumn{1}{c|}{49/96} &
  \multicolumn{1}{c|}{36/71} &
  \multicolumn{1}{c|}{33/57} &
  30/65 &
  \multicolumn{1}{c|}{43/81} &
  \multicolumn{1}{c|}{27/53} &
  \multicolumn{1}{c|}{25/31} &
  5/27 \\ \hline
\textbf{P\% (with PFL)} &
  \multicolumn{1}{c|}{\textbf{67.89\%}} &
  \multicolumn{1}{c|}{\textbf{58.04\%}} &
  \multicolumn{1}{c|}{\textbf{54.81\%}} &
  \multicolumn{1}{c|}{51.76\%} &
  \textbf{58.82\%} &
  \multicolumn{1}{c|}{\textbf{71.58\%}} &
  \multicolumn{1}{c|}{\textbf{58.00\%}} &
  \multicolumn{1}{c|}{54.83\%} &
  \textbf{46.15\%} \\
\textbf{P\% (without PFL)} &
  \multicolumn{1}{c|}{55.56\%} &
  \multicolumn{1}{c|}{51.04\%} &
  \multicolumn{1}{c|}{50.70\%} &
  \multicolumn{1}{c|}{\textbf{57.89\%}} &
  46.15\% &
  \multicolumn{1}{c|}{53.09\%} &
  \multicolumn{1}{c|}{50.94\%} &
  \multicolumn{1}{c|}{\textbf{80.65\%}} &
  18.52\% \\ \hline
\end{tabular}
\begin{tablenotes}
\item{
These repair results were extracted from latest works~\cite{AlphaRepair,Recoder,DEAR}. X/Y: correct patches / plausible patches; P\%: the percentage of correct patches.}
\end{tablenotes}
\end{table}


\textbf{Finding 5:}
\textbf{\textit{From the repair results, the \textit{patch overfitting problem} is prevalent for either repair technique. Here, an interesting observation is that a more accurate FL might further alleviate the \textit{patch overfitting problem}.}}
We observe that using perfect fault localization (PFL)~\cite{AlphaRepair} and not using PFL (using off-the-shelf FL tools) lead to significant differences in patch overfitting on the same test benchmark.
Specifically, to clearly present the impact of FL on repair quality, we selected the repair results of APR tools on the most widely used Defects4J V1.2.0 and presented them uniformly.
Table~\ref{tab:tableRepairResult} shows that most APR tools based on PFL were better at patch overfitting (higher percentage of correct patches). This indicates that more accurate FL is crucial for repair quality, which not only helps to promote APR tools to repair more defects, but also further alleviates the patch overfitting. It also points out the way to further optimize the repair quality, i.e., by optimizing the FL technique to mitigate the patch overfitting problem.
However, limited by the paper's type, we did not explore the deeper factors behind how other factors (e.g., patch generation, patch validation.) affect the repair results. We suggest future work to explore these topics.


\textbf{Finding 6:} 
\textbf{\textit{In terms of actual results on Defects4J, the most widely used current benchmark, the learning-based APR tools outperform previous traditional APR tools.}} 
Next, we attempt to analyze the reasons for the superior performance of the learning-based solutions. 
The learning-based APR technique is a data-driven solution. Its powerful repair ability is due to the huge amount of open-source project data (e.g., GitHub) available in the world today. 
The vast amount of data provides the empirical knowledge that neural models use to learn, resulting in powerful AI APR tools. 
But data-driven techniques have never been as accurate and appropriate as humans. 
Its interpretability is still lacking, resulting in actual industrial deployments that may not work well with developers. 
However, it is undeniable that we see powerful potential in data-driven APR tools, and more researchers will devote themselves to learning-based APR techniques. 


Next, we present an in-depth analysis of learning-based APR works and describe our findings.

\begin{figure*}
    \centering
    \includegraphics[width=1\textwidth]{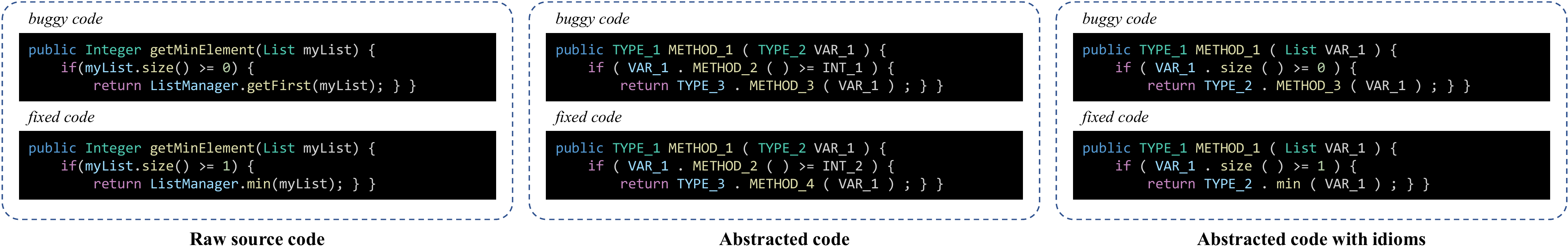}
    \caption{Code abstraction example~\cite{145}.}
    \label{fig:CodeAbs}
    \vspace{-3ex}
\end{figure*}

\textbf{Finding 7:}
\textbf{Pre-processing.
\textit{Different pre-processing strategies in APR techniques have yet to be systematically studied.}
}
\label{find:Pre_processing}
Learning-based APR works use different code pre-processing approaches. However, these approaches are not compared under the same benchmark, and it is unclear whether these approaches can be practical for LLMs-based APR techniques.

We observe that for APR techniques~\cite{49,145,136,DEAR,SeqTrans,TransR} using traditional models (not LLMs), they often rename the variable names, method names, etc., in the raw source code uniformly. For example, Tufano \textit{et al.}~\cite{145} call their approach \textit{code abstraction}, Li \textit{et al.}~\cite{DEAR,136} call them \textit{alpha-rename}, and Chi \textit{et al.}~\cite{SeqTrans} call them \textit{normalization abstraction}. For convenience, we refer to these pre-processing approaches collectively as \textit{Code Abs}.
To better understand the code form of \textit{Code Abs}, we show an example in Fig.~\ref{fig:CodeAbs}.
Early work mainly used \textit{Code Abs} approaches to alleviate the OOV problem and facilitate models learning generic repair patterns.
However, in later LLM-based APR works~\cite{VulRepair}, \textit{Code Abs} has been discontinued. One of the reasons behind this is that LLMs usually use subword-level tokenizers, alleviating the OOV problem.
However, no studies have explored the impact of \textit{Code Abs} on LLM repair in-depth, and we do not know whether \textit{Code Abs} can further improve the repair result of LLMs.
Here, we have explored this through preliminary experiments. We followed the implementation of the LLM-based APR work~\cite{VulRepair,LLM2Code} and the \textit{Code Abs} scheme of Tufano \textit{et al.}~\cite{145} and employed several SOTA LLMs~\cite{LLM2Code} for fine-tuning on the BFP dataset~\cite{145}. The result (see Table~\ref{tab:tableLLM2CodeAbs}) shows that \textit{Code Abs} does not significantly improve the repair performance of most LLMs.
We think there are two main reasons why \textit{Code Abs} does not work well on LLMs: First, as Chen \textit{et al.}~\cite{197} argue, the code abstraction process may lose some special semantic information of method names and variable names, which prevents the models from learning this critical information; Second, LLMs are pre-trained on a large amount of raw source code, which learns the unprocessed source code knowledge, so they are better suited to work on unprocessed raw source code in downstream tasks. Therefore, this also inspires us that no additional code processing steps need to be used for LLMs repair.

In addition, we observe that different APR efforts employ different code abstraction details. For example, Tufano \textit{et al.}~\cite{145} rename variable names, method names and literals, and they preserve idioms; SequenceR~\cite{49} replaces only tokens that do not appear in the vocabulary; SeqTrans~\cite{SeqTrans} renames only variable names and literals; TransRepair~\cite{TransR} replaces the function name, variable name and self-defined struct with the identifiers "<funcN>", "<varN>" and "<typeN>" for normalization; DLFix~\cite{136} renames variables but retains the variable types; and so on. 
While these different \textit{Code Abs} approaches have not been compared under the same benchmark, we cannot know the impact of these differences in technical details on the same APR tool, which requires more empirical studies to explore this issue.

\begin{table}[]
\tiny
\caption{The repair accuracy of LLM-based APR tools with/without Code Abs}
\label{tab:tableLLM2CodeAbs}
\resizebox{1.0\columnwidth}{!}{
\begin{tabular}{|c|cccc|c|c|cccc|}
\hline
\multirow{2}{*}{\textbf{Dataset}} &
  \multicolumn{4}{c|}{\textbf{Repair Accuracy}} &
   &
  \multirow{2}{*}{\textbf{Dataset}} &
  \multicolumn{4}{c|}{\textbf{Repair Accuracy}} \\ \cline{2-5} \cline{8-11} 
 &
  \multicolumn{1}{c|}{CodeBERT} &
  \multicolumn{1}{c|}{GraphCodeBERT} &
  \multicolumn{1}{c|}{PLBART} &
  CodeT5 &
   &
   &
  \multicolumn{1}{c|}{CodeBERT} &
  \multicolumn{1}{c|}{GraphCodeBERT} &
  \multicolumn{1}{c|}{PLBART} &
  CodeT5 \\ \cline{1-5} \cline{7-11} 
BFP\_small with Code Abs &
  \multicolumn{1}{c|}{16.90\%} &
  \multicolumn{1}{c|}{17.46\%} &
  \multicolumn{1}{c|}{\textbf{19.40\%}} &
  21.80\% &
   &
  BFP\_medium with Code Abs &
  \multicolumn{1}{c|}{\textbf{9.73\%}} &
  \multicolumn{1}{c|}{10.36\%} &
  \multicolumn{1}{c|}{\textbf{13.17\%}} &
  \multicolumn{1}{c|}{13.95\%}
  \\ \cline{1-5} \cline{7-11} 
BFP\_small without Code Abs &
  \multicolumn{1}{c|}{\textbf{17.48\%}} &
  \multicolumn{1}{c|}{\textbf{17.62\%}} &
  \multicolumn{1}{c|}{17.93\%} &
  \textbf{25.18\%} &
   &
  BFP\_medium without Code Abs &
  \multicolumn{1}{c|}{9.41\%} &
  \multicolumn{1}{c|}{\textbf{10.45\%}} &
  \multicolumn{1}{c|}{10.13\%} &
  \multicolumn{1}{c|}{\textbf{16.03\%}}
   \\ \hline
\end{tabular}
}
\end{table}

\textbf{Finding 8: Code Representation.} 
\label{find:Code Representation}
\textbf{\textit{Token Sequence is the most widely used type of code representation. In addition, Tree (AST), Graph, and other representations are also used.}}
Token Sequence is widely used because it is one of the most concise and intuitive ways to represent code. 
Code elements are treated as word elements and passed to the neural network to facilitate programming language processing using NLP approaches without tedious processing steps.
Although the Token Sequence approach is simple, it ignores the strict code structure and is hard to capture and record the complex dependencies between code elements. 
As a result, this input information may limit the model's ability to capture comprehensive and rich code features, ultimately impacting the quality of the generated code.
In contrast, representations such as Tree~\cite{DEAR,136} and Graph~\cite{Recoder,156,163} can record the syntactic structure of the program code and its dependencies such as data-flow and control-flow, which in principle can capture a richer representation of the code. 
It is important to note that we do not evaluate which code representation is better or worse here. 
We think that different code representation approaches can work effectively on APR tasks, e.g., many works~\cite{CIRCLE,AlphaRepair} have shown superior performance on repair tasks considering only the simple Token Sequence approach. 
Specifically, we believe that the code representation approach needs to be suitably improved in conjunction with the programming language repair scenario to perform well on a given task. In the future, researchers will continue to focus on research on code representation approaches suitable for program repair scenarios to capture programming language features more richly.

\begin{figure*}
    \centering
    \includegraphics[width=1\textwidth]{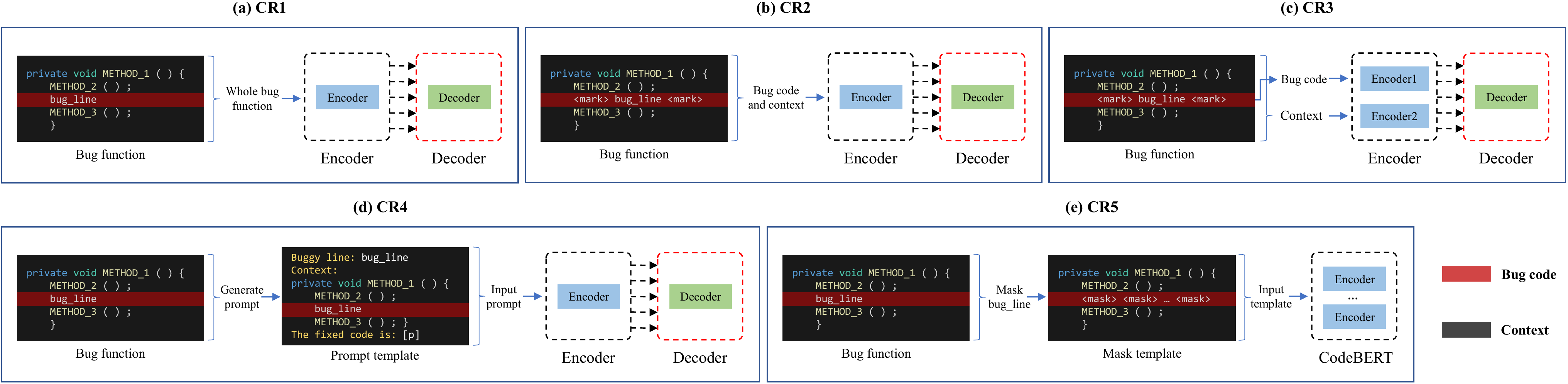}
    \caption{Five typical input code representation ways.}
    \label{fig:CR}
    \vspace{-3ex}
\end{figure*}

\textbf{Finding 9: Context-aware.}
\label{find:Context-aware}
\textbf{\textit{Context-aware patch generation strategies are widely used in learning-based APR techniques. However, the different ways of input code representation for context-aware strategies have yet to be compared in depth.}}

In short, context-aware strategies refer to guiding patch generation by combining bug code and its contextual information, but different strategies have significant differences in code representation. 
Here we focus on the most widely used code representation in the form of Token Sequence.
We depict five typical code representation ways~\cite{VulRepair,197,49,137,161,145,160,CIRCLE,AlphaRepair} in Fig.~\ref{fig:CR}, which we abbreviate as \textbf{CR1}, \textbf{CR2}, \textbf{CR3}, \textbf{CR4} and \textbf{CR5}.
\textbf{CR1:} Contextual information used for fixes as early as the work of Tufano \textit{et al.}~\cite{145}. The model input is just a piece of source code, but the bug location is not marked. The model does not distinguish between the bug code and the context when translating.
\textbf{CR2:} Chen \textit{et al.}~\cite{197,49} distinguish between bug location and context by adding markers to the bug line locations. Although the model input is still token sequences, the special markers help the model identify the bug locations and thus achieve accurate fixes.
\textbf{CR3:} Unlike Chen \textit{et al.} treat the context and the bug code as one whole input to the encoder, Jiang \textit{et al.}~\cite{137,161} feed the bug code and the context information as two parts of the input to different encoders for separate processing. The insight behind this is that these two parts of information may have different weights in the view of the model. The model should translate the bug code in conjunction with the contextual information instead of processing the context and the bug code uniformly.
\textbf{CR4 and CR5:} CR4 and CR5 are input forms designed for specific models. CIRCLE~\cite{CIRCLE} borrows the pre-training task for the T5 model to set up the prompt template input (CR4) for training. AlphaRepair's~\cite{AlphaRepair} input form (CR5) is designed to accommodate zero-shot learning, so it needs to incorporate CodeBERT's masked language modeling (MLM) pre-training task to design the mask template so that the LLM can directly predict the token at the mask location.
However, these different ways of CR were not compared under a unified baseline, and we have no way of knowing the impact of the different CRs on the results. In addition, it is an open issue how to design appropriate prompt templates for LLMs with specific repair tasks, which still need more guiding insights. Therefore, future empirical studies are needed to bridge this gap and to design more appropriate input code representations in conjunction with repair tasks.

\textbf{Finding 10: Code Tokenization.
\textit{The subword-level tokenizer is suitable for PL scenarios, but it also has limitations.}
}
\label{find:Code Tokenization}
In terms of code tokenization techniques, most early works used the word-level tokenizer~\cite{145,49,197,TransR,DEAR,136,137}, and in recent years the subword-level tokenizer seems to be more popular~\cite{SeqTrans,VulRepair,AlphaRepair,CIRCLE,160,161,162}.
We try to analyze the reasons behind this trend. 

Earlier word-level tokenizer techniques faced the OOV problem. This is because PLs have many custom variable names, method names, and other special tokens, so word-level tokenizers will generate huge vocabularies, which makes it difficult for neural network models to work effectively~\cite{145,49}. Therefore, researchers have proposed various technical details to alleviate this problem. 
For example, Tufano \textit{et al.}~\cite{145} used \textit{code abstraction} to alleviate OOV, Chen \textit{et al.}~\cite{49,197} introduced the \textit{copy mechanism} to alleviate OOV (similarly, TransRepair~\cite{TransR} introduced the \textit{pointer mechanism}), Li \textit{et al.}~\cite{DEAR,136} used the \textit{alpha-rename} method, and CoCoNut~\cite{137} reduced the size of vocabularies by setting rules such as camel and underscore to split complex words.

In recent years, many works have used subword-level tokenizers (e.g., BPE/BBPE/SentencePiece) to accommodate large PL vocabularies~\cite{SeqTrans,VulRepair,AlphaRepair,CIRCLE,160,161,162}. This approach splits a complex word into multiple subwords to alleviate the OOV problem. Therefore it is more suitable for PL tasks. However, the subword-level tokenizer may generate a longer input sequence, e.g., a word "\textit{getMinElement}" is split into "\textit{get}", "\textit{Min}" and "\textit{Element}".
However, current neural network models still face this problem of long sequence processing~\cite{145}, and it is known that the processing ability of LLMs decreases as the input/output sequences grow~\cite{VulRepair}. 
Thus the subword tokenizer may lead to longer input/output sequences, which may affects the repair ability of the model.

In summary, while suitable for repair tasks, the subword-level tokenizer also leads to long sequence problems. Therefore, further improving the tokenization approach and simplifying the code representation are still directions to be worked on.
In addition, different subword tokenization schemes have not been compared uniformly in APR work, and we do not know the differences in the impact of these different technical details on the repair results. Therefore, empirical studies are still needed to explore to guide the selection of appropriate subword tokenizers.

\begin{table}[]
\tiny
\caption{APR works on the Defects4J with different parameter settings when performing patch generation}
\label{tab:tableBeamSize}
\begin{tabular}{|c|c|c|c|c|c|c|c|c|c|}
\hline
Parameter Setting           & SequenceR & DLFix  & CoCoNut & Recoder & CURE  & DEAR  & RewardRepair & CIRCLE & AlphaRepair \\ \hline
beam size                 & 50 & - & 1000 & 100 & 1000 & - & 200 & 250 & 5/25 \\ \hline
running-time limit (hour) & -  & 5 & 6    & 5   & -    & 5 & -   & -   & 5    \\ \hline
Number of candidate patches & -         & - & 20000   & 100     & 10000 & - & 200   & 1000   & 40/5000     \\ \hline
\end{tabular}
\end{table}

\textbf{Finding 11: Patch Generation.}
\label{find: Patch Generation}
\textbf{\textit{During the patch generation phase, the size of the candidate patch space is inconsistent across APR tools, which may lead to unfair performance comparisons.}}
Specifically, we observe that during the patch generation phase, different APR tools set different parameters for the number of patch candidates or set different runtime limits. We summarize the configuration parameters used by some APR tools for patch candidate generation, as shown in Table~\ref{tab:tableBeamSize}. Different APR tools differ in the time limit for patch generation and the size of the number of candidate patches. Generally speaking, a larger search space and a longer runtime increase the chances of generating the correct patches~\cite{160}. Therefore these different configuration differences may result in different APR tools being unfairly compared. We suggest that future work use uniform parameter configurations to compare the performance of different APR tools fairly.



\textbf{Finding 12: Post-processing.
\textit{Different post-processing strategies in APR techniques have yet to be systematically studied.}
}
\label{find: Post-processing}
Different APR efforts have used different post-processing (i.e., patch filtering and ranking) strategies~\cite{DEAR,136,161,AlphaRepair,RING}. Yet, there is a lack of uniform comparisons to reveal the impact of these different strategies on repair results under the same benchmark. It is not clear which filtering and ranking strategies are better or worse.

Specifically, we observe that most learning-based APR tools use the beam search strategy to generate and rank candidate patches. (The details of the beam search are not discussed here, interested readers can refer to the related work~\cite{161}.)
However, the beam search strategy is not a patch filtering strategy, nor is it specifically designed for PL tasks. Therefore, it cannot filter out some patches containing syntax errors nor incorporate PL knowledge to guide patch ranking.
To solve this problem, many researchers have incorporated different post-processing strategies.
For example, Li \textit{et al.}~\cite{DEAR,136} used a program analysis filter to filter patches that did not conform to the PL rules and proposed a CNN-based binary classification model as a patch re-ranking strategy.
Jiang \textit{et al.}~\cite{161} proposed a \textit{code-aware beam-search} strategy, where they designed two techniques (\textit{valid-identifier check} and \textit{length control}) to filter and rank patches.
Xia \textit{et al.}~\cite{AlphaRepair} also used a patch re-ranking strategy, where they used CodeBERT to compute the scores of mask positions to infer the scores of individual predictions.
Joshi \textit{et al.}~\cite{RING} averaged the log probabilities of tokens selected during the decoding process and sorted the candidates in descending order of their averages.
However, a current issue is that these different patch filtering and ranking strategies have not been compared uniformly. Therefore, empirical studies are urgently needed to explore this issue and provide valid insights for selecting and setting appropriate post-processing strategies.

Furthermore, it is worth noting that the post-processing strategy is not only applicable to learning-based techniques, but it is also necessary for other traditional APR techniques as it can effectively improve the patch space to alleviate the patch overfitting problem. However, many early traditional APR techniques (e.g., GenProg, PAR, ARJA, etc.) do not use an additional post-processing step~\cite{217}. In particular, ARJA-e~\cite{217} notes the importance of the post-processing step and thus mitigates the patch overfitting problem through a specially designed post-processing strategy. In addition, HERCULES~\cite{13} uses the machine learning technique to rank and prune candidate patches.
Overall, this reflects the lack of effective post-processing strategies in traditional APR techniques and illustrates the importance of post-processing strategies to improve repair quality. 
Therefore, how to design appropriate post-processing strategies for traditional APR techniques is also something that needs to be explored in the future.


\textbf{Finding 13: Patch Validation.}
\label{find: Patch Validation}
\textbf{\textit{In terms of patch validation strategies, the current validation strategies still need improvement.}}
Current APR techniques mostly use test suits (T.S.) and hand checks (H.C.) as validation methods for small-scale test benchmarks~\cite{49,136,137,Recoder,161,DEAR,AlphaRepair,CIRCLE,TRANSFER}, but this method is inefficient and suffers from patch overfitting problems.
Specifically, test suites are not entirely reliable validation means. For example, weak test suites~\cite{224} can still introduce patch overfitting problems, leading to the screening of incorrect patches or patches with side effects. Therefore, manual review is still required to determine whether patches are correct. In summary, the existing patch validation strategy affects the APR tool repair quality (patch overfitting problem) and repair efficiency (inefficient manual checking).
Therefore, it is a future trend to propose a more optimized patch validation solution and integrate it into the APR tool. For example, introducing patch correctness assessment techniques (Section~\ref{sec:Patch Correctness Assessment}) or automated validation tools~\cite{226} in the patch validation phase to improve validation efficiency and patch quality.
In addition, for large-scale datasets not equipped with test suites, APR efforts typically use exact match (E.M.) to calculate repair accuracy~\cite{SeqTrans,197,VulRepair,162,DEAR,136,154,49,145}. However, there may be multiple repair solutions for a single bug, so the exact match may not be an optimal validation strategy.


\begin{figure*}
    \centering
    \includegraphics[width=1\textwidth]{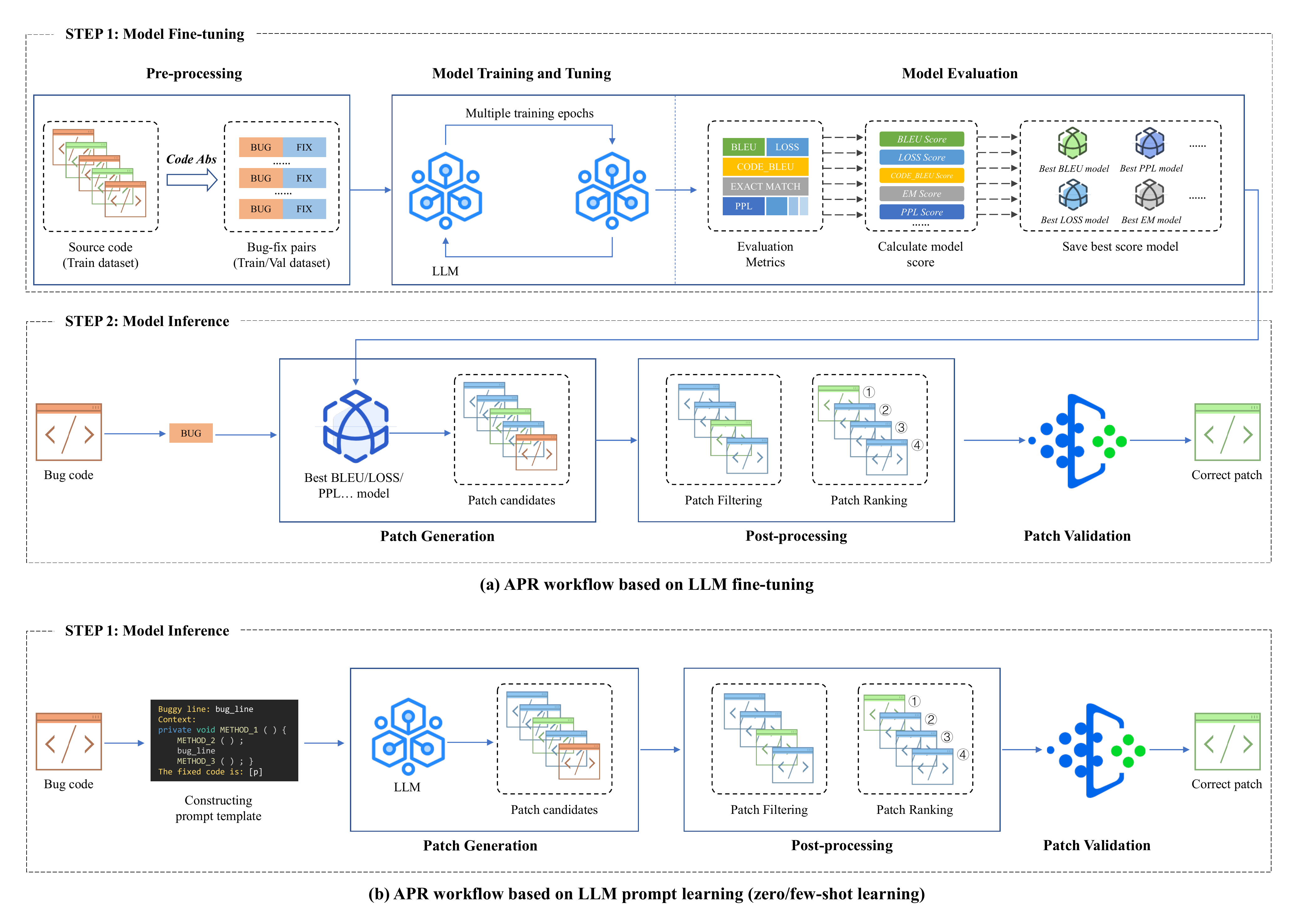}
    \caption{Workflow of LLM applied to APR tasks.}
    \label{fig:LLM2APR}
    \vspace{-3ex}
\end{figure*}

\begin{table}[]
\tiny
\caption{Impact of the best model using different evaluation metrics on repair accuracy}
\label{tab:tableBestModel}
\begin{tabular}{|c|cccc|}
\hline
\textbf{LLM}                             & \multicolumn{4}{c|}{CodeT5}            \\ \hline
\textbf{fine-tuning and testing dataset} & \multicolumn{4}{c|}{VulRepair dataset} \\ \hline
\textbf{Model Evaluation Metric} & \multicolumn{1}{c|}{Best Loss Model}  & \multicolumn{1}{c|}{Best BLEU Model} & \multicolumn{1}{c|}{Best PPL Model} & Last Model \\ \hline
\textbf{Repair Accuracy}         & \multicolumn{1}{c|}{44\% (VulRepair)} & \multicolumn{1}{c|}{55\%}            & \multicolumn{1}{c|}{35\%}           & 55\%       \\ \hline
\end{tabular}
\end{table}

\textbf{Finding 14: Large Language Model.}
\label{find: LLM}
\textbf{\textit{By analyzing the neural network models used in APR research, we observe that learning-based solutions have begun to enter the era of LLMs.}}
In particular, LLMs-based APR tools have outperformed traditional models in repair results~\cite{AlphaRepair}. This is because the LLMs have learned PL knowledge in the pre-training stage, which enables them to better capture and understand the characteristics of PLs and can be widely used in many downstream tasks related to PLs.

Through the review, there are two main ideas for the current LLMs-based solution: (1) Fine-tuning; (2) Prompt learning. As shown in Fig.~\ref{fig:LLM2APR}, we provide the workflow for both types of scenarios.

(1) \textbf{Fine-tuning}~\cite{VulRepair,SeqTrans,197,162,161}. This type of repair scheme fine-tunes the pre-trained LLM on the bug, vulnerability, or error repair dataset so that the LLM can obtain the corresponding repair knowledge and work effectively on the downstream task (APR task). 
We depict the workflow of the fine-tuning-based APR scheme in Fig.~\ref{fig:LLM2APR} (a), and notably, most of the learning-based APR works also conform to this workflow.
Specifically, the raw source code data first goes through the \textbf{pre-processing} phase, where the source code is processed into a format that suits the task requirements with the help of various \textit{Code Abs} methods.
Then the pre-processed training and validation datasets are fed to the model for the \textbf{model training and tuning} phases.
At the same time, after each round of model training, \textbf{model evaluation} is carried out simultaneously by setting various evaluation metrics~\cite{LLM2Metric} (e.g., \textit{BLEU}, \textit{CodeBLEU}, \textit{Loss}, etc.) to derive the best score model. 
After the model fine-tuning step, the best score model will be used in the subsequent model inference step.
In the \textbf{patch generation} phase, the best score model will generate a set of candidate patches for the input buggy code.
However, the candidate patch space may contain many meaningless or non-compilable patches, and the presence of these patches may affect the efficiency of patch validation. Therefore, the researchers use the \textbf{post-processing} phase to perform patch filtering and ranking strategies to filter invalid patches and rank the candidates with the highest probability of correct patches at the front of the validation queue.
Finally, the \textbf{patch validation} phase typically employs test suites and manual checks to obtain the correct patches.

In-depth analysis, we found that technical details in the pre-processing, model evaluation, and post-processing phases may affect LLMs' repair ability and quality. For example, our preliminary experiments (Table~\ref{tab:tableLLM2CodeAbs}) suggest that \textit{Code Abs} may lower the repair effectiveness of LLMs. In addition, we also found that different evaluation metric settings are a focus. For example, VulRepair~\cite{VulRepair} used \textit{loss} scores to evaluate the best model when fine-tuning CodeT5 and obtained a repair accuracy of 44\%. However, our preliminary experiments (Table~\ref{tab:tableBestModel}) show that using \textit{BLEU} scores as the evaluation metric can improve repair accuracy by up to 55\%.
Similarly, different post-processing strategies may have an impact on the repair results.

Overall, these different technical details can affect the repair capability of LLM. 
However, there is still a lack of empirical studies to explore these issues in depth.
As discussed in \textbf{Finding 7 and 12}, various pre-processing and post-processing strategies have been born during the development of learning-based APR techniques. And the AI field has proposed a variety of evaluation metrics~\cite{LLM2Metric}.
It is unclear how to select and set suitable technical details to further improve the LLM-based APR techniques.

(2) \textbf{Prompt learning} (zero/few-shot learning)\cite{AlphaRepair,MMAPR,RING,LLM2APR_fan, LLM2APR_xia, LLM2Vul}. Compared with the fine-tuning scheme, prompt learning emphasizes the LLM's code understanding and generation capabilities. 
As shown in Fig.~\ref{fig:LLM2APR} (b), the prompt learning scheme does not require an additional training step.
The scheme's core is to set a suitable prompt template to guide the LLM to generate output that meets the target expectations. Representative works like AlphaRepair~\cite{AlphaRepair}, RING~\cite{RING}, MMAPR~\cite{MMAPR}, and many empirical studies~\cite{LLM2APR_fan, LLM2APR_xia, LLM2Vul}. These works have shown that prompt learning of LLMs can be applied to bug, error, and vulnerability repair tasks. However, the problem of setting prompt templates is still an area worthy of exploration. More work is needed to provide insights to guide how to set and choose suitable repair templates for different LLMs.

In general, applying LLMs to promote APR research is a future trend. In this process, opportunities and challenges coexist. How to design the more advanced LLMs and how to set up and adjust the technical details to better exploit the potential of the LLMs in the APR task is still a future goal.

\begin{table}[]
\tiny
\caption{APR works evaluated on Defects4J using different train datasets}
\label{tab:tableDatasetDiff}
\resizebox{1.0\columnwidth}{!}{
\begin{tabular}{|c|ccccccccc|ccc|}
\hline
\textbf{Test Benchmark} &
  \multicolumn{9}{c|}{\textbf{Defects4J V1.2.0}} &
  \multicolumn{3}{c|}{\textbf{Defects4J V2.0.0}} \\ \hline
\textbf{APR Tool} &
  \multicolumn{1}{c|}{SequenceR} &
  \multicolumn{1}{c|}{DLFix} &
  \multicolumn{1}{c|}{CoCoNut} &
  \multicolumn{1}{c|}{Recoder} &
  \multicolumn{1}{c|}{CURE} &
  \multicolumn{1}{c|}{RewardRepair} &
  \multicolumn{1}{c|}{DEAR} &
  \multicolumn{1}{c|}{CIRCLE} &
  AlphaReapir &
  \multicolumn{1}{c|}{Recoder} &
  \multicolumn{1}{c|}{RewardRepair} &
  AlphaRepair \\ \hline
\textbf{Train Dataset} &
  \multicolumn{1}{c|}{BFP+CodRep} &
  \multicolumn{1}{c|}{BigFix} &
  \multicolumn{1}{c|}{CoCoNut} &
  \multicolumn{1}{c|}{Recoder} &
  \multicolumn{1}{c|}{CoCoNut} &
  \multicolumn{1}{c|}{CoCoNut+MegaDiff+CodRep} &
  \multicolumn{1}{c|}{CPatMiner} &
  \multicolumn{1}{c|}{CoCoNut} &
  CodeSearchNet &
  \multicolumn{1}{c|}{Recoder} &
  \multicolumn{1}{c|}{CoCoNut+MegaDiff+CodRep} &
  CodeSearchNet \\ \hline
\textbf{\#Bug-fix Pairs} &
  \multicolumn{1}{c|}{35,578} &
  \multicolumn{1}{c|}{25,936} &
  \multicolumn{1}{c|}{3,241,966} &
  \multicolumn{1}{c|}{103,585} &
  \multicolumn{1}{c|}{2.73 million} &
  \multicolumn{1}{c|}{3,241,966+240,306+24,969} &
  \multicolumn{1}{c|}{44,154} &
  \multicolumn{1}{c|}{400,000} &
  2 million &
  \multicolumn{1}{c|}{103,585} &
  \multicolumn{1}{c|}{3,241,966+240,306+24,969} &
  2 million \\ \hline
\end{tabular}
}
\begin{tablenotes}
\item{
Note that CodeSearchNet is not a bug-fix pairs dataset. It consists of 2 million (comment, code) pairs that are used to pre-train LLMs.}
\end{tablenotes}
\end{table}

\begin{table}[]
\tiny
\caption{Results of dataset overlap between train dataset and test benchmark}
\label{tab:tableDatasetOverlap}
\resizebox{1.0\columnwidth}{!}{
\begin{tabular}{|c|ccccccccccccccccc|c|}
\hline
\multirow{2}{*}{\textbf{Dataset}} &
  \multicolumn{17}{c|}{\textbf{Defects4J}} &
  \multirow{2}{*}{\textbf{ALL}} \\ \cline{2-18}
 &
  \multicolumn{1}{c|}{Chart} &
  \multicolumn{1}{c|}{Closure} &
  \multicolumn{1}{c|}{Lang} &
  \multicolumn{1}{c|}{Math} &
  \multicolumn{1}{c|}{Mockito} &
  \multicolumn{1}{c|}{Time} &
  \multicolumn{1}{c|}{Cli} &
  \multicolumn{1}{c|}{Codec} &
  \multicolumn{1}{c|}{Collections} &
  \multicolumn{1}{c|}{Compress} &
  \multicolumn{1}{c|}{Csv} &
  \multicolumn{1}{c|}{Gson} &
  \multicolumn{1}{c|}{JacksonCore} &
  \multicolumn{1}{c|}{JacksonDatabind} &
  \multicolumn{1}{c|}{JacksonXml} &
  \multicolumn{1}{c|}{Jsoup} &
  JxPath &
   \\ \hline
\textbf{BFP} &
  \multicolumn{1}{c|}{-} &
  \multicolumn{1}{c|}{-} &
  \multicolumn{1}{c|}{-} &
  \multicolumn{1}{c|}{-} &
  \multicolumn{1}{c|}{-} &
  \multicolumn{1}{c|}{-} &
  \multicolumn{1}{c|}{-} &
  \multicolumn{1}{c|}{-} &
  \multicolumn{1}{c|}{-} &
  \multicolumn{1}{c|}{-} &
  \multicolumn{1}{c|}{-} &
  \multicolumn{1}{c|}{-} &
  \multicolumn{1}{c|}{-} &
  \multicolumn{1}{c|}{-} &
  \multicolumn{1}{c|}{-} &
  \multicolumn{1}{c|}{-} &
  - &
  - \\ \hline
\textbf{CodRep} &
  \multicolumn{1}{c|}{-} &
  \multicolumn{1}{c|}{-} &
  \multicolumn{1}{c|}{17} &
  \multicolumn{1}{c|}{26} &
  \multicolumn{1}{c|}{-} &
  \multicolumn{1}{c|}{-} &
  \multicolumn{1}{c|}{-} &
  \multicolumn{1}{c|}{5} &
  \multicolumn{1}{c|}{3} &
  \multicolumn{1}{c|}{8} &
  \multicolumn{1}{c|}{2} &
  \multicolumn{1}{c|}{-} &
  \multicolumn{1}{c|}{-} &
  \multicolumn{1}{c|}{-} &
  \multicolumn{1}{c|}{-} &
  \multicolumn{1}{c|}{-} &
  - &
  61 \\ \hline
\textbf{BigFix} &
  \multicolumn{1}{c|}{-} &
  \multicolumn{1}{c|}{-} &
  \multicolumn{1}{c|}{-} &
  \multicolumn{1}{c|}{-} &
  \multicolumn{1}{c|}{-} &
  \multicolumn{1}{c|}{-} &
  \multicolumn{1}{c|}{-} &
  \multicolumn{1}{c|}{-} &
  \multicolumn{1}{c|}{-} &
  \multicolumn{1}{c|}{-} &
  \multicolumn{1}{c|}{-} &
  \multicolumn{1}{c|}{-} &
  \multicolumn{1}{c|}{-} &
  \multicolumn{1}{c|}{-} &
  \multicolumn{1}{c|}{-} &
  \multicolumn{1}{c|}{-} &
  - &
  - \\ \hline
\textbf{CPatMiner} &
  \multicolumn{1}{c|}{-} &
  \multicolumn{1}{c|}{-} &
  \multicolumn{1}{c|}{-} &
  \multicolumn{1}{c|}{-} &
  \multicolumn{1}{c|}{-} &
  \multicolumn{1}{c|}{2} &
  \multicolumn{1}{c|}{2} &
  \multicolumn{1}{c|}{-} &
  \multicolumn{1}{c|}{-} &
  \multicolumn{1}{c|}{-} &
  \multicolumn{1}{c|}{2} &
  \multicolumn{1}{c|}{-} &
  \multicolumn{1}{c|}{1} &
  \multicolumn{1}{c|}{3} &
  \multicolumn{1}{c|}{-} &
  \multicolumn{1}{c|}{-} &
  - &
  10 \\ \hline
\textbf{CoCoNut 2006} &
  \multicolumn{1}{c|}{-} &
  \multicolumn{1}{c|}{-} &
  \multicolumn{1}{c|}{2} &
  \multicolumn{1}{c|}{-} &
  \multicolumn{1}{c|}{-} &
  \multicolumn{1}{c|}{-} &
  \multicolumn{1}{c|}{-} &
  \multicolumn{1}{c|}{-} &
  \multicolumn{1}{c|}{-} &
  \multicolumn{1}{c|}{-} &
  \multicolumn{1}{c|}{-} &
  \multicolumn{1}{c|}{-} &
  \multicolumn{1}{c|}{-} &
  \multicolumn{1}{c|}{-} &
  \multicolumn{1}{c|}{-} &
  \multicolumn{1}{c|}{-} &
  - &
  2 \\ \hline
\textbf{MegaDiff} &
  \multicolumn{1}{c|}{-} &
  \multicolumn{1}{c|}{54} &
  \multicolumn{1}{c|}{-} &
  \multicolumn{1}{c|}{26} &
  \multicolumn{1}{c|}{6} &
  \multicolumn{1}{c|}{4} &
  \multicolumn{1}{c|}{-} &
  \multicolumn{1}{c|}{4} &
  \multicolumn{1}{c|}{-} &
  \multicolumn{1}{c|}{-} &
  \multicolumn{1}{c|}{-} &
  \multicolumn{1}{c|}{-} &
  \multicolumn{1}{c|}{-} &
  \multicolumn{1}{c|}{4} &
  \multicolumn{1}{c|}{-} &
  \multicolumn{1}{c|}{-} &
  1 &
  99 \\ \hline
\textbf{Recoder} &
  \multicolumn{1}{c|}{-} &
  \multicolumn{1}{c|}{4} &
  \multicolumn{1}{c|}{-} &
  \multicolumn{1}{c|}{4} &
  \multicolumn{1}{c|}{-} &
  \multicolumn{1}{c|}{-} &
  \multicolumn{1}{c|}{-} &
  \multicolumn{1}{c|}{-} &
  \multicolumn{1}{c|}{-} &
  \multicolumn{1}{c|}{-} &
  \multicolumn{1}{c|}{-} &
  \multicolumn{1}{c|}{-} &
  \multicolumn{1}{c|}{-} &
  \multicolumn{1}{c|}{-} &
  \multicolumn{1}{c|}{-} &
  \multicolumn{1}{c|}{-} &
  - &
  8 \\ \hline
\textbf{CodeSearchNet} &
  \multicolumn{1}{c|}{-} &
  \multicolumn{1}{c|}{3} &
  \multicolumn{1}{c|}{7} &
  \multicolumn{1}{c|}{3} &
  \multicolumn{1}{c|}{3} &
  \multicolumn{1}{c|}{10} &
  \multicolumn{1}{c|}{9} &
  \multicolumn{1}{c|}{-} &
  \multicolumn{1}{c|}{-} &
  \multicolumn{1}{c|}{1} &
  \multicolumn{1}{c|}{-} &
  \multicolumn{1}{c|}{1} &
  \multicolumn{1}{c|}{-} &
  \multicolumn{1}{c|}{-} &
  \multicolumn{1}{c|}{-} &
  \multicolumn{1}{c|}{11} &
  - &
  48 \\ \hline
\end{tabular}
}
\begin{tablenotes}
\item{
The results in the table reflect the dataset overlap that exists between each dataset and Defects4J. For example, CodRep overlaps with 61 bugs in Defects4J.}
\end{tablenotes}
\end{table}

\textbf{Finding 15: Dataset.
\textit{The training and testing data used in the APR research exposed potential problems, which have not received widespread attention.}
}
\label{find: Dataset}
By analyzing the train datasets and test benchmarks of the DL-based APR works, we obtained three critical observations: first, the unfair experimental data evaluation; second, the overlap problem between train and test data; and third, the significant dataset quality problem.

First, unfair experimental evaluation.
Looking at the training and testing datasets, we found many unfair comparisons. Specifically, many works test on the same test benchmark, yet their training data were inconsistent.
As shown in Table~\ref{tab:tableDatasetDiff}, many APR tools test on Defects4J, yet their train datasets differ.
It is well known that data-driven APR techniques gain repair knowledge from training data, and different train datasets will result in different empirical knowledge for APR tools. Therefore, the differences due to train datasets may lead to unfair experimental evaluation. Therefore, we suggest that future work choose uniform training data to develop evaluation comparisons.

Second, the dataset overlap problem (data leakage).
This problem refers to the presence of overlapping samples in the train dataset and test benchmark, which may make the model learn the repair knowledge of the test samples during the training phase, resulting in the illusion of obtaining better repair results.
In recent years, the overlap problem has also gained the attention of researchers. For example:
CoCoNut~\cite{137} distinguishes whether there is overlap with the test set by year; Recoder~\cite{Recoder} uses AST matching to filter out train dataset that overlap with the test benchmark; TRANSFER~\cite{TRANSFER} removed four projects that overlapped with Defects4J; AlphaRepair~\cite{AlphaRepair} also found data overlap between CodeBERT's pretraining data and Defects4J.
However, we do not know which data overlap with Defects4J in the current large-scale training dataset.
To shed light on this issue and provide a reference for subsequent work to remove the overlapping data, we conducted preliminary experiments to explore the overlap of each train dataset with Defects4J. 
Similar to Xia \textit{et al.}~\cite{AlphaRepair}, we use exact matching to detect the presence of Defects4J's bug/fix function in the training data.
Table~\ref{tab:tableDatasetOverlap} provides the preliminary results of the dataset overlap. In addition, we also mark which data in the train dataset overlap with the test benchmark for subsequent work to remove the overlap data.
(Due to page limitations, we have provided detailed information on the website~\footnote{https://github.com/huangkNIPC/APR-Survey/tree/main/dataset\_overlap}.)
As shown in Table~\ref{tab:tableDatasetOverlap}, multiple datasets were exposed to overlap problems. We hope that this preliminary experiment will attract the attention of researchers and explore the broader dataset overlap problem in depth. In the future, using finer-grained data overlap detection to analyze the overlap problems on more benchmarks will remain a key topic for APR research.

Third, the dataset quality problem.
In our manual inspection of the dataset, we found invalid or incorrect data samples in some datasets. For example, we found 38 (38/8,482=0.45\%) invalid samples in the dataset used by VulRepair~\cite{VulRepair}, which did not provide corresponding fix codes. In addition, there were also 671,497 (671,497/3,241,966=20.71\%) invalid or incorrect samples in CoCoNut~\cite{137}, and the bug codes and fix codes in these samples were identical.
These potential dataset quality problems may affect the model's learning of repair knowledge, and the invalid training samples may increase the model training overhead or even cause side effects to the model training.

Overall, we call on the APR community to focus on dataset-related issues in the future, as these issues have been neglected in historical studies. We hope that future empirical studies will use uniform training data to compare APR efforts fairly. In addition, we suggest that the APR community collectively explore dataset overlap issues and dataset quality issues, which can provide insights into the construction of high-quality training data and the appropriate selection of training and testing benchmarks.

\textbf{Finding 16: Multi-Type Fixes.
\textit{
Multi-type repair has been achieved; however, it remains to be explored whether the different defect problems can contribute to each other.
}
}
\label{find:MT Multi-Type Fixes}
Most APR tools support multi-type fixes (e.g., fixing multiple CWE types~\cite{VulRepair,197,SeqTrans}). However, current APR tools do not have the ability to fix different defect problems simultaneously (i.e., fix error/bug/vulnerability at the same time), and it is unclear whether the repair knowledge of different defect issues can contribute to each other.

Specifically, most APR tools are specially designed for different defect repair scenarios. Their repair targets are either error~\cite{TransR,143}, bug~\cite{DEAR,CIRCLE}, or vulnerability~\cite{197,VulRepair}. Thus, an error repair tool may not have the bug repair ability. This makes it possible for developers to use different APR tools to solve different defect problems in the real fixing process. Therefore, the possibility of implementing a unified APR tool to integrate the ability to fix different defect problems is a path worth exploring.
Essentially, whether it is a compilation error or a software bug, they are both PL correction tasks and can even be fixed using the same workflow (NMT) implementation.

In addition, whether repair knowledge among different defect problems can contribute to each other has yet to be explored in depth.
The closest case is that VRepair~\cite{197} and SeqTrans~\cite{SeqTrans} have facilitated vulnerability repair by combining bug repair knowledge with transfer learning. This stimulates us to consider that there is currently a lack of correlation and facilitation between different repair problems. A deeper study of the relationship between different defect problems may improve APR tools' repair ability. Like meta-learning, it is an open issue whether we can leverage empirical knowledge from multiple defect problem domains simultaneously to facilitate fixing.

In summary, we believe that building unified APR tools to support multiple defect problem repair scenarios is a future trend. Furthermore, it is still a worthwhile exploration to leverage repair knowledge from different defect problem domains to mutually facilitate repair ability.

\textbf{Finding 17: Multi-Language Fixes.}
\label{find:Multi-Language Fixes}
\textbf{\textit{Learning-based APR techniques make multi-language repair possible, yet the root causes of defects behind different languages are not well exploited.}} 
Since most traditional APR techniques require hard-coding of rules or templates, they require designing different implementations of the tool for different PLs (e.g., the original implementation of GenProg was evaluated on C, and its alternative implementation jGenProg supports Java~\cite{ASTOR}), which makes them difficult to make an APR tool with ML repair ability simultaneously.

Fortunately, DL-based APR techniques allow an APR tool to learn and gain empirical knowledge of repairing multiple PLs simultaneously by leveraging the powerful learning capabilities of neural networks.
We found that DL-based APR tools supporting multi-language repair ability usually work in three ways: the first one is by training multiple models supporting different PLs so that APR tools can obtain multi-language repair ability~\cite{137}; the second one is by using multi-task learning or continual learning to learn repair knowledge in different PLs so that one model (one tool) acquires multi-language repair ability~\cite{CIRCLE}; the other one is multi-language repair by using LLMs that support PLs~\cite{AlphaRepair}, thanks to the fact that these LLMs are pre-trained on multiple PLs, which makes them inherently knowledgeable in multiple PLs.

However, whether different PLs' repair knowledge can contribute to each other has not been deeply explored. For example, the same type of bugs may appear in different PLs, but the root cause behind defects is the same. Therefore, it is worth exploring whether the model can acquire high-level bug repair knowledge of different PLs by learning the in-depth root cause of defects. Learning the deep root causes of defects will enable DL-based APR techniques to further break through the limitations of PLs to enhance repair ability.

\begin{figure*}
    \centering
    \includegraphics[width=1\textwidth]{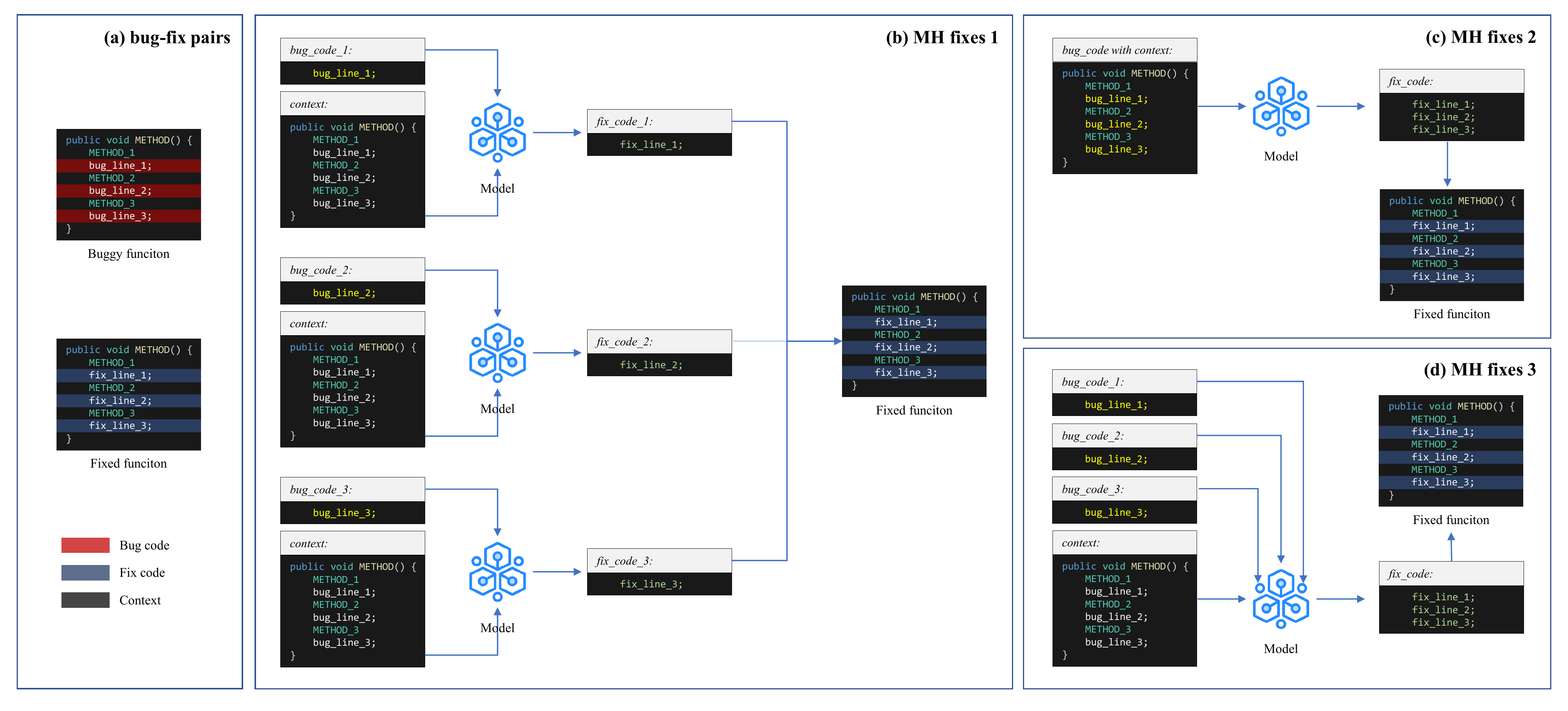}
    \caption{Three multi-hunk fixes approaches.}
    \label{fig:MH_fixes}
    \vspace{-3ex}
\end{figure*}

\textbf{Finding 18: Multi-hunk Fixes.
\textit{The multi-hunk repair issue was not completely resolved.}
}
\label{find:Multi-hunk Fixes}
The multi-hunk repair issue has been alleviated during the research process of learning-based APR techniques, and the SOTA DL-based APR tool (DEAR~\cite{DEAR}) has surpassed the APR tools of traditional techniques (HERCULES~\cite{13}) in MH repair ability. However, after revisiting MH APR works, we found that many challenges remain.
There are three main solutions currently used by learning-based APR techniques to address MH repair (we provide an overview in Fig.~\ref{fig:MH_fixes}):

(1) Iterative repair~\cite{TransR}. As shown in Fig.~\ref{fig:MH_fixes} (b), the iterative approach fixes one bug location at a time, generates multiple candidate patches for different bug locations through multiple iterations, and finally combines multiple candidate patch sets to obtain the final patch space. The advantage of this scheme is that it can extend the single-hunk APR tools to support multi-hunk fixes~\cite{DEAR}. However, this approach has significant drawbacks in two major ways.
First, iterative fixes generate a combined solution of multiple candidate patch sets, which can lead to a huge search space \cite{13}. For example, if there are 3 bug hunks, iteratively generate 10 candidate patches for each bug hunk. By combining the patch sets of the 3 bug hunks, then the final patch space is 10×10×10, which further increases the cost of patch validation.
Second, iterative fixes may use incorrect contextual content, which may interfere with the model to predict the correct fix \cite{DEAR}. For example, in Fig.~\ref{fig:MH_fixes} (b) we are trying to fix \textit{bug\_line\_1}, but at this time the context has \textit{bug\_line\_2} and \textit{bug\_line\_3} as the correct context information, which is obviously an incorrect assumption. We hope that the context information is the correct code implementation in addition to the \textit{bug\_line\_1}.

(2) Holistic repair~\cite{197}. As shown in Fig.~\ref{fig:MH_fixes} (c), holistic fixes identify the bug locations in the context with special mark tokens, then input the marked bug code and the context as a whole to the model. The model's output is a one-time prediction of patches for different bug locations. Holistic repair is a more concise solution than iterative repair, widely used by VRepair~\cite{197} and VulRepair~\cite{VulRepair}.
However, this approach does not input the bug code and context separately into different encoders or assign different weights, which makes the model unable to process the bug code and context separately. In addition, MMAPR~\cite{MMAPR}, which is based on few-shot learning, also adopts a similar approach. It provides multiple bug locations for the LLM in its prompt template, but it is still unclear how the black-box LLM handles and fixes the defects in multiple locations internally.

(3) Synchronous repair~\cite{DEAR}. As shown in Fig.~\ref{fig:MH_fixes} (d), the model learns the fixing transformations of multiple associated bug locations simultaneously and can implement the fixing behavior of multiple bug locations simultaneously.
DEAR~\cite{DEAR}, one of the most advanced multi-hunk repair tools, uses a divide-and-conquer strategy to simultaneously learn the fixing behavior of multiple bug locations in a bug. 
Specifically, it designs a two-layer tree RNN model, which learns contextual knowledge and the repair behavior of multiple fix sub-trees separately. Such an approach allows it to avoid the limitation of iterative repair to learn the incorrect contextual features and the shortcoming of treating bugs and contexts with uniform weights.
However, the approach requires the source code to be converted into the AST to obtain the corresponding bug-fix subtree and contextual features. Some compilation errors may not be converted into correct AST since they may carry syntactic structure errors, so this scheme is difficult to extend to syntax error fixes.
In addition, DEAR's scheme learns one-to-one bug-fix subtree fixing relationships, which means that the bug location and the fix location correspond to one-to-one. However, the bug location and the fix location are not the same for some special bugs. For example, it may be necessary to add fixing behavior at multiple non-contiguous locations for a single bug location. DEAR may not suit a repair scenario where one bug hunk corresponds to multiple fix hunks.

In conclusion, MH repair is still a long-term challenge for APR research, and we still need to continuously improve MH APR solutions to accommodate more complex defect repair scenarios and further enhance the repair ability of APR tools.

\textbf{Finding 19: Multi-fault Fixes.}
\label{find:Multi-fault Fixes}
\textbf{\textit{As reviewed in the Table~\ref{tab:table2-3-2}, the multi-fault (or multi-defect) repair is still an open issue.}} 
Even though the multi-hunk repair problem has been alleviated in the learning-based APR research process, the multi-fault repair has not been specifically studied.

We observed that most error repair efforts~\cite{TransR,142,143,150,163,164} support multi-error repair and explored this ability. However, the multi-bug repair is still not explored in depth in bug repair efforts, and we do not know the effectiveness of these APR tools in solving the multi-bug problem.
This is because the complexity of different defect (fault) problems is different. 
Compared to software bugs, programming errors are a relatively simple class of defect problems. For example, compilation errors may have a few complex logical dependencies between multiple errors. 
Therefore many error repair tools take an iterative approach to fixing multiple errors.
In principle, bug repair tools can also iteratively fix multiple bugs in a program, treating them as multi-hunk bugs. However, logical bugs are more complex defect problems that require test cases for defect triggering and patch validation, and there are likely to be complex dependencies between multiple bugs. Therefore, for bug repair, the implementation of multi-bug repair needs to be built based on multi-hunk repair ability.
In addition, as neural network models usually have a limited input/output length, they cannot handle excessively long input sequences. This also means they have difficulty handling defect scenarios with excessively long contextual information. 
Whereas multi-hunk and multi-defect repair scenarios often have complex contextual information, which may exceed the maximum input length of the model, resulting in the model not capturing the full contextual information.
Therefore, models' input/output length limitations also limit the repair ability.

In conclusion, given that the current APR tools do not explore the multi-bug repair ability, we suggest that more empirical studies explore this issue and that APR works should also consider this repair ability when conducting experimental evaluations. 
In addition, future work also should explore ways to alleviate the limitations imposed by the model's limited input/output length.





\textbf{Finding 20: Repair Quality and Repair Efficiency.
\textit{There is a lack of systematic research on repair quality and repair efficiency issues.}
}
\label{find:Repair Quality and Repair Efficiency}
We do not provide a uniform perspective for presenting repair quality and repair efficiency, not because they are unimportant (on the contrary, they are two critical topics), but because only a small part of the work in the APR research history has explored these elements.


While exploring repair quality, most works have analyzed the patch quality of APR tools only regarding patch overfitting~\cite{CIRCLE,AlphaRepair,DEAR,Recoder,136,161}. However, this is incomplete. For example, the extent to which developers accept patches~\cite{73}, the possibility of patches introducing side effects~\cite{10}, the ease of maintenance of patches~\cite{77}, etc., all reflect repair quality issues.
Specifically, some earlier works~\cite{62,10} assessed repair quality by analyzing whether patches might introduce new defects, and other researchers~\cite{73,77,74,79,154} analyzed repair quality by manually assessing patch acceptability, maintainability, or understandability.
However, there is no uniform assessment metric system for repair quality assessment. In addition, there is still a lack of large-scale empirical studies to explore the differences in repair quality among different families of APR works. We still need a deeper understanding of repair quality issues.

During the exploration of repair efficiency, we cannot provide a uniform view for fair comparison because different APR tools may operate under different experimental settings. In a previous study, Liu \textit{et al.}~\cite{40} assessed the repair efficiency of traditional APR techniques and explored the issue of time cost. In addition, another study~\cite{127} evaluated the money cost of APR tools to repair bugs in a unified cloud environment.
Recently, Martinez \textit{et al.}~\cite{APR2energy} started to explore the energy consumption of APR, which provides a new perspective to analyze repair efficiency.
However, in-depth research on the repair efficiency of recent learning-based APR tools is lacking. Future empirical studies should comprehensively explore and compare the issue of repair efficiency among each type of APR technique by setting up a uniform experimental benchmark environment.

Since repair quality and repair efficiency require a uniform experimental setting and manual review to produce accurate results, this is beyond the scope of a survey and needs to be investigated in an empirical study.
Overall, there is a lack of uniform metrics and principles to guide the assessment of repair quality and efficiency. 
Based on this, we provide some assessment perspectives on repair quality and efficiency in Section~\ref{sec:2-3-evaluation criteria}, which aims to provide a reference for subsequent work.
Future work should reveal the strengths and weaknesses of APR techniques in terms of repair quality and comprehensively compare the differences in repair efficiency between different APR techniques. This will help us to continuously improve APR techniques and accelerate the pace of industrial deployment of APR techniques.

\textbf{Finding 21: 
Learning Paradigm. 
\textit{Learning paradigms suitable for APR tasks have yet to be explored in depth.}}
\label{find:Learning Paradigm}
By analyzing the paradigms of repair tasks used by learning-based APR techniques, NMT techniques are currently the most popular repair paradigm~\cite{DEAR,136,137,161,49,197,SeqTrans,160}. However, whether there are more suitable paradigms for repair tasks remains to be explored.

We believe that NMT is a suitable processing paradigm for APR tasks, but it is not necessarily an optimal paradigm for repair tasks. Specifically, the NMT task requires many labeled samples (bug-fix pairs) to train the model to learn empirical knowledge. However, crawling and labeling many samples is tedious, especially since vulnerability fixes also suffer from the small sample problem~\cite{SeqTrans,197}. 
In addition, current NMT repair solutions only learn about repair knowledge from bug-fix pairs, while other source code information within the same project (e.g., code outside of the bug method and fix method) is not fully utilized. This results in the vision of the NMT approach being able to observe only a limited part of a software project rather than being able to take a global view of the whole project to guide fixes.
In general, although the NMT solution is practical, it has the above shortcomings. How to alleviate the above limitations remains an open issue.

In contrast to the NMT approach, which requires many bug-fix pairs to train for the support model, pre-trained models only need to provide the correct code to achieve unsupervised learning. For example, the MLM pre-training task of the BERT family of models learns the relationship between code elements by masking some tokens.
With the help of the pre-trained model of the MLM task, the repair task can be regarded as the \textit{cloze task}. Predict fix codes by combining contextual content instead of translating bug codes to fix codes like NMT tasks.
A recent case is that AlphaRepair's repair approach~\cite{AlphaRepair} does not use NMT tasks but treats repair tasks as cloze tasks. It predicts the correct code by incorporating the contextual content and achieves the best results on Defects4J. The insight behind this is that AlphaRepair uses CodeBERT, a pre-training LLM. The success of AlphaRepair shows the great potential of LLMs in APR research, and the pre-training tasks of these LLMs are also effective for repair tasks. This also shows that the NMT paradigm may not be the optimal solution for the APR task, and the new learning paradigm is worth exploring in the future.

\section{Related Advance Work}
\label{sec:Related Work}

Previous section has reviewed the APR techniques. 
However, for the advancement of APR, relevant studies also play an important role in the detailed technical aspects. 
Next, we describe the latest advances and representative works in related areas.
This section is intended to facilitate the reader's knowledge of related techniques and incorporate new methods into APR research in the future.

\subsection{APR Evaluation}
\label{sec:APR Evaluation}

APR evaluation is a specialized study to evaluate and analyze APR techniques. 
It aims to analyze and evaluate performance metrics, working principles, and repair quality, dig deeper into the problems and technical pain points in the APR field, and point the way for future research. 
Next, we present some typical APR evaluation efforts and point out topics that can be further explored in the future.

Back in 2014, the work of Monperrus~\cite{74} focused on the evaluation criteria for automated repair: understandability, correctness, and completeness. 
Whereas early works~\cite{42,48,62,73,124} used the number of defects fixed to evaluate the effectiveness of APR tools, later works~\cite{11,13,43,44,45,77,83,102,124,166} added correctness, reasonableness, maintainability, and other metrics to help compare the performance of APR tools. 
Their evaluations are only used to compare their APR tools with others, and no systematic, large-scale experimental evaluations have been conducted. Next, we focus on large-scale evaluation works.

In 2019, Durieux \textit{et al.}~\cite{169} analyzed the repairability of APR tools in different benchmarks. 
Importantly, their evaluation metrics revealed the \textit{benchmark overfitting problem}, which suggests that APR tools should be tested on multiple benchmarks to reflect true performance.

In 2020, Liu \textit{et al.}~\cite{51} focused on repair effectiveness, repair efficiency, and the impact of FL. They found that FL has a significant impact on both effectiveness and efficiency. 
These evaluation metrics are designed to explore the repair efficiency problem and provide key insights for improving efficiency.
Later, Liu \textit{et al.}~\cite{40} proposed a more comprehensive set of evaluation metrics, which to reduce the biases in the evaluation metrics of prior works
to evaluate the performance of the repair tools in a fairer and more detailed way. 

In 2021, Aleti \textit{et al.}~\cite{170} introduced machine learning to reveal APR performance by learning software features. 
They proposed an evaluation framework, Explaining Automated Program Repair (E-APR). 
E-APR identified nine significant features.
These features are used to reveal the strengths and weaknesses of each APR technique, which is a great help for future APR improvement efforts.

Overall, the establishment of evaluation metrics is a process of continuous addition and improvement.
These efforts have provided strong support for an unbiased comparison of the performance of APR tools. Importantly, they dig deeper into important factors affecting program repair, which provides insights for the continuous improvement of APR techniques. 
However, previous evaluation efforts have focused on traditional APR techniques, and learning-based APR techniques have not been studied in depth.
As such, we hope that future work will focus on this gap.

\subsection{Fault Localization}
\label{sec:Fault Localization}


The study of FL has received the attention of researchers before APR research.
FL techniques~\cite{175} aim to discover and analyze the location and causes of faults, which fall into two main categories.

Dynamic FL techniques, which usually analyze the dynamic execution information of a program to determine the location of defects, such as spectrum-based fault localization (SBFL)~\cite{179}. 
These methods allow fine-grained FL for programs, but the drawback is obvious that they are very resource intensive for programs. 
Static FL techniques, these techniques determine the fault location by semantic or syntactic analysis at the bug report or source code level, but they cannot achieve fine-grained detection. However, it is used for fast location due to its low resource consumption and fast detection. Typical techniques such as information retrieval-based fault localization (IRFL)~\cite{182}.

For APR research, FL techniques determine the quality of repair from the beginning, and more advanced FL techniques will guide APR tools to find more errors and achieve fixes. 
The multiple fault localization techniques (MFL)~\cite{178} and even FL techniques that combine dynamic and static methods~\cite{184} have emerged, and immediate defect prediction and localization frameworks~\cite{185}. We also expect that more new FL techniques will emerge to assist the APR field.

\subsection{Patch Correctness Assessment}
\label{sec:Patch Correctness Assessment}

The \textit{patch overfitting problem} has long plagued researchers in the APR field. 
It has become a difficult task to filter the truly valid and correct patches from many candidate patches. 
In recent years, many works have focused on patch correctness~\cite{36,202,203,204,205}, and the concept of \textit{Automated Patch Correctness Assessment} (APCA) technique was introduced in the literature~\cite{202}.

Existing APCA techniques are mainly classified into static and dynamic approaches, where static approaches sort and filter patches by analyzing their static characteristics, and dynamic approaches often identify correct patches with the help of automated test generation tools. 
In the work of Xiong \textit{et al.}~\cite{203} in 2018, patch correctness was identified by observing the similarity of the behavior of patching programs as they passed the tests, which successfully prevented 56.3\% of incorrect patches. 
In 2020, Tian \textit{et al.}~\cite{204} proposed to use deep representational learning for patch correctness prediction, which also showed good results.
However, the effectiveness and the reasons behind these APCA tools are unknown, and there is an urgent need to explore them. Therefore, several empirical studies~\cite{202,203,204,205} have also analyzed the effectiveness of APCA techniques. In 2020, Le \textit{et al.}~\cite{202} conducted an experimental analysis of nine APCA techniques. 
They found that all of them had shortcomings.
They propose to combine multiple methods to improve the performance of patch correctness assessment, which is also a direction worth exploring in the future by combining dynamic and static methods.

Manual patch assessment is a complex, time-consuming and biased task, so automated patch assessment methods are a much-needed solution. Currently, the emergence of APCA techniques has positive implications for patch evaluation, helping to assess the effectiveness of APR techniques and improving the repair quality. 
In further development, it is worthwhile to expect the integration of APCA techniques into APR tools to improve repair performance.

\subsection{Programming Language Model}
\label{sec:PL Model}

In learning-based APR techniques, neural network models serve as one of the critical components that affect the performance of APR tools. With the above review, 
we note that these state-of-the-art efforts \cite{157,160,161} focus on enabling models to learn PL knowledge better to improve the repair quality.
APR research has shifted from NL models to PL models to work better on PL repair tasks. 

In contrast to NL models, PL models need to be designed with features such as the strict syntactic structure and logical relationships of PLs and also take into account dependencies such as data flow and control flow of code statements.
In recent years, large PL models (or large language model trained on code, LLMC) have been extensively studied~\cite{CodeModel}. 
Importantly, LLMCs have been successfully applied to APR tasks and have shown effective results~\cite{CodexVul,AlphaRepair,LLM2APR_fan,LLM2APR_xia,LLM2Vul,MMAPR,RING}.
These works show that PL models have shown promise for APR. 

Thanks to the efforts of AI researchers, LLMs possess even better code processing capabilities. 
Therefore, prioritizing the development and design of LLMCs will further advance various PL-related tasks and provide greater technical support for APR research.

\section{Challenges and Directions}
\label{sec:Challenges and Directions}
 

After reviewing the evolution of APR techniques (Section ~\ref{sec:Search-based}-\ref{sec:Learning-based}) and conducting a thorough analysis of the SOTA works (Section~\ref{sec:Summary and Discussion}), we have identified the key challenges and future directions in the development of APR research.


\subsection{Existing Challenges}
\label{sec:Existing Challenges}

\subsubsection{Patch Overfitting Problem}
\label{Patch Overfitting Problem}

APR tools usually use test cases to verify that the patches have been successfully fixed. 
But test cases usually do not cover all tests completely, and even if the patch passes all test cases, it does not mean that the patch is completely correct. 
This is known as the \textit{patch overfitting problem}~\cite{4,6}, also known as the \textit{patch correctness problem}~\cite{44,5}. 
As discussed in \textbf{Finding 5}, patch overfitting is a long-standing problem.
Numerous analytical evaluations~\cite{36,41,79,206,207} have been carried out by researchers to study this problem.
Researchers are attempting to mitigate the \textit{patch overfitting problem} in terms of both enhanced test suite techniques~\cite{208,209} and optimization of the patch generation process~\cite{42,44,160}.
In general, the \textit{patch overfitting problem} is accompanied by many influencing factors. Fault localization accuracy, patch generation capability, post-processing strategies, and patch validation techniques all directly or indirectly affect the degree of patch overfitting. Therefore, solving patch overfitting is not an overnight task, requiring researchers to improve many technical details to mitigate the overfitting problem gradually.
Nowadays, there is still a lack of mature and effective solutions, and how to mitigate the \textit{patch overfitting problem} is still a future challenge.

\subsubsection{Repair Quality Problem}
\label{Repair Quality Problem}

The repair quality issue~\cite{6,79} can be understood as whether the APR tools introduce new defects during the repair process, or as the \textit{acceptability}, \textit{understandability}, \textit{maintainability}, and \textit{effectiveness} of the generated patches. 
The level of research on this issue varies at each stage of the APR development history, but all tend to be increasingly more comprehensive.
Weimer \textit{et al.}~\cite{48} describe repair quality as the ability required to compile, fix defects, and avoid damage. Le Goues \textit{et al.}~\cite{62} studied repair quality by whether new defects are introduced. Jin \textit{et al.}~\cite{211} and Liu \textit{et al.}~\cite{212} studied repair quality of concurrent software by looking at the introduction of deadlocks. Assiri \textit{et al.}~\cite{77} studied repair quality in terms of both repair correctness and repair maintainability. 
Manish \textit{et al.}~\cite{41} proposed some directions for high-quality fixes by investigating the impact of improved fault localization and test suites. 
Up to now, there is still a lack of more comprehensive and in-depth studies on the APR repair quality issue, and we believe that future in-depth studies on this issue will provide valuable insights to advance APR research.

\subsubsection{Repair Efficiency Problem}
\label{Repair Efficiency Problem}

The repair efficiency issue illustrates the money and time cost of the APR tools in repairing defects, and the percentage of effective fixes to candidate fixes.
However, there is still a lack of large-scale in-depth studies on the repair costs of APR tools.
Only a few efforts have focused on this topic.
For example, 
Le Goues \textit{et al.}~\cite{127} deployed APR tools on a large server and evaluated its repair cost. 
Liu \textit{et al.}~\cite{40,51} explored the differences in patch generation efficiency of traditional APR tools.
Recently, Martinez \textit{et al.}~\cite{APR2energy} explored the energy consumption of APR tools through a large-scale empirical study, revealing the differences in energy consumption behind bug fixing by different APR tools from the green software perspective, which provides rich insights into the repair cost problem. However, they only explored traditional APR techniques, and the recent learning-based APR techniques have yet to be explored in depth.
Besides, invalid patches generated by APR tools also affect the repair efficiency. 
Qi \textit{et al.}~\cite{76} pointed out that GP methods may generate meaningless patches. Even data-driven methods can generate invalid patches~\cite{161}.
Unfortunately, the generation of invalid patches is unavoidable by all repair techniques.
In short, the repair efficiency issue limits the pace of large-scale deployment of APR tools, and continuous optimization of repair efficiency is still a long-term goal.

\subsubsection{Generic Fixes Problem}
\label{Generic Fixes Problem}

Cross-language, cross-program, cross-category and large-scale program fixing are often major issues that limit the scalability of APR tools. As shown in Table~\ref{tab:table2-3-2}, 
most APR tools only support single language fixes and single hunk fixes,
and they are not tested in complex multi-fault fixes scenarios.
Therefore, APR tools are always limited in scalability and generality.~\cite{6}
Fortunately, the data-driven APR techniques offer the possibility to implement multi-language, multi-type, multi-hunk, and multi-fault fixes. 
However, as discussed in \textbf{Finding 16-19}, there are still many challenges in learning-based APR techniques for MT, ML, MH, and MF fixes.
Besides, there is a lack of general-purpose large-scale repair systems. Therefore, research on general-purpose APR techniques is still ongoing efforts, and we urgently need APR tools with a more robust repair ability to overcome complex repair tasks. 

\subsubsection{Dataset Quality Problem}
\label{Dataset Quality Problem}

APR datasets are the important basis for evaluating the performance of APR tools. 
For APR research, especially data-driven APR techniques, the dataset quality greatly impacts the model training. 
However, we reveal many problems with learning-based APR work on datasets in \textbf{Finding 15}.
The latest work also pointed out some problems in datasets. Yang \textit{et al.}~\cite{214} found that even on the widely used Defects4J~\cite{215}, the patch data is not optimized enough. They further streamlined the original patches and proposed a \textit{Dataset Purification} technique to assist this process. 
In addition, another problem is \textit{benchmark overfitting}~\cite{169}, the ability of APR tools to repair outside of experimental datasets is still questionable. Fortunately, there are many works dedicated to proposing new datasets
(as shown in Appendix~\ref{sec:APR datasets}),
which helps to better evaluate the actual performance of APR tools.
At present, the optimization and exploration of APR datasets are just beginning, and we expect more work to be done in the future. Not only proposing new datasets with larger sizes but also identifying and correcting dataset quality issues.

\subsubsection{Industrial Deployment Problem}
\label{Industrial Deployment Problem}

APR tools were created to address the problem of industrial-scale repair and increase human productivity. However, the application of APR tools has remained in an experimental setting, and only in recent years have there been attempts to apply it industrially. 

In 2017, Fujitsu explored industrial APR by proposing Elixir~\cite{11}, an AI-based automated repair tool for Java, to be deployed in industrial environments, and later improved~\cite{14}. In 2019, they presented the new APR tool HERCULES~\cite{13}, which addresses the multi-line bug fixes.

In 2018, Facebook announced Getafix~\cite{9}. Getafix does not use test cases but instead resorts to static analysis for FL and patch validation, which adapts to the requirements of industrial for fast fixes. Later, they introduced SapFix~\cite{10} to provide fixes, which was the first time automated end-to-end fixing was deployed in a continuous integration manner for industrial systems.

In 2019, Alibaba deployed the APR tool ProFL~\cite{15} on Alipay, which uses the results of fault repair to help with FL. In 2020, the Ali team launched Precfix~\cite{16}, a semi-automated patch recommendation system. Which extracts a large number of repair templates from a large industrial code base, and is currently deployed in various key Ali businesses. 

\hkr{In 2021, Baudry \textit{et al.}~\cite{155} proposed R-Hero, an APR robot based on continul learning (CL). 
It integrates a complete code commit and repair process.
This constant stream of data will serve as a knowledge base for CL.
In the context of the knowledge-driven era of big data, R-Hero is a milestone of sorts, 
a self-improving software system, 
and helps automate the process of program repair one step further.}

Although APR has been used in industry, it is not known how well it really works. This problem was explored in an empirical report~\cite{14}. 
The results show that the status of industrial applications is not promising.
Next, we summarized the difficulties in industrial application:

\begin{enumerate}
\item{\bf Lack of Bug-Exposing Test Cases.} Mainstream APR tools use test cases to locate bugs, but real environments usually lack comprehensive test case coverage~\cite{14}. After a new defect is discovered, it is not common to have a complete set of test cases to aid in the fix, which means that APR tools based on test cases may not be suitable for real industrial deployments.
\item{\bf Ease of Use of APR Tools.} Real industrial environments require developers to easily use repair tools to accomplish quick fixes, but most APR tools are prototypes or sets of methods. Properly integrating APR tools into industrial development environments and making them easily available to developers is a practical issue to consider for industrial repair~\cite{6}.
\item{\bf Patch Interpretability.} Machine-generated patches are difficult and time-consuming to review because they lack appropriate comments. 
Therefore, providing explainable fixes will help developers review and choose whether to accept the patch quickly~\cite{6}.
\item{\bf Large-scale Project Bug Fixes.} In industrial software systems, some large projects may be developed in multiple PLs simultaneously, which requires APR tools to have the ability to fix multiple PLs. Moreover, most projects are large programs with many code lines, but most APR tools only fix small code segments. And large-scale projects often have complex dependencies. Extending APR tools to work on these projects remains to be explored.
\item{\bf Non-source Bug Fixes.} Many real-world bugs are not source-level bugs~\cite{192}, and they may require the removal or addition of related files to achieve the fix operation. Current APR tools do not yet have the ability to fix non-source-level (e.g., file-level) bugs.
\item{\bf Cross file fixes.} Real-world bugs can involve multiple source files, which requires modifying multiple files to fix a bug. However, current APR tools have not been explored in cross-file fix scenarios.
\end{enumerate}

Overall, we have observed the great promise that APR techniques show in industrial applications. 
However, industrial APR deployment is still a difficult and long process. Although researchers have made an initial exploration, there are still many challenges, such as the requirements for APR tools for high-precision detection, high responsiveness, multi-location repair, multi-domain repair, cross-language repair, and side-effect-free repair, etc. 
In the future, more cooperation between researchers and industrial developers is needed to help the progress of APR industrial applications.

\subsection{Improvement Directions}
\label{Improvement Direction}

\subsubsection{Exploring the Fitness Function}
\label{Exploring the Fitness Function}

In search-based techniques, the fitness function is an important factor that affects the search ability~\cite{79}. 
The fitness is a specific measure of a program's ability to pass test cases, and finding a program with a high fitness means getting a good fix. 
Theoretically, improvements to the fitness function (e.g., using a fine-grained fitness function) will help discover candidate fixes from a large search space faster~\cite{217}. However, different results emerged in recent studies~\cite{218},~\cite{219}. Their experiments showed that whether or not using a finer-grained fitness function did not make a significant difference to results. The reason behind this remains unknown, and we believe that further exploration of the fitness function is necessary to find out why the fitness function does or does not affect the repair results, which will provide useful insights into whether the search-based method can be enhanced by improving the fitness function.

\subsubsection{Search Strategy Optimization}
\label{Search Strategy Optimization}

Consistent with the aim of improving the fitness function, using a more optimized search strategy can improve the patch search speed and speed up the acquisition of candidate patches for search-based APR techniques. It is an effective way to improve the repair efficiency. Previous works (Section~\ref{sec:Search-based}) have emerged with a range of search strategies such as GP methods, random search, target value search, code similarity, etc., which would help to better generate and discover candidate patches from the search space. We hope that more works in the future will propose new search strategies and perform deeper optimizations.

\subsubsection{Program Representation Form}
\label{Program Representation Form}

Our review found that how to better capture and understand code representations is a key factor influencing data-driven APR tools. During the repair process, the code representation is an important way for the model to understand the source code information. 
While the traditional approach extracts source code information from the token sequence or AST level, researchers have now proposed to characterize the PL in the form of change graphs~\cite{125}, data flow~\cite{156}, control flow~\cite{59}, bytecode~\cite{166}, etc., with effective results. 
Overall, researchers can work on extracting concise contextual features and rich code dependencies, which will further improve the quality of fixes and even solve more complex bugs. 
We hope that future work will propose forms of representation that better fit the APR scenario and better incorporate the syntactic and semantic features of the PL, which is the basis for both understanding the PL and fixing bugs.

\subsubsection{FL Techniques Research}
\label{FL Techniques Research}

For APR, more advanced FL will help accurately locate defects, which is a prerequisite for a successful repair and a primary stage to improve the repair quality. 
The literature~\cite{110} indicates that the accuracy of FL is a key factor in helping repair success, and Liu \textit{et al.}~\cite{51} shows that better FL will increase the probability of generating the correct patch.
However, as shown in Table~\ref{tab:table2-3-2}, most APR tools directly use existing FL tools or simply couple FL techniques into APR tools, which does not closely link localization and repair. 
Besides, current FL techniques are still not satisfactory in detection accuracy. Some researchers~\cite{43,51,221} are exploring how to improve or more deeply integrate FL techniques into the repair process to enhance the results. 
We think the focus on FL should receive as much attention as patch generation techniques. Achieving more efficient and accurate FL remains a goal to strive for in the future.



\subsubsection{Test Techniques Improvement}
\label{Test Techniques Improvement}

For APR tools based on test cases, test cases can be used to assist in FL, guide patch synthesis, and validate patch candidates. 
Recent works have also shown that patch G\&V may encounter weak test suite problems~\cite{155},~\cite{224}. Since flaky tests exist in reality, this may lead to the generation of patches that seem reasonable but are incorrect, affecting the quality of the generated patches. Therefore, some researchers want to improve patch quality by enhancing test case techniques~\cite{208,209}; others want to find new ways to remedy test case deficiencies or replace test cases with new ones~\cite{225,5}. Thus, designing high-quality test cases and proposing more optimal testing techniques will also help alleviate the \textit{patch overfitting problem}.

\subsubsection{Patch Quality Enhancement}
\label{Patch Quality Enhancement}

Generating high-quality patches will directly affect the final repair quality. 
The improvement of patch quality lies not only in generating effective patches but also in reducing the generation of invalid patches. 
In the literature~\cite{21}, the principle of “simple is better” is proposed for patch repair, where simpler patches will help fixers to understand them and prevent the introduction of new defects to a certain extent. 
In addition, patch quality improvement also requires addressing the \textit{patch overfitting problem}, and proposing new or improved patch validation techniques. For example, adding formal specifications, designing enhanced test suites, and even using the APCA techniques are all useful ways to improve patch quality. 
Therefore, future work can focus on improving patch design principles and patch validation evaluation.


\subsubsection{OSS Knowledge and AI Technology Application}
\label{AI Technology Application}
By revisiting the evolution of APR techniques, it is observed that traditional APR techniques have undergone nearly two decades of development (2005-2022), while learning-based solutions have only gained widespread attention in recent years (2016-2022). However, they have surpassed previous traditional techniques in terms of repair outcomes. This benefits from the vast amount of data supported by the OSS community and the rapid development of AI technology. In the future, researchers need to focus on these elements. The first is for the extraction and application of OSS knowledge, and how to make extensive use of the OSS community's high-quality repair knowledge to better guide APR research. In addition, how to apply AI techniques to APR research also needs to be explored in depth. For example, future research still needs to propose more advanced LLMs to handle PL tasks in the context of PL scenarios; future research needs to combine AI techniques with traditional APR techniques (e.g., TRANSFER~\cite{TRANSFER}, DeepRepair~\cite{87}, etc.) to promote each other; future research needs to apply new methods (e.g., ensemble learning~\cite{137}, reinforcement learning~\cite{150}, continual learning~\cite{155}, transfer learning~\cite{197}, etc.) from the AI field to APR techniques to propose novel repair solutions.

\subsubsection{Hybrid Technology Development}
\label{Hybrid Technology Development}

By analyzing the development of APR at the technical level, we have observed that APR is a multi-disciplinary cross-section of research. It has become mainstream for APR tools to integrate a mix of techniques simultaneously. These APR tools use a mixture of several schemes of heuristics, constraints, templates, or learning to propose new solutions. The advantage of this approach is that it cleverly avoids the problems associated with using a single type of technique. 
Also, the hybrid intersection of multiple research areas is fueling the development of APR, such as software engineering, system security, and artificial intelligence. 
We advocate multi-disciplinary cross-sectional research. In the future, the development of multiple hybrid technologies will be an important means to improve the repair quality and repair efficiency.

\subsubsection{Lightweight Context-aware Strategy}
\label{Lightweight context-aware strategy}

Previous works~\cite{VulRepair,145,49,143} have shown that the processing capability of neural networks decreases with the growth of input/output sequences. In other words, existing models still have challenges in processing long sequences. 
In addition, as discussed in \textbf{Finding 19}, the input/output length limit also affects the repair ability of the model.
And one idea to alleviate this problem is to simplify the input (i.e., the contextual content) of the model. Precisely, we can remove the parts of the contextual content that are not related to defects by combining dependency analysis (e.g., data flow and control flow) and thus achieve the goal of simplifying the contextual content. We suggest that future works focus on this content in the pre-processing step to propose specialized lightweight context-aware strategies.

\subsubsection{Advanced Code Tokenization Techniques.}
\label{Advanced Code Tokenization Techniques}
As discussed in \textbf{Finding 10}, the subword-level tokenization technique alleviates the OOV problem but makes the input sequence longer. Long sequences still limit the repair ability of models, so proposing more effective tokenization techniques to alleviate this deficiency is a direction to explore. Future tokenizers should be specially designed to consider the special features of PLs fully. 
Firstly, the length of the processed sequences should be minimized based on the mitigation of OOV. Secondly, information such as the special syntactic structure of PLs should be recorded as much as possible.

\subsubsection{Practical Post-processing Strategy.}
\label{Practical post-processing strategy}

As discussed in \textbf{Finding 12}, the post-processing step is an important component of both traditional APR techniques and the latest learning-based solutions. However, no specific studies have emerged to explore practical post-processing strategies, and it is still unclear what post-processing strategies should be designed or used for different APR techniques. We suggest that future works focus on this topic, not only to propose practical post-processing strategies, but also to make post-processing strategies that can be easily integrated into different APR techniques.

\subsubsection{Patch Validation Improvement}
\label{Patch Verification Efficiency Improvement}

As discussed in \textbf{Finding 13}, the current validation strategies still need improvement.
The improvement of patch validation techniques lies in two aspects, one is to improve the efficiency , and the other is to improve the accuracy.
First, previous works~\cite{102,124,166} have tried to improve efficiency by various means.
In recent years, a series of techniques have been proposed to speed up patch validation, such as UniAPR~\cite{226}, a bytecode-level patch validation framework, and SeAPR~\cite{50}, a patch validation technique using the unified debugging idea. 
These patch validation techniques can work with APR tools to improve efficiency by reducing the number of patch executions.
Second, the improved accuracy of validation helps obtain high-quality fixes. 
Patch validation techniques have been a continuous direction of inquiry in terms of test cases, static analysis, formal specifications, and other testing tools that are constantly emerging. 
A recent work~\cite{115} used static analysis and fuzzy testing to improve patch validation accuracy.
In short, compared with the current APR research that focuses more on patch generation techniques, the future improvement of patch validation techniques will be a direction worth exploring. 
We propose to improve the efficiency and accuracy of patch validation by proposing new techniques and using a combination of multiple validation techniques.


\subsubsection{Prompt Engineering for LLM-based APR}
\label{Prompt Engineering for LLM-based APR}

As discussed in \textbf{Finding 14}, LLM is the future trend, and prompt learning schemes have been successfully applied to APR research. This technical path provides a new paradigm for learning-based APR research, i.e., zero/few-shot learning. This solution focuses on designing suitable prompt templates for different scenarios (i.e., different defect problems, different LLMs, etc.) to make the LLM work effectively. In short, this requires researchers to focus on exploring prompt engineering. An in-depth study of prompt engineering for APR will provide insights to guide the better application of LLM to repair tasks.

\subsubsection{Fine-tuning Settings for LLM-based APR}
\label{Fine-tuning Settings for LLM-based APR}


In contrast to prompt learning-based solutions~\cite{LLM2APR_fan,LLM2APR_xia,LLM2Vul}, APR research based on fine-tuning LLMs still lacks large-scale empirical studies for exploration. As discussed in \textbf{Finding 14}, it is still unclear how to choose appropriate code representations, model evaluation metrics, and other technical details in the fine-tuning process. Therefore, future work is needed to bridge this gap. We should not only explore how to choose appropriate technical details further to improve the repair capability of the fine-tuning workflow, but also comprehensively examine the shortcomings and strengths of fine-tuning-based solutions in terms of repair ability, repair quality, and repair efficiency. These insights can guide further improvements in LLMs.

\subsubsection{LLM for Mroe PL Task}
\label{LLM for mroe PL task}

As illustrated in Fig.~\ref{fig:LLM2APR}, the application of the LLM in APR workflow is mainly in the patch generation phase, where it generates candidate fixes. However, the potential of PL-based LLMs (or LLMCs) goes far beyond patch generation and can be applied to a wide range of PL-related downstream tasks. 
Therefore, LLMs can also be used for tasks such as fault localization, patch filtering and ranking, and patch validation. We believe that LLMs can perform well in these tasks too.
Additionally, LLMs can be integrated into traditional APR techniques. For example, like TRANSFER~\cite{TRANSFER}, we can fine-tune the LLM to guide the template matching selection process or even for automated template mining.
Moreover, LLMs can be employed for code similarity-related tasks, which can further facilitate APR efforts based on code similarity search. In conclusion, the potential of LLMs in APR research is yet to be fully explored, and utilizing LLMs for more PL-related tasks is a promising direction for the future.

\subsubsection{Dataset Research}
\label{Dataset Research}

Data and knowledge are powerful drivers of technology. 
Up to now, defect datasets have been widely used for defect prediction, localization, and repair. 
Therefore, research on datasets is also important. 
As discussed in \textbf{Finding 15}, even though APR has been extensively studied and developed in the past decade or so, the dataset quality is still a concern. 
Based on this, future work can be in-depth in three aspects: 
First, it is necessary to conduct a large-scale data quality assessment analysis to reveal the dataset overlap problem.
Second, we hope to further optimize and improve the data based on the current datasets to obtain a better data source (e.g., removing dataset noise and invalid or incorrect repair examples). 
Finally, what most researchers are currently working on, i.e., to propose more comprehensive and more optimized datasets to cover the defects more comprehensively and provide better repair examples.

\subsubsection{Integrating Multi-domain Repair Knowledge}
\label{Integrating Multi-domain Repair Knowledge}

As discussed in \textbf{Finding 16}, while most APR tools support multi-type fixes, they are not typically designed to handle multiple defect problems simultaneously. This is because existing APR techniques are generally tailored to specific problem domains and do not draw upon repair knowledge from multiple domains at the same time. However, we are inspired by the transfer learning solutions of VRepair~\cite{197} and SeqTrans~\cite{SeqTrans}, which leverage bug repair knowledge to facilitate vulnerability repair.
Hence, it would be an intriguing topic to investigate whether bug, error, and vulnerability repair knowledge from different problem domains can complement each other. Meta-learning is a promising approach to enhancing the problem-solving capability of another domain by acquiring empirical knowledge from multiple domains. Therefore, exploring the potential of meta-learning~\cite{MetaLearning} in facilitating knowledge transfer between different problem domains is a valuable direction for future research.

\subsubsection{Defect Root Cause Understand}
\label{Defect Root Cause Understand}

As discussed in \textbf{Finding 17}, multi-language fixes have been addressed. However, existing multi-language solutions require learning the corresponding repair knowledge from multiple PLs. One of the key factors is that the deep root causes of defects are not extracted and utilized by APR techniques. This means that existing APR techniques are still unable to cross-fertilize repair knowledge from multiple languages (e.g., can repair knowledge learned from C be applied to Java repair?). We believe that a deeper understanding of the root causes of defects will help to overcome the limitations of PLs, and thus enable APR tools to acquire deeper defect repair knowledge. Therefore, we emphasize the importance of understanding the root causes of defects, which can further enhance the multi-language repair ability of APR tools by learning the deep root causes and repair knowledge of defect problems.
Moreover, meta-learning is also applicable to multi-language repair scenarios, and applying this technique to multi-language repair is a promising direction to explore.

\subsubsection{Researching the Relationship between Bugs and Fixes}
\label{Researching the relationship between bugs and fixes}
As discussed in \textbf{Finding 18}, multi-hunk bug fixing remains a long-term challenge. One thing hindering this process is a special type of multi-hunk bugs, i.e., bugs where a single bug location corresponds to multiple fix locations, making it difficult for current APR tools to handle this type of complex bug. In other words, this also reflects that current research has not delved into the in-depth relationship between bugs and fixes. It is still unclear how many fixes a bug needs and where these fixes should be placed. Therefore, we suggest that future work should focus on the deeper relationship between bugs and fixes to help guide the direction of multi-hunk fixes.

\subsubsection{Exploring Multi-fault fixes}
\label{Exploring Multi-fault fixes}

As discussed in \textbf{Finding 19}, the multi-fault fixing ability of APR tools has not been thoroughly investigated, and such complex repair scenarios place higher demands on the repair ability of APR tools. In real-world scenarios, multiple bugs may exist in the same source file, and these bugs may interact with each other. Theoretically, existing APR tools can fix individual bugs one by one. However, it is still a question whether fixing one of the bugs will impact the others. In the future, we hope that the issue of multi-fault fixing will be taken seriously, and dealing with bug fixing in such complex scenarios will further advance the pace of the industrial deployment of APR.

\subsubsection{Establish Uniform Assessment Criteria for Repair Quality and Repair Efficiency}

As discussed in \textbf{Finding 20}, previous works on APR are still incomplete or lack uniform metrics for assessing repair quality and efficiency during the evaluation phase, which limits our understanding of these issues (repair quality and efficiency). We suggest that future work focuses on addressing this gap. One important theme is to propose uniform assessment criteria for repair quality and efficiency so that subsequent APR studies can follow consistent metrics in the experimental phase. As this paper presents multiple perspectives on repair quality and efficiency, we hope they will serve as a reference for future research. However, the perspectives presented in this paper may not be exhaustive, and we encourage further work to explore this topic more deeply.

\subsubsection{Exploring New Paradigms for Program Repair}
\label{Exploring new paradigms for program repair}

As discussed in \textbf{Finding 21}, NMT is a suitable paradigm for repair tasks. However, it may not be the most appropriate learning paradigm. As shown in AlphaRepair~\cite{AlphaRepair} works, the MLM task is equally effective in solving the program repair problem. This approach directly applies the code understanding and generation capabilities of the large model itself, requiring no additional fine-tuning steps. In the long run, LLM will be the future. Therefore, it is time for us to revisit the learning paradigm of APR research in the context of LLMs. We believe that future work does not necessarily have to follow the NMT workflow, and incorporating LLMs to design appropriate new learning paradigms should be the focus of future research.

\subsubsection{Conducting More Comprehensive Empirical Studies}








In Section~\ref{discussion}, we have emphasized the need for more empirical studies to explore various topics necessary to gain a deeper understanding of the current problems encountered in developing APR techniques and provide guidance for further advances in APR. In the following, we describe the relevant topics for empirical studies.

(1) Empirical studies on patch overfitting problem. As discussed in \textbf{Finding 5}, the factors influencing patch overfitting are not explored in-depth. Future empirical studies should focus on this issue to explore the deeper factors affecting patch overfitting and provide insights for continuous improvement of repair quality. Additionally, future works should also compare the performance of learning-based APR techniques to traditional techniques on the overfitting problem to understand the advantages and shortcomings of learning-based APR efforts.

(2) Empirical studies on the technical details of learning-based solutions. As discussed in \textbf{Finding 7, 9-12}, current learning-based APR techniques employ different technical details at different steps. However, these technical details are not compared under the same benchmark, and it is unclear which scheme settings are more effective. Future empirical studies should evaluate and compare different pre-processing strategies, post-processing strategies, context-aware strategies, and other details to provide better guidance on the detailed design of APR techniques at each process. Additionally, future empirical studies should set uniform parameters and datasets in a unified experimental setting to fairly compare the performance of different APR tools.

(3) Empirical studies on LLM-based APR. As discussed in \textbf{Finding 14}, it is unclear whether the technical details of previous learning-based schemes are effective for LLM-based APR. There is still a lack of empirical studies to explore how the technical details can be improved to enhance LLMs' repair capabilities sustainably. Therefore, future empirical studies should investigate the impact of different technical details on the results under fine-tuning schemes. Moreover, future empirical studies should explore the effectiveness of the LLM on different repair tasks (errors, bugs, vulnerabilities) to reveal the strengths and weaknesses of the LLM in handling different problems and provide insights for continuous improvement of the LLM. Additionally, an in-depth exploration of prompt engineering designed for repair tasks is also a needed focus for future empirical studies.

(4) Empirical studies on multi-fault fixes. As discussed in \textbf{Finding 19}, special repair scenarios such as multi-fault fixes have not been explored in depth, and we still do not know the repair ability of APR techniques for such complex repair scenarios. Based on this, we suggest that future empirical studies should not only explore multi-language, multi-type, and multi-hunk repair scenarios but also focus on such multi-fault repair scenarios to gain insight into the repair ability of current APR techniques.

(5) Empirical studies on repair quality and efficiency. As discussed in \textbf{Finding 20}, there is a lack of in-depth exploration into APR techniques' repair quality and efficiency. Future empirical studies should propose uniform assessment metrics for these topics and explore them comprehensively. Investigating repair quality will reveal differences in repair quality among different APR techniques and provide insights into addressing shortcomings. Moreover, research on repair efficiency will help understand the cost consumption of APR tools in repairing bugs in real-world environments, further facilitating the industrial deployment of APR techniques.

Overall, we have highlighted the necessity for more empirical studies. Despite the rapid development of the APR field in recent years (with nearly dozens of new works emerging each year), timely reviews and critical assessments of advanced techniques have not been keeping up. Thus, we urge the APR community to undertake more empirical studies to comprehensively evaluate and scrutinize the current progress of APR techniques, thereby fostering a positive development and reflection of the APR community.
\section{Conclusion}
\label{sec:Conclusion}

Obviously, APR techniques have made a huge impact on our scientific research as well as industrial production, helping to free mankind from heavy manual restoration. 
As a hot research area, APR has grown by leaps and bounds, especially in recent years with the addition of deep learning.
The techniques go beyond solving software bugs, programming errors, or security vulnerabilities; it aims to delve deeper into the processing of programming languages and bring together knowledge from multiple domains to build automated systems for productivity improvement.

By reviewing the development history of APR techniques, this paper successively sorts out the four major categories of APR techniques, presenting a clear development lineage of APR techniques in a linear order. 
We revisit the development process of APR techniques, emphasize the importance of learning-based solutions in the future APR field, and provide a comprehensive discussion of APR techniques, which presents the development trends and technical details of APR techniques in detail. Finally, we present the future challenges and directions of APR, highlighting the role of LLM in facilitating APR techniques and the importance of more empirical studies.


\appendix

\section{APR Works}
\label{sec:APR work}

See Table \ref{tab:tableA1}, Table \ref{tab:tableA2}, Table \ref{tab:tableA3} and Table \ref{tab:tableA4}.
We present the details of the APR works in a table
format.


\begin{table}[h]
\tiny
\caption{Summary of some search-based APR works}
\label{tab:tableA1}
\resizebox{1.0\columnwidth}{!}{

}
\end{table}

\vspace{-3ex}
\begin{table}[h]
\tiny
\caption{Summary of some constraint-based APR works}
\label{tab:tableA2}
\resizebox{1.0\columnwidth}{!}{
%
}
\end{table}

\vspace{-3ex}
\begin{table}[h]
\tiny
\caption{Summary of some template-based APR works}
\label{tab:tableA3}
\resizebox{1.0\columnwidth}{!}{
%
}
\end{table}


\begin{table}[h]
\tiny
\caption{Summary of some learning-based APR works}
\label{tab:tableA4}
\resizebox{1.0\columnwidth}{!}{
%
 \\ \hline
\end{tabular}
}
\begin{tablenotes}
\item{
Data Pre.: Data Pre-processing; 
Code Rep.: Code Representation; 
Patch Post.: Patch Post-processing (i.e. Patch Filtering and Ranking); 
Patch Val.: Patch Validation; 
N: No; 
N.A.: Not Available or Unknown; 
Cod Abs.: Code Abstraction;
Alpha Re.: Alpha-Rename;
Token Seq.: Token Sequence;
E.M.: Exact Match; 
T.S.: Test Suite; 
H.C.: Hand Check.}
\end{tablenotes}
\begin{tablenotes}
\item{
Where we need to add explanations to some of the column names:
}
\item{
The “\textit{Defect Type}” is a classification of the objects is fixed by APR tools. “Error” refers to programming errors, which are simple errors caused by the inexperience of programming beginners, usually syntax errors that fail to compile. Typical errors are mainly extracted from programming websites or students' programming assignments. “Bug” refers to real-world software bugs, which are potential defects in the software system caused by the developer during the development process and affect the functionality or other aspects of the software system. These bugs can often be successfully compiled and thus may be overlooked by developers. Unlike programming errors, software bugs are extracted from real-world projects.
“Vul” refers to security bugs, i.e. vulnerabilities.
}
\item{
The “\textit{Bug Line}” is a classification of the complexity of bugs resolved by APR tools (i.e., the number of lines of code that need to be changed to fix a bug). “Single” refers to \textit{single-hunk bugs}, which require the APR tool to edit a single location or a series of consecutive locations to complete the fix. “Multi” refers to \textit{multi-hunk bugs}, which are bug fixes that require the APR tool to edit in multiple discrete locations to fix a multi-hunk bug.
}
\item{
The “\textit{Result}” mainly records the performance of the APR tool on each benchmark, and there are two main forms of records here. The “x/y” format, i.e., \textit{the number of correct patches}/\textit{the number of plausible patches}, shows the results of fixes on benchmarks with test cases.
The “z\%” format refers to the \textit{Repair Accuracy} on each benchmark, and this presentation is mainly in the form of fixed results from some bug repair tools on massive datasets. These datasets are usually not equipped with test cases, so only the ratio of correct patch fixes to all bugs is considered to represent the fix results.
Note that error repair tools have two ways to calculate \textit{Repair Accuracy}~\cite{TransR}, \textit{Single Repair} and \textit{Full Repair}. Given that most efforts have used \textit{Full Repair}, we only show \textit{Full Repair} results here.
}
\end{tablenotes}

\end{table}

\section{APR Datasets}
\label{sec:APR datasets}

See Table \ref{tab:tableB}. We investigated and summarized these relevant datasets for APR tasks.

\begin{table}[]
\tiny
\caption{APR datasets}
\label{tab:tableB}
\resizebox{1.0\columnwidth}{!}{
\begin{tabular}{|cllcrl|}
\hline
\multicolumn{1}{|c|}{\textbf{Time}} &
  \multicolumn{1}{c|}{\textbf{Dataset}} &
  \multicolumn{1}{c|}{\textbf{Language}} &
  \multicolumn{1}{c|}{\textbf{Defect Type}} &
  \multicolumn{1}{c|}{\textbf{Defect Num.}} &
  \multicolumn{1}{c|}{{\color[HTML]{000000} \textbf{Url}}} \\ \hline
 &
   &
   &
   &
   &
   \\ \hline
\multicolumn{6}{|c|}{\textbf{Small-scale test benchmarks}} \\ \hline
\multicolumn{1}{|c|}{2012} &
  \multicolumn{1}{l|}{ICSE 2012 dataset} &
  \multicolumn{1}{l|}{C/C++} &
  \multicolumn{1}{c|}{Bug} &
  \multicolumn{1}{r|}{105} &
  {\color[HTML]{000000} N.A.} \\ \hline
\multicolumn{1}{|c|}{2014} &
  \multicolumn{1}{l|}{Defects4J} &
  \multicolumn{1}{l|}{Java} &
  \multicolumn{1}{c|}{Bug} &
  \multicolumn{1}{r|}{835} &
  {\color[HTML]{0563C1} https://github.com/rjust/defects4j} \\ \hline
\multicolumn{1}{|c|}{2015} &
  \multicolumn{1}{l|}{ManyBugs} &
  \multicolumn{1}{l|}{C} &
  \multicolumn{1}{c|}{Bug} &
  \multicolumn{1}{r|}{185} &
  {\color[HTML]{0563C1} https://repairbenchmarks.cs.umass.edu/} \\ \hline
\multicolumn{1}{|c|}{2015} &
  \multicolumn{1}{l|}{IntroClass} &
  \multicolumn{1}{l|}{C} &
  \multicolumn{1}{c|}{Bug} &
  \multicolumn{1}{r|}{998} &
  {\color[HTML]{0563C1} https://repairbenchmarks.cs.umass.edu/} \\ \hline
\multicolumn{1}{|c|}{2016} &
  \multicolumn{1}{l|}{IntroClassJava} &
  \multicolumn{1}{l|}{Java} &
  \multicolumn{1}{c|}{Bug} &
  \multicolumn{1}{r|}{297} &
  {\color[HTML]{0563C1} https://github.com/Spirals-Team/IntroClassJava} \\ \hline
\multicolumn{1}{|c|}{2017} &
  \multicolumn{1}{l|}{CodeFlaws} &
  \multicolumn{1}{l|}{C} &
  \multicolumn{1}{c|}{Bug} &
  \multicolumn{1}{r|}{3,902} &
  {\color[HTML]{0563C1} https://codeflaws.github.io/} \\ \hline
\multicolumn{1}{|c|}{2017} &
  \multicolumn{1}{l|}{DBGBench} &
  \multicolumn{1}{l|}{C} &
  \multicolumn{1}{c|}{Bug} &
  \multicolumn{1}{r|}{27} &
  {\color[HTML]{0563C1} https://dbgbench.github.io/} \\ \hline
\multicolumn{1}{|c|}{2017} &
  \multicolumn{1}{l|}{DroixBench} &
  \multicolumn{1}{l|}{Java} &
  \multicolumn{1}{c|}{Bug} &
  \multicolumn{1}{r|}{24} &
  {\color[HTML]{0563C1} https://droix2017.github.io/} \\ \hline
\multicolumn{1}{|c|}{2017} &
  \multicolumn{1}{l|}{QuixBugs} &
  \multicolumn{1}{l|}{Java/Python} &
  \multicolumn{1}{c|}{Bug} &
  \multicolumn{1}{r|}{40} &
  {\color[HTML]{0563C1} https://github.com/jkoppel/Quixbugs} \\ \hline
\multicolumn{1}{|c|}{2018} &
  \multicolumn{1}{l|}{Bugs.jar} &
  \multicolumn{1}{l|}{Java} &
  \multicolumn{1}{c|}{Bug} &
  \multicolumn{1}{r|}{1,158} &
  {\color[HTML]{0563C1} https://github.com/bugs-dot-jar/bugs-dot-jar} \\ \hline
\multicolumn{1}{|c|}{2019} &
  \multicolumn{1}{l|}{Bears} &
  \multicolumn{1}{l|}{Java} &
  \multicolumn{1}{c|}{Bug} &
  \multicolumn{1}{r|}{251} &
  {\color[HTML]{0563C1} https://github.com/bears-bugs/bears-benchmark} \\ \hline
\multicolumn{1}{|c|}{2019} &
  \multicolumn{1}{l|}{BugsJS} &
  \multicolumn{1}{l|}{JS} &
  \multicolumn{1}{c|}{Bug} &
  \multicolumn{1}{r|}{453} &
  {\color[HTML]{0563C1} https://bugsjs.github.io/} \\ \hline
\multicolumn{1}{|c|}{2019} &
  \multicolumn{1}{l|}{BugSwarm} &
  \multicolumn{1}{l|}{Java/Python} &
  \multicolumn{1}{c|}{Bug} &
  \multicolumn{1}{r|}{3,091} &
  {\color[HTML]{0563C1} https://github.com/BugSwarm/bugswarm} \\ \hline
\multicolumn{1}{|c|}{2019} &
  \multicolumn{1}{l|}{Defexts} &
  \multicolumn{1}{l|}{Kotlin/Groovy/Scala} &
  \multicolumn{1}{c|}{Bug} &
  \multicolumn{1}{r|}{654} &
  {\color[HTML]{0563C1} https://github.com/ProdigyXable/defexts} \\ \hline
\multicolumn{1}{|c|}{2020} &
  \multicolumn{1}{l|}{BugsInPy} &
  \multicolumn{1}{l|}{Python} &
  \multicolumn{1}{c|}{Bug} &
  \multicolumn{1}{r|}{493} &
  {\color[HTML]{0563C1} https://github.com/soarsmu/BugsInPy} \\ \hline
\multicolumn{1}{|c|}{2020} &
  \multicolumn{1}{l|}{TANDEM} &
  \multicolumn{1}{l|}{Java/C/SQL/…} &
  \multicolumn{1}{c|}{Bug} &
  \multicolumn{1}{r|}{125} &
  {\color[HTML]{0563C1} https://github.com/belene/tandem} \\ \hline
\multicolumn{1}{|c|}{2021} &
  \multicolumn{1}{l|}{AndroR2} &
  \multicolumn{1}{l|}{Android} &
  \multicolumn{1}{c|}{Bug} &
  \multicolumn{1}{r|}{90} &
  {\color[HTML]{0563C1} https://zenodo.org/record/4646313} \\ \hline
\multicolumn{1}{|c|}{2021} &
  \multicolumn{1}{l|}{Defects4Cpp} &
  \multicolumn{1}{l|}{C/C++} &
  \multicolumn{1}{c|}{Bug} &
  \multicolumn{1}{r|}{210} &
  {\color[HTML]{0563C1} https://github.com/Suresoft-GLaDOS/defects4cpp} \\ \hline
\multicolumn{1}{|c|}{2019} &
  \multicolumn{1}{l|}{Refactory} &
  \multicolumn{1}{l|}{Python} &
  \multicolumn{1}{c|}{Error} &
  \multicolumn{1}{r|}{1,783} &
  {\color[HTML]{0563C1} https://github.com/githubhuyang/refactory} \\ \hline
\multicolumn{1}{|c|}{2022} &
  \multicolumn{1}{l|}{RING} &
  \multicolumn{1}{l|}{6 PLs} &
  \multicolumn{1}{c|}{Error} &
  \multicolumn{1}{r|}{1,200} &
  {\color[HTML]{0563C1} https://github.com/microsoft/prose-benchmarks/} \\ \hline
\multicolumn{1}{|c|}{2022} &
  \multicolumn{1}{l|}{MMAPR} &
  \multicolumn{1}{l|}{Python} &
  \multicolumn{1}{c|}{Error} &
  \multicolumn{1}{r|}{286} &
  {\color[HTML]{000000} N.A.} \\ \hline
\multicolumn{1}{|c|}{2019} &
  \multicolumn{1}{l|}{Ponta's dataset} &
  \multicolumn{1}{l|}{Java} &
  \multicolumn{1}{c|}{Vul} &
  \multicolumn{1}{r|}{1,068} &
  {\color[HTML]{0563C1} https://github.com/SAP/project-kb} \\ \hline
\multicolumn{1}{|c|}{2020} &
  \multicolumn{1}{l|}{Big-Vul} &
  \multicolumn{1}{l|}{C/C++} &
  \multicolumn{1}{c|}{Vul} &
  \multicolumn{1}{r|}{3,754} &
  {\color[HTML]{0563C1} https://github.com/ZeoVan/MSR\_20\_Code\_vulnerability\_CSV\_Dataset} \\ \hline
\multicolumn{1}{|c|}{2021} &
  \multicolumn{1}{l|}{CVEfixes} &
  \multicolumn{1}{l|}{C/C++} &
  \multicolumn{1}{c|}{Vul} &
  \multicolumn{1}{r|}{5,495} &
  {\color[HTML]{0563C1} https://github.com/secureIT-project/CVEfixes} \\ \hline
\multicolumn{1}{|c|}{2022} &
  \multicolumn{1}{l|}{CrossVul} &
  \multicolumn{1}{l|}{40 PLs} &
  \multicolumn{1}{c|}{Vul} &
  \multicolumn{1}{r|}{5,131} &
  {\color[HTML]{0563C1} https://zenodo.org/record/4734050} \\ \hline
\multicolumn{1}{|c|}{2022} &
  \multicolumn{1}{l|}{Vul4J} &
  \multicolumn{1}{l|}{Java} &
  \multicolumn{1}{c|}{Vul} &
  \multicolumn{1}{r|}{79} &
  {\color[HTML]{0563C1} https://github.com/tuhh-softsec/vul4j} \\ \hline
 &
   &
   &
   &
   &
   \\ \hline
\multicolumn{6}{|c|}{\textbf{Large-scale train datasets}} \\ \hline
\multicolumn{1}{|c|}{2016} &
  \multicolumn{1}{l|}{ETH-Py150} &
  \multicolumn{1}{l|}{Python} &
  \multicolumn{1}{c|}{Bug} &
  \multicolumn{1}{r|}{150K} &
  {\color[HTML]{0563C1} https://www.sri.inf.ethz.ch/py150} \\ \hline
\multicolumn{1}{|c|}{2016} &
  \multicolumn{1}{l|}{BugAID} &
  \multicolumn{1}{l|}{JavaScript} &
  \multicolumn{1}{c|}{Bug} &
  \multicolumn{1}{r|}{105,133} &
  {\color[HTML]{0563C1} http://salt.ece.ubc.ca/software/bugaid} \\ \hline
\multicolumn{1}{|c|}{2018} &
  \multicolumn{1}{l|}{BFP} &
  \multicolumn{1}{l|}{Java} &
  \multicolumn{1}{c|}{Bug} &
  \multicolumn{1}{r|}{123,804} &
  {\color[HTML]{0563C1} https://sites.google.com/view/learning-fixes/data} \\ \hline
\multicolumn{1}{|c|}{2018} &
  \multicolumn{1}{l|}{MSR-VarMisuse} &
  \multicolumn{1}{l|}{C\#} &
  \multicolumn{1}{c|}{Bug} &
  \multicolumn{1}{r|}{7407 files} &
  {\color[HTML]{0563C1} https://aka.ms/iclr18-prog-graphs-dataset} \\ \hline
\multicolumn{1}{|c|}{2019} &
  \multicolumn{1}{l|}{ManySStuBs4J} &
  \multicolumn{1}{l|}{Java} &
  \multicolumn{1}{c|}{Bug} &
  \multicolumn{1}{r|}{153,652} &
  {\color[HTML]{0563C1} https://zenodo.org/record/3653444} \\ \hline
\multicolumn{1}{|c|}{2019} &
  \multicolumn{1}{l|}{CodRep} &
  \multicolumn{1}{l|}{Java} &
  \multicolumn{1}{c|}{Bug} &
  \multicolumn{1}{r|}{58,069} &
  {\color[HTML]{0563C1} https://github.com/KTH/CodRep-competition/} \\ \hline
\multicolumn{1}{|c|}{2020} &
  \multicolumn{1}{l|}{Review4Repair} &
  \multicolumn{1}{l|}{Java} &
  \multicolumn{1}{c|}{Bug} &
  \multicolumn{1}{r|}{55,060} &
  {\color[HTML]{0563C1} https://zenodo.org/record/4445747} \\ \hline
\multicolumn{1}{|c|}{2020} &
  \multicolumn{1}{l|}{CoCoNut} &
  \multicolumn{1}{l|}{C/Java/Python/JavaScript} &
  \multicolumn{1}{c|}{Bug} &
  \multicolumn{1}{r|}{24,471,491} &
  {\color[HTML]{0563C1} https://github.com/lin-tan/CoCoNut-Artifact/releases} \\ \hline
\multicolumn{1}{|c|}{2020} &
  \multicolumn{1}{l|}{BigFix} &
  \multicolumn{1}{l|}{Java} &
  \multicolumn{1}{c|}{Bug} &
  \multicolumn{1}{r|}{26K} &
  {\color[HTML]{0563C1} https://github.com/AutomatedProgramRepair-2021/dear-auto-fix} \\ \hline
\multicolumn{1}{|c|}{2021} &
  \multicolumn{1}{l|}{CPatMiner} &
  \multicolumn{1}{l|}{Java} &
  \multicolumn{1}{c|}{Bug} &
  \multicolumn{1}{r|}{44K} &
  {\color[HTML]{0563C1} https://github.com/AutomatedProgramRepair-2021/dear-auto-fix} \\ \hline
\multicolumn{1}{|c|}{2021} &
  \multicolumn{1}{l|}{MegaDiff} &
  \multicolumn{1}{l|}{Java} &
  \multicolumn{1}{c|}{Bug} &
  \multicolumn{1}{r|}{240,306} &
  {\color[HTML]{0563C1} https://github.com/monperrus/megadiff} \\ \hline
\multicolumn{1}{|c|}{2021} &
  \multicolumn{1}{l|}{Recoder} &
  \multicolumn{1}{l|}{Java} &
  \multicolumn{1}{c|}{Bug} &
  \multicolumn{1}{r|}{103,585} &
  {\color[HTML]{0563C1} https://github.com/pkuzqh/Recoder} \\ \hline
\multicolumn{1}{|c|}{2022} &
  \multicolumn{1}{l|}{Dataset\_pr} &
  \multicolumn{1}{l|}{Java} &
  \multicolumn{1}{c|}{Bug} &
  \multicolumn{1}{r|}{408,091} &
  {\color[HTML]{0563C1} https://github.com/mxx1219/TRANSFER} \\ \hline
\multicolumn{1}{|c|}{2021} &
  \multicolumn{1}{l|}{TFix dataset} &
  \multicolumn{1}{l|}{JavaScript} &
  \multicolumn{1}{c|}{Bug/Error} &
  \multicolumn{1}{r|}{104,804} &
  {\color[HTML]{0563C1} https://github.com/eth-sri/TFix} \\ \hline
\multicolumn{1}{|c|}{2017} &
  \multicolumn{1}{l|}{DeepFix dataset} &
  \multicolumn{1}{l|}{C} &
  \multicolumn{1}{c|}{Error} &
  \multicolumn{1}{r|}{6,971} &
  {\color[HTML]{0563C1} https://bitbucket.org/iiscseal/deepfix} \\ \hline
\multicolumn{1}{|c|}{2018} &
  \multicolumn{1}{l|}{SynFix dataset} &
  \multicolumn{1}{l|}{Python} &
  \multicolumn{1}{c|}{Error} &
  \multicolumn{1}{r|}{74,818} &
  {\color[HTML]{000000} N.A.} \\ \hline
\multicolumn{1}{|c|}{2018} &
  \multicolumn{1}{l|}{TRACER dataset} &
  \multicolumn{1}{l|}{C} &
  \multicolumn{1}{c|}{Error} &
  \multicolumn{1}{r|}{16,985} &
  {\color[HTML]{0563C1} https://sites.google.com/view/transrepair/source-code-and-dataset} \\ \hline
\multicolumn{1}{|c|}{2020} &
  \multicolumn{1}{l|}{SPoC} &
  \multicolumn{1}{l|}{C++} &
  \multicolumn{1}{c|}{Error} &
  \multicolumn{1}{r|}{18,356} &
  {\color[HTML]{0563C1} https://sumith1896.github.io/spoc} \\ \hline
\multicolumn{1}{|c|}{2021} &
  \multicolumn{1}{l|}{Github-Python} &
  \multicolumn{1}{l|}{Python} &
  \multicolumn{1}{c|}{Error} &
  \multicolumn{1}{r|}{3M} &
  {\color[HTML]{0563C1} https://github.com/michiyasunaga/bifi} \\ \hline
\multicolumn{1}{|c|}{2019} &
  \multicolumn{1}{l|}{Ponta's dataset} &
  \multicolumn{1}{l|}{Java} &
  \multicolumn{1}{c|}{Vul} &
  \multicolumn{1}{r|}{1,068} &
  {\color[HTML]{0563C1} https://github.com/SAP/project-kb} \\ \hline
\multicolumn{1}{|c|}{2020} &
  \multicolumn{1}{l|}{Big-Vul} &
  \multicolumn{1}{l|}{C/C++} &
  \multicolumn{1}{c|}{Vul} &
  \multicolumn{1}{r|}{3,754} &
  {\color[HTML]{0563C1} https://github.com/ZeoVan/MSR\_20\_Code\_vulnerability\_CSV\_Dataset} \\ \hline
\multicolumn{1}{|c|}{2021} &
  \multicolumn{1}{l|}{CVEfixes} &
  \multicolumn{1}{l|}{C/C++} &
  \multicolumn{1}{c|}{Vul} &
  \multicolumn{1}{r|}{5,495} &
  {\color[HTML]{0563C1} https://github.com/secureIT-project/CVEfixes} \\ \hline
\multicolumn{1}{|c|}{2022} &
  \multicolumn{1}{l|}{CrossVul} &
  \multicolumn{1}{l|}{40 PLs} &
  \multicolumn{1}{c|}{Vul} &
  \multicolumn{1}{r|}{5,131} &
  {\color[HTML]{0563C1} https://zenodo.org/record/4734050} \\ \hline
\multicolumn{1}{|c|}{2019} &
  \multicolumn{1}{l|}{CodeSearchNet} &
  \multicolumn{1}{l|}{6 PLs} &
  \multicolumn{1}{c|}{-} &
  \multicolumn{1}{r|}{2M} &
  {\color[HTML]{0563C1} https://github.com/github/CodeSearchNet} \\ \hline
\end{tabular}
}
\end{table}

\bibliographystyle{ACM-Reference-Format}
\bibliography{sample-base}


\end{document}